\documentstyle[prd,aps,epsfig,eqsecnum,amssymb,amsmath,floats]{revtex}
\tighten
\begin{document}
\thispagestyle{empty}

\vskip 2.5 cm

\begin{center}
\section*{}

{\LARGE\bf
Atmospheric Muon Flux at Sea Level, \\ [0.2cm]
    Underground and Underwater}
\\[0.5cm]
{\bf E.~V.~Bugaev\and${}^1$ A.~Misaki\and${}^2$
            V.~A.~Naumov\and${}^{3,4}$
            T.~S.~Sinegovskaya\and${}^3$ S.~I.~Sinegovsky\and${}^3$
            and N.~Takahashi${}^5$}   \\[0.2cm]
${}^1$Institute for Nuclear Research, Russian Academy of
               Science, Moscow 117312, Russia                    \\
         ${}^2$National Graduate Institute for Policy Studies,
                                                  Urawa 338, Japan \\
         ${}^3$Department of Theoretical Physics, Physics Faculty,
               Irkutsk State University, Irkutsk 664003, Russia  \\
         ${}^4$Istituto Nazionale di Fizica Nucleare, Sezione di
               Firenze, Firenze 50125, Italy                     \\
         ${}^5$Department of Electronic and Information System
               Engineering, Faculty of Science and Technology,
               Hirosaki University, Hirosaki 036-8561,  Japan \\[0.2cm]
\end{center}
The vertical sea-level muon spectrum at energies above 1 GeV and the muon
intensities at depths up to 18 km w.e. in different rocks and in water are
calculated. The results are particularly collated with a great
body of the ground-level, underground, and underwater muon data.
In the hadron-cascade calculations, we take into
account the logarithmic growth with energy of inelastic cross sections and
pion, kaon, and nucleon generation in pion-nucleus collisions.
For evaluating the prompt muon contribution to the muon flux, we
apply the two phenomenological approaches to the charm production
problem:  the recombination quark-parton model and the quark-gluon
string model.  We give simple fitting formulas describing our
numerical results. To solve the muon transport equation at large
depths of a homogeneous medium, we use a semianalytical method,
which allows the inclusion of an arbitrary (decreasing) muon spectrum
at the medium boundary and real energy dependence of muon energy
losses.  Our analysis shows that at the depths up  to 6--7 km w.e.,
essentially all underground data on the muon flux correlate with each
other and with the predicted one for conventional ($\pi,K$)-muons, to
within 10\,\%.  However, the high-energy sea-level muon data as well
as the data at high depths are contradictory and cannot be
quantitatively described by a single nuclear-cascade model.
\vskip 0.5 cm
\begin{center}
{\bf A short version was published in Phys. Rev. D58, 05401 (1998).}
\end{center}

\clearpage
\preprint{}
\draft
\title{Atmospheric Muon Flux at Sea Level, Underground, and Underwater}
\author{E.~V.~Bugaev\and${}^1$
        A.~Misaki\and${}^2$
        V.~A.~Naumov\and${}^{3,4,5}$
        T.~S.~Sinegovskaya\and${}^4$
        S.~I.~Sinegovsky\and${}^4$ and
        N.~Takahashi~${}^6$
       }
\address{${}^1$Institute for Nuclear Research, Russian Academy of
                                  Science, Moscow 117312, Russia \\
         ${}^2$National Graduate Institute for Policy Studies,
                                           Urawa 338-8570, Japan \\
         ${}^3$Laboratory of Theoretical Physics,
               Irkutsk State University, Irkutsk 664003, Russia  \\
         ${}^4$Department of Theoretical Physics, Physics Faculty,
               Irkutsk State University, Irkutsk 664003, Russia  \\
         ${}^5$Istituto Nazionale di Fisica Nucleare, Sezione di
                                   Firenze, Firenze 50125, Italy \\
         ${}^6$Department of Electronic and Information System
                 Engineering, Faculty of Science and Technology, \\
                  Hirosaki University, Hirosaki 036-8561, Japan
        }
\pagenumbering{roman}
\setcounter{page}{1}
\tableofcontents
\clearpage
\listoffigures
\listoftables
\maketitle
\begin{abstract}
  The vertical sea-level muon spectrum at energies above 1 GeV and the muon
intensities at depths up to 18 km w.e. in different rocks and in water are
calculated. The results are particularly collated with a great
body of the ground-level, underground, and underwater muon data.
In the hadron-cascade calculations, we take into
account the logarithmic growth with energy of inelastic cross sections and
pion, kaon, and nucleon generation in pion-nucleus collisions.
For evaluating the prompt muon contribution to the muon flux, we
apply the two phenomenological approaches to the charm production
problem:  the recombination quark-parton model and the quark-gluon
string model.  We give simple fitting formulas describing our
numerical results. To solve the muon transport equation at large
depths of a homogeneous medium, we use a semianalytical method,
which allows the inclusion of an arbitrary (decreasing) muon spectrum
at the medium boundary and real energy dependence of muon energy
losses.  Our analysis shows that at the depths up  to 6--7 km w.e.,
essentially all underground data on the muon flux correlate with each
other and with the predicted one for conventional ($\pi,K$) muons, to
within 10\,\%.  However, the high-energy sea-level muon data as well
as the data at high depths are contradictory and cannot be
quantitatively described by a single nuclear-cascade model.
\end{abstract} \pacs{PACS Number(s): 13.85.Tp,13.85.\_t, 96.40\_z,
96.40.Tv} \widetext

\pagenumbering{arabic}
\setcounter{page}{1}
\protect\section{Introduction}
\label{sec:Int}

The flux of cosmic-ray muons in the atmosphere, underground, and
underwater provides a way of testing the inputs of nuclear cascade
models, that is, parameters of the primary cosmic-ray flux (energy
spectrum, chemical composition) and particle interactions at high
energies. In particular, measurements of the muon energy spectra,
angular distributions and the depth-intensity relation (DIR) have
much potential for yielding information about the mechanism of charm
production in hadron-nucleus collisions at energies beyond the
reach of accelerator experiments. This information is a subject
of great current interest for particle physics~\cite{DeRujula93}
and yet is a prime necessity in high-energy and very high-energy
neutrino astronomy~\cite{Gaisser95}. Indeed, the basic and
unavoidable background for many future astrophysical experiments
with full-size underwater or underice neutrino telescopes will be an
effect of the atmospheric neutrino flux of energies from about 1~TeV
to tens of PeV. However, in the absence of a generally recognized
and tried model for charm hadroproduction, the current estimates of
the $\nu_\mu$ and (most notably) $\nu_e$ backgrounds have
inadmissibly wide scatter even at multi-TeV neutrino energies, which
shoots up with energy. At $E_\nu \sim 100$ TeV, different estimates
of the $\nu_\mu$ and $\nu_e$ spectra vary within {\em a few orders
of magnitude} (see Refs.~\cite{Gaisser95,Bugaev89a,Zas93-94,%
Thunman95-96} for reviews and references).

The present state of the art of predicting the atmospheric neutrino
flux seems to be more satisfactory at energies below a few TeV.
However, the theory meets more rigid requirements on accuracy of
the calculations here~\cite{ANproblem}: for an unambiguous treatment
of the current data on up-going (atmospheric neutrino induced) muon
flux, the neutrino flux must be calculated with a 10\,\% accuracy at
least, whereas the uncertainties in the required input data (primary
spectrum, cross sections for light meson production, etc{}.) hinder
to gain these ends. Because of this, a vital question is a
normalization of the calculated (model-dependent) atmospheric
neutrino flux and the muon flux is perhaps the only tool for such a
normalization. The point is that atmospheric muons and neutrinos are
generated in just the same processes. Therefore, the accuracy of the
neutrino flux calculation can be improved by forcing the poorly
known input parameters of the cascade model to fit the data on the
muon flux.

The sea-level muon data obtained by direct measurements with
magnetic spectrometers are crucial but still insufficient for this
purpose. The fact is that numerous sea-level measurements (see e.g.
Refs.~\cite{MARS63,Baber68,Bateman71,Allkofer71,Nandi72,%
            MARS75,MARS77,Rastin84,MASS93,L3Cosmic93,EAS-TOP95}
for the vertical muon flux, Refs.~\cite{MUTRON84,DEIS85} for
near-horizontal flux, and Ref.~\cite{Allkofer84} for a compilation
of the data) are in rather poor agreement to one another, even
though each of the experiments by itself typically has very good
statistical accuracy.  This is true to a greater or lesser extent
everywhere over the whole energy region accessible to the
ground-based installations.

On the other hand, a quite representative array of data on cosmic-ray
muon DIR in rock and, to a lesser extent, in water has been
accumulated. Underground muon experiments may number in the
tens in a span of sixty years (see Refs.~%
\cite{Wilson38,Clay39,Bollinger50,Randall51,Avan55,KGF64a,KGF65-71,%
      Castagnoli65,Stockel69,Bergamasco71,Crookes73,Utah,MtBlanc78,%
      ERPM,Castagnoli86,BNO87,BNO90,KGF90a,Frejus89-90,Frejus96,%
      NUSEX90,SOUDAN90,Kasahara95,MACRO95,LVD95,LVD97,LVD98}
and also~\cite{OldRev,Bugaev70,Crouch87} for reviews and further
references). It should be noted that the results of many early
measurements, specifically those performed at shallow depths, have
not lost their significance today, considering that modern
experiments principally aim at greater depths. Underwater muon
experiments have over 30 years of history~%
\cite{Higashi66,Davitaev70-74,Rogers84,Fyodorov85,DUMAND90,%
NESTOR93,Baikal93,Baikal95,AMANDA99} and it is believed that they
will gain in importance with the progress of high-energy neutrino
telescopes.

It may be somewhat unexpected but the underground data
are more self-consistent in comparison with ground-level data,
at least for depths to about 6 km w.e. (corresponding roughly to
3--4~TeV of muon energy at sea level) and hence they provide a
useful check on nuclear cascade models. There is a need to
piece together all these data in order to extract some physical
inferences thereof. Also, it would be useful to correlate the
underground and underwater data with the results of the mentioned
direct measurements of the sea-level muon spectrum~%
\cite{MARS63,Baber68,Bateman71,Allkofer71,Nandi72,MARS75,MARS77,%
      Rastin84,MASS93,L3Cosmic93,EAS-TOP95}
as well as with the data deduced by indirect routes~%
\cite{MtBlanc78,BNO87,BNO90,MACRO95,KGF64b,KGF86,Collapse85,%
      KGF90b,BNO92,Frejus94,MSU94}.

It is the purpose of this paper to discuss the above-mentioned
data on the vertical muon flux at sea level, underground, and
underwater in the context of a single calculation, with emphasis
on the prompt muon problem. The implementation of the results to
the normalization of the high-energy atmospheric neutrino flux
will be discussed elsewhere~\cite{Kl3}.

In Section~\ref{sec:Cascade} we discuss the employed model for the
primary spectrum and composition as well as the nuclear-cascade
model for production and propagation of high-energy nucleons, pions,
and kaons in the atmosphere.
Some required formulas for the atmospheric muon flux are given in
Section~\ref{sec:Muons}; at the end of that Section, we give a
simple parametrization for the calculated vertical spectrum of
conventional ($\pi,K$) muons at sea level.
The models for charm hadroproduction, those are used in the present
work to make an estimate of the prompt-muon (PM) contribution, are
the concern of Section~\ref{sec:PM}; the recombination
quark-parton model is considered with some details. At the end of
this Section, we present simple parametrizations for the predicted
differential and integral PM spectra.
In Section~\ref{sec:SLM} we compare our predictions for the vertical
muon spectra (differential and integral) with the direct and
indirect data at sea level.
Section~\ref{sec:Prop} is concerned with muon propagation through
matter. Calculation of the muon intensity at large depths is a
rather nontrivial problem even though the muon energy spectrum at
the medium boundary is assumed to be known; we briefly sketch our
semianalytical approach to that problem.
The comparison between the calculated muon DIR and the aforecited
underground and underwater data is fully considered in
Section~\ref{sec:DIR}.
In Appendix~\ref{app:Decay} we give the model formulas for the
spectra of muons from inclusive semileptonic decays of a $D$ meson
and $\Lambda_c$ hyperon in the lab{}. frame.
In Appendix~\ref{app:MuInt} we present a summary for the
differential cross sections of the muon--matter interactions (direct
$e^+e^-$ pair production, bremsstrahlung, photonuclear interaction)
as well as (for completeness sake) Sternheimer's formula for
ionization energy loss.
Our conclusions are presented in Section~\ref{sec:Concl}.

\protect\section{Nuclear-cascade model}
\label{sec:Cascade}

\protect\subsection{Primary spectrum and composition}
\label{ssec:Prim}

For energies above 1\,TeV we use the semiempirical model for the
integral primary spectrum proposed by Nikol'sky
{\em et al.}~\cite{Nikolsky84} (from here on we will call it
``NSU model''):
\begin{equation}\label{NSUint}
F(\geq E_0) = F_0 E_0^{-\gamma}\sum_AB_A
\left(1+\delta_A\frac{E_0}{A}\right)^{-\ae}.
\end{equation}
Here $E_0$ is the energy per particle in GeV, $F_0 = 1.16$
cm$^{-2}$s$^{-1}$sr$^{-1}$, $\gamma = 1.62\,(\pm 0.03)$, and
$\ae = 0.4$. The $\delta_A$'s specify the region of the
``knee'' in the primary spectrum. We adopt $\delta_p=6\times10^{-7}$
and $\delta_{A \geq 4} = 10^{-5}$. These values correspond to
the hypothesis which attributes the change in the energy spectrum of
the primaries at $E_0 \gtrsim 10^3$~TeV to photodesintegration of
nuclei with pion photoproduction by photons with energy $\sim 70$~eV
inside the cosmic ray sources. The chemical composition is given
with the following values for $B_A$:
$B_1    = 0.40\,(\pm 0.03)$,
$B_4    = 0.21\,(\pm 0.03)$,
$B_{15} = 0.14\,(\pm 0.03)$,
$B_{26} = 0.13\,(\pm 0.03)$, and
$B_{51} = 0.12\,(\pm 0.04)$
for the five standard groups of nuclei. The numerical values of $A$
indicate the average atomic weights in the groups.
The corresponding differential spectrum is given by
\begin{equation}\label{NSUdif}
\frac{d F}{d E_0} = \gamma F_0 E_0^{-(\gamma+1)}
\sum_AB_A\left(1+\delta_A\frac{E_0}{A}\right)^{-\ae}
\left[1+\frac{\ae\delta_AE_0/A}{\gamma(1+\delta_AE_0/A)}\right].
\end{equation}

The NSU approximation has been deduced from an analysis of
fluctuations in the relative number of electrons and muons in
extensive air showers and corresponds to the data on absolute
intensities of primary protons and various nuclei at energies
$E_0 \geq 1,\,10^3,\,10^6$ TeV/particle, and also to the data on the
shape of the integral spectrum in the vicinity of the knee (see
Ref.~\cite{Nikolsky84} for specific sources of the data).

The model, on the whole, fits the modern data on the primary spectrum
and composition from about 100~GeV/particle up to 100~EeV/particle.
Specifically, at $E_0 \lesssim 10^3$~TeV/particle, the model fits
reasonably well the recent results of the COSMOS satellite
experiment~\cite{SOKOL93}, the JACEE balloon
experiment~\cite{JACEE91}, and the BASJE air-shower
experiment~\cite{Teshima93}. On the other hand, there is a strong
discrepancy between the NSU model and the recent data of the Japan
balloon-borne emulsion chamber experiment~\cite{Ichimura93}, which
indicates a milder knee shape than that found in the previous
experiments, although the data of Ref.~\cite{Ichimura93} for the
nuclear composition agree with the NSU model at $E_0 \gtrsim
10$~TeV/particle. The data for the spectrum and composition are most
inconsistent in the vicinity of the knee
[$(10^2\div10^4)$~TeV/particle]. Scanty experimental data favor a
pure proton composition at $E_0\gtrsim10^4$~TeV/particle rather than
almost fixed composition predicted by the NSU model. In the
connection it should be noted that an essential contribution to the
deep underground flux of muons, in particular, ones originated from
the decay of charmed hadrons (at depths below $\sim10$~km~w.e.), is
given by the primaries with energies from the knee region. Thus the
long-standing problem of the knee is closely allied to the PM
problem. At the same time, the total intensity of underground/water
muons is scarcely affected by the region $E_0\gg10^4$~TeV/particle.
Thus we will not discuss here the problem of the primary spectrum
and composition at super-high energies (see
Refs.~\cite{Teshima93,PrimSpRev} for current reviews).

\protect%
\subsection{Nuclear cascade at high energies: Basic assumptions}
\label{ssec:Ass}

Our nuclear-cascade calculations at high energies are based on the
analytical model of Ref.~\cite{Vall86} which describes well all
available experimental data on hadron spectra for various atmospheric
depths and for energies from about 1~TeV up to about 600~TeV. The
processes of regeneration and overcharging of nucleons, and charged
pions, as well as production of kaons, nucleons, and charmed
particles in pion--nucleus collisions have been properly accounted
for. Let us outline the basic assumptions of the model.

(i) The nuclear component of the primary spectrum is replaced with a
superposition of free nucleons. Eq.~(\ref{NSUdif}) transforming to
the equivalent nucleon spectrum yields the following
differential energy spectra of protons and neutrons:
\begin{eqnarray*}
\frac{dF_p}{dE_N} &\equiv& {\cal D}_p^0(E_N) = {\cal D}_1(E_N)
                      +\frac{1}{2}\sum_{A\geq4}{\cal D}_A(E_N), \\
\frac{dF_n}{dE_N} &\equiv& {\cal D}_n^0(E_N) =
                       \frac{1}{2}\sum_{A\geq4}{\cal D}_A(E_N)
\end{eqnarray*}
Here $E_N$ is the nucleon energy (in GeV),
\[
{\cal D}_A(E_N) =
  \frac{C_A{\cal D}_0E_N^{-(\gamma+1)}}{(1+\delta_AE_N)^{\ae}}
  \left[1+\frac{\ae\delta_AE_N}{\gamma(1+\delta_AE_N)}\right],
\]
${\cal D}_0 = \gamma B_1F_0 = 0.75$\,%
cm${}^{-2}$s${}^{-1}$sr${}^{-1}$(GeV/nucleon)${}^{-1}$, and
$C_A = A^{1-\gamma}B_A/B_1$ ($A = 1,\;4,\;15,\;26,\;51$).
Outside the knee region we use the asymptotic formulas:
\begin{equation} \label{Power}
{\cal D}_A(E_N) = \left\{
\begin{aligned} C_A{\cal D}_0E_N^{-(\gamma+1)}
                         & \text{ for }  E_N \ll E_N^{(1)}, \\
1.25\delta_A^{-\ae}C_A{\cal D}_0E_N^{-{(\gamma+\ae+1)}}
                         & \text{ for }  E_N \gg E_N^{(2)},
\end{aligned}\right.
\end{equation}
where $E_N^{(1)} = 6.5/\delta_A$~GeV/nucleon and
      $E_N^{(2)} = 0.6/\delta_A$~GeV/nucleon. A numerical procedure
is applied to smooth out the calculated spectra of secondary
hadrons at energies around the knee region.

(ii) We assume a logarithmic growth with energy of the total
inelastic cross sections $\sigma_{iA}^{\rm inel}$ for interactions
of a hadron $i$ with a nuclear target $A$. Such a dependence arises
from a model for elastic amplitude of hadron--hadron collisions,
based on the conception of double pomeron with the supercritical
intercept~\cite{Vall86}.  For simplicity sake we will use also
another consequence of this model:  the asymptotic equality of the
inelastic cross sections for any hadron. Thereby
\begin{equation} \label{sigma}
\sigma_{iA}^{\rm inel}(E) = \sigma_{iA}^0
+\sigma_A\ln\left(\frac{E}{E_1}\right) \qquad (i = N,\pi,K,\ldots)
\end{equation}
at $E \geq E_1 = 1$~TeV. The following values of the parameters are
adopted: $\sigma_A =  19$~mb, $\sigma_{NA}^0 = 275$~mb ($N = p,n$),
$\sigma_{\pi^\pm A}^0 = 212$~mb, $\sigma_{KA}^0 = 183$~mb
($K = K^\pm, K^0,\overline{K}{}^0$).

(iii) It is assumed that Feynman scaling holds in the
fragmentation region of the inclusive processes $iA\rightarrow fX$,
where $i=p,n,\pi^\pm$, $f=p,n,\pi^\pm,K^\pm,K^0,\overline{K}{}^0$,
and $A$ is the ``air nucleus''. So the normalized invariant
inclusive cross sections $\left(E/\sigma_{iA}^0\right)
d^3\sigma_{iA\rightarrow fX}/d^3p$ are energy independent at large
$x$ (where $x$ is the ratio of the final particle energy to that of
the initial one). Let us denote
\[
{\cal W}_{fi}(x) = \frac{\pi}{\sigma_{iA}^0}
                   \int_0^{(p_T^{\max})^2}\frac{E}{p_L}
      \left(E\frac{d^3\sigma_{iA\rightarrow fX}}{d^3p}\right)dp_T^2.
\]
Then the fractional moments (``$Z$-factors'') defined by
\begin{equation}\label{Z}
Z_{fi}(\gamma) = \int_0^1x^{\gamma-1}{\cal W}_{fi}(x)dx
\end{equation}
are constant inside the regions with constant exponent $\gamma$ (that
is outside the knee energy region in the primary spectrum).
Table \ref{tab:Z1} shows fractional moments $Z_{fi}(\gamma)$ for the
two values of $\gamma$ in the case where the incident particle $i$
is a proton or $\pi^+$ meson and $f = p,n,\pi^\pm,K^\pm,K_L^0$. The
moments for $i=n$ and $\pi^-$ can be derived using the well-known
isotopic relations for the inclusive cross sections. To calculate
the $Z$-factors for all reactions except $\pi A\rightarrow NX$ and
$\pi A\rightarrow KX$, we used a parametrization of ISR data put
forward by Minorikawa and Mitsui~\cite{Minorikawa78-84}. The
quantities $Z_{N\pi}$ and $Z_{K\pi}$ were calculated from the two
central moments, $\langle x \rangle$ and $\langle x^2 \rangle$, for
the inclusive distributions obtained by Anisovich
{\em et al.}~\cite{Anisovich84} in the framework of quasinuclear
quark model.
\begin{table}[htb]
\protect\caption[Fractional moments of inclusive distributions of
                 nucleons, pions, and kaons]
                {Fractional moments $Z_{fi}(\gamma)$ of inclusive
                 distributions of nucleons, pions, and kaons for
                 the two values of $\gamma$.
\label{tab:Z1}}
\center{\begin{tabular}{cccccccc}
       &      &      &      & $f$  &      &      & \\ \cline{2-8}
  $i$  & $p$  & $n$  &$\pi^+$&$\pi^-$& $K^+$&$K^-$&$K_L^0$ \\ \hline
       &      &      & & $\gamma = 1.62$& &      &         \\
  $p$  &0.1990&0.0763&0.0474&0.0318&0.0067&0.0023& 0.0045  \\
$\pi^+$&0.0070&0.0060&0.1500&0.0552&0.0120&0.0120& 0.0100  \\ \hline
       &      &      & & $\gamma = 2.02$& &      &         \\
  $p$  &0.1980&0.0585&0.0257&0.0162&0.0039&0.0012& 0.0026  \\
$\pi^+$&0.0060&0.0040&0.1480&0.0346&0.0100&0.0100& 0.0080  \\
\end{tabular}}
\end{table} 

(iv) The kaon regeneration (i.e. the processes $K^\pm A \rightarrow
K^\pm X$, $K^{\pm }A\rightarrow K^0X$, etc{}.) is disregarded in our
calculations. Also, we neglect the nucleon and pion production in
kaon--nucleus collisions as well as pion production in kaon decays,
which makes it possible to split up the total system of the transport
equations into nucleon-pion part and kaon one. Our estimations show
that the inclusion of the aforementioned effects will cause the muon
flux to increase, but no more than by a few per cent. It is clear
that similar effects for charmed particles are completely negligible.

(v) At the stage of nuclear cascade (but, of course, not at the muon
production stage) the decay of $\pi^\pm$ mesons (critical energy
$E_\pi^{\rm cr} \simeq 0.12$~TeV) is neglected for directions close
to vertical at pion energies $\gtrsim 1$~TeV. This approximation
greatly simplifies the description of the pion regeneration and the
production of nucleons, kaons, and charmed particles in
pion--nucleus collisions.

\protect\subsection{Nucleon-pion cascade equations}
\label{ssec:Pi-N}

In line with the above-listed assumptions, the $4\times4$ system of
transport equations for the nucleon-pion part of the cascade can
be written
\begin{equation}\label{PiN}
\left[\frac{\partial}{\partial h}
       +\frac{1}{\lambda_i(E)}\right]{\cal D}_i(E,h) =
\sum_j\frac{1}{\lambda_j^0}\int_0^1
{\cal W}_{ij}(x){\cal D}_j\left(\frac{E}{x},h\right)\frac{dx}{x^2},
\end{equation}
($i,j = p,n,\pi^+,\pi^-$) with the boundary conditions
\[
{\cal D}_p(E,0)       = {\cal D}_p^0(E), \quad
{\cal D}_n(E,0)       = {\cal D}_n^0(E), \quad
{\cal D}_{\pi^+}(E,0) = {\cal D}_{\pi^-}(E,0) = 0.
\]
Here ${\cal D}_i(E,h)$ is the differential energy spectrum of
particles $i$ at the atmospheric depth $h$,
\[
\lambda_i(E) = \frac{1}{N_0\sigma_{iA}^{\rm inel}(E)}, \qquad
\lambda_i^0  = \frac{1}{N_0\sigma_{iA}^0},
\]
and $N_0$ is the number of target nuclei in 1~g of air.

The solution to the system~(\ref{PiN}) can be found as an expansion
in powers of the dimensionless parameter $h/\lambda_A$, where
$\lambda_A = 1/(N_0\sigma_A) \simeq 14.5\lambda_N^0$. Within the
power-behaved regions of the primary spectrum described by
Eq.~(\ref{Power}), the solution is of the form
\begin{eqnarray*}
{\cal D}_p(E,h)      & = &\frac{1}{2}\left[N^+(E,h)+N^-(E,h)\right],
                                                             \quad
{\cal D}_n(E,h)        =  \frac{1}{2}\left[N^+(E,h)-N^-(E,h)\right],
                                                              \\
{\cal D}_{\pi^+}(E,h)& = &\frac{1}{2}\left[\Pi^+(E,h)+
                                           \Pi^-(E,h)\right],\quad
{\cal D}_{\pi^-}(E,h)  =  \frac{1}{2}\left[\Pi^+(E,h)-
                                           \Pi^-(E,h)\right],
\end{eqnarray*}
where
\begin{eqnarray*}
N^\kappa(E,h)   & = & \frac{{\cal D}_p^0(E)+
    \kappa{\cal D}_n^0(E)}{2j^\kappa}
    \sum_{\kappa'}\left(j^\kappa+\kappa'\right)
    \exp\left[-\frac{h}{\Lambda_{N\pi}^{\kappa\kappa'}(E)}\right]
        \left[1+{\cal O}\left(\frac{h}{\lambda_A}\right)\right],\\
\Pi^\kappa(E,h) & = & \frac{{\cal D}_p^0(E)+
    \kappa{\cal D}_n^0(E)}{2j^\kappa}Z_{\pi N}^\kappa(\gamma)
    \left(\frac{\Lambda_\kappa}{\lambda_N^0}\right)
    \sum_{\kappa'}(-\kappa')
    \exp\left[-\frac{h}{\Lambda_{N\pi}^{\kappa\kappa'}(E)}\right]
        \left[1+{\cal O}\left(\frac{h}{\lambda_A}\right)\right],
\end{eqnarray*}
\[
\frac{1}{\Lambda_{N\pi}^{\kappa\kappa'}(E)} =
\frac{1+\kappa'j^\kappa(E)}{2\Lambda_N  ^\kappa(E)}+
\frac{1-\kappa'j^\kappa(E)}{2\Lambda_\pi^\kappa(E)} \quad
 (\kappa,\kappa' = \pm),
\]
\[
j^\kappa(E)  = \sqrt{1+\frac{Z_{\pi N}^\kappa
       Z_{N\pi}^\kappa \Lambda_\kappa^2}{ \lambda_N^0\lambda_\pi^0}}
         \simeq     1+\frac{Z_{\pi N}^\kappa
       Z_{N\pi}^\kappa \Lambda_\kappa^2}{2\lambda_N^0\lambda_\pi^0},
\]
\[
\frac{1}{\Lambda_\kappa} =
                 \frac{1-Z_{NN}    ^\kappa}{2\lambda_N  ^0}-
                 \frac{1-Z_{\pi\pi}^\kappa}{2\lambda_\pi^0}, \qquad
\frac{1}{\Lambda_i^\kappa(E)} = \frac{1}{\lambda_i(E)}-
                 \frac{Z_{ii}^\kappa}{\lambda_i^0},
\]
\[
Z_{NN}^\kappa     = Z_{pp}        +\kappa Z_{np},        \qquad
Z_{\pi\pi}^\kappa = Z_{\pi^+\pi^+}+\kappa Z_{\pi^+\pi^-},
\]
\[
Z_{\pi N}^\kappa  = Z_{\pi^+p}+\kappa Z_{\pi^+n},        \qquad
Z_{N\pi} ^\kappa  = Z_{p\pi^+}+\kappa Z_{p\pi^-}.
\]
The functions $\Lambda_{N\pi}^{\kappa\kappa'}(E)$ can be treated as
the generalized absorption ranges. Not counting the processes of
nucleon-antinucleon pair production by pions, the formulas for
$\Lambda_{N\pi}^{\kappa\kappa'}(E)$ are very simple:
\[
\Lambda_{N\pi}^{\kappa+}(E) = \Lambda_N  ^\kappa(E), \qquad
\Lambda_{N\pi}^{\kappa-}(E) = \Lambda_\pi^\kappa(E),
\]
and thus
\[
 N ^\kappa(E,h)\propto
               \exp\left[-\frac{h}{\Lambda_N^\kappa(E)}\right],\quad
\Pi^\kappa(E,h)\propto
               \exp\left[-\frac{h}{\Lambda_\pi^\kappa(E)}\right]
              -\exp\left[-\frac{h}{\Lambda_N  ^\kappa(E)}\right].
\]
The ${\cal O}\left(h/\lambda_A\right)$ corrections were calculated
in Ref.~\cite{Vall86} and it was demonstrated that they became
important at $h > 500-600$~g/cm${}^2$. However these corrections are
of no significance for present purposes, because the greater part of
the atmospheric muon flux is generated on the depths
$h\lesssim300$~g/cm${}^2$.

\protect\subsection{Kaon production and transport}
\label{ssec:Kaons}

Kaon decay cannot be neglected even at very high energies; as a
result the differential energy spectra of kaons,
${\cal D}_K(E,h,\vartheta)$, depend on zenith angle $\vartheta$.
In line with approximation (iv) of Section~\ref{ssec:Ass} and
assuming isothermality of the atmosphere, the transport equation
for kaons may be written as
\begin{equation} \label{Ktrans}
\left[\frac{\partial}{\partial h}+\frac{1}{\lambda_K(E)}
+\frac{E_K^{\rm cr}(\vartheta)}{Eh}\right]{\cal D}_K(E,h,\vartheta)
                 = G_K(E,h), \quad (K = K^\pm, K_L^0),
\end{equation}
where
$E_K^{\rm cr}(\vartheta) = m_K H_0\sec\vartheta/\tau_K$ is the
kaon critical energy (at $\vartheta\lesssim75^\circ$), $m_K$ and
$\tau_K$ are the kaon mass and lifetime, and $H_0\simeq6.44$~km is
the parameter of the isothermal atmosphere. The source function
$G_K(E,h)$ describes kaon production in $NA$ and $\pi A$ collisions.
Taking into account the explicit form of the nucleon and pion
spectra outside the knee region (see Section~\ref{ssec:Pi-N}),
we have
\begin{eqnarray} \label{Gkaon}
G_K(E,h) & = &  \sum_{i=p,n,\pi^+,\pi^-}\frac{1}{\lambda_i^0}
\int_0^1{\cal W}_{Ki}(x){\cal D}_i\left(\frac{E}{x},h\right)
                            \frac{dx}{x^2} \nonumber \\  &\simeq&
\frac{1}{2}\sum_\kappa\left[\frac{Z_{KN}^\kappa(\gamma_h)}
                             {\lambda_N^0}N^\kappa(E,h)+
                 \frac{Z_{K\pi}^\kappa(\gamma_h)}
                             {\lambda_\pi^0}\Pi^\kappa(E,h)\right],
\end{eqnarray}
where
\[
Z_{KN}  ^\kappa(\gamma_h)=Z_{Kp}(\gamma_h)
      +\kappa Z_{Kn}(\gamma_h), \qquad
Z_{K\pi}^\kappa(\gamma_h)=Z_{K^+\pi^+}(\gamma_h)
      +\kappa Z_{K^+\pi^-}(\gamma_h),
\]
and $\gamma_h=\gamma+h/\lambda_A$.

Upon integrating Eq.~(\ref{Ktrans}) with the source
function~(\ref{Gkaon}) and neglecting the weak $h$-dependence of
the kaon $Z$-factors we obtain
\begin{eqnarray}
{\cal D}_K(E,h,\vartheta)                                &  =   &
\int_0^h\exp\left[-\frac{(h-h')}{\lambda_K(E)}\right]
\left(\frac{h'}{h}\right)^{E_K^{\rm cr}(\vartheta)/E}\!
                              G_K(E,h')\,dh' \nonumber\\ &\simeq&
      \Gamma\left(\varepsilon_K(\vartheta)\right)
      \exp\left[-\frac{h}{\lambda_K(E)}\right]\sum_\kappa
 \left[Z_{KN}^\kappa(\gamma)\left(\frac{h}{\lambda_N^0}\right)
            N_K^\kappa(E,h,\vartheta)
    +Z_{K\pi}^\kappa(\gamma)\left(\frac{h}{\lambda_\pi^0}\right)
               \Pi_K^\kappa(E,h,\vartheta)\right], \label{Kapp}
\end{eqnarray}
\begin{eqnarray*}
N_K^\kappa(E,h,\vartheta)   & = &
\frac{{\cal D}_p^0(E)+\kappa{\cal D}_n^0(E)}{4j^\kappa}
                  \sum_{\kappa'}\left(j^\kappa+\kappa'\right)
              \overstar{\gamma}\left(\varepsilon_K(\vartheta),
                  \frac{h}{\Lambda_K^{\kappa\kappa'}(E)}\right)
      \left[1+{\cal O}\left(\frac{h}{\lambda_A}\right)\right], \\
\Pi_K^\kappa(E,h,\vartheta) & = &
      \frac{{\cal D}_p^0(E)+\kappa{\cal D}_n^0(E)}{4j^\kappa}
      Z_{\pi N}^\kappa(\gamma)\left(\frac{\Lambda_\kappa}
     {\lambda_N^0}\right)\sum_{\kappa'}(-\kappa')\overstar{\gamma}
      \left(\varepsilon_K(\vartheta),\frac{h}
     {\Lambda_K^{\kappa\kappa'}(E)}\right)
      \left[1+{\cal O}\left(\frac{h}{\lambda_A}\right)\right].
\end{eqnarray*}
Here $\varepsilon_K(\vartheta) = E_K^{\rm cr}(\vartheta)/E+1$,
$\Gamma$ is the gamma-function, $\overstar{\gamma}$ is the
incomplete gamma-function,
\[
\overstar{\gamma}(\varepsilon,z) = \frac{1}{\Gamma(\varepsilon)}
                                 \int_0^1t^{\varepsilon-1}e^{-zt}dt,
\]
and
\[
\frac{1}{\Lambda_K^{\kappa\kappa'}(E)} =
\frac{1}{\Lambda_{N\pi}^{\kappa\kappa'}(E)}-\frac{1}{\lambda_K(E)}.
\]
The approximate solution~(\ref{Kapp}) is valid at $E\lesssim40$~TeV
(with $\gamma=1.62$) and $E\gtrsim2\times10^3$~TeV (with
$\gamma=2.02$). The ${\cal O}\left(h/\lambda_A\right)$ corrections
are small at $h\lesssim500$~g/cm${}^2$ and the derived solution will
suffice for our purpose.

\protect%
       \subsection{Nuclear cascade at low and intermediate energies}
\label{ssec:NClow}

For the ``low-energy part'' of the nuclear cascade
($E_0 \lesssim 1$~TeV/particle), we adopt the relevant results of
Refs.~\cite{NB,BN} obtained within a rather circumstantial
nuclear-cascade model. The model includes the effects of strong
scaling violation in hadron--nucleus and nucleus--nucleus
collisions, ionization energy losses of charged particles,
temperature gradient of the stratosphere, geomagnetic cutoffs and
solar modulation of the primary spectrum.
The computational results were verified considering a great body of
data on the secondary nucleons, mesons, and muons in wide ranges of
geographical latitudes and altitudes in the atmosphere. The model
was also tested using low-energy data on contained events
observed with several underground neutrino detectors.

Since the key features of the model were discussed in several papers
(see e.g. Ref.~\cite{ANproblem} and references therein), we shall not
dwell upon the question here. Only one point needs to be made.
The geomagnetic effects for the sea-level muon flux are sizable up
to about 5 GeV. However, later on, we are going to deal with the
muon data at high latitudes that are insensitive to the geomagnetic
cutoff. The same all the more true of the solar modulation effects.

\protect\section{Conventional muon flux}
\label{sec:Muons}

Our calculation of the muon production and propagation through the
atmosphere is based on the standard continuos loss approximation.
Our interest is in the muon flux at momenta $p\gtrsim1$~GeV/c. Thus
the ${\cal O}\left(m_\mu^2/p^2\right)$ effects can be neglected.
For simplicity, the nonisothermality of the atmosphere will
be ignored in the formulas which follow (see Ref.~\cite{BN} for the
corresponding corrections).

Let ${\cal D}_\mu(E,h,\vartheta)$ be the differential energy
spectrum of muons at depth $h$ and zenith angle $\vartheta$
and $\beta_\mu(E) = -dE/dh = a_\mu(E)+b_\mu(E)E$ be the rate of
the muon energy loss due to ionization ($a_\mu(E)$) and radiative
and photonuclear interactions in the air ($b_\mu(E)E$). The muon
transport equation is
\begin{equation}\label{m1}
\left[\frac{\partial}{\partial h}+\frac{E_\mu^{\rm cr}(\vartheta)}
         {Eh}\right]{\cal D}_\mu(E,h,\vartheta) =
         \frac{\partial}{\partial E}\left[\beta_\mu(E)
   {\cal D}_\mu(E,h,\vartheta)\right]+G_\mu^{\pi,K}(E,h,\vartheta)
\end{equation}
with
\begin{eqnarray}\label{GpiK}
G_\mu^{\pi,K}(E,h,\vartheta) & = & \sum_{M=\pi^\pm,K^\pm}
B(M_{\mu2})\frac{E_M^{\rm cr}(\vartheta)}{hE}
\left(1-\frac{m_\mu^2}{m_M^2}\right)^{-1}\int_{m_\mu^2/m_M^2}^1
{\cal D}_M\left(\frac{E}{x},h,\vartheta\right)dx \nonumber\\
                             &   & +\sum_{K=K^\pm,K_L^0}
B(K_{\mu3})\frac{E_K^{\rm cr}(\vartheta)}{hE}
                                 \int_{x_K^-}^{x_K^+}F_K^\mu(x)
{\cal D}_K\left(\frac{E}{x},h,\vartheta\right)dx,
\end{eqnarray}
Here $E_\mu^{\rm cr}(\vartheta) = m_\mu H_0\sec\vartheta/\tau_\mu
\simeq 1.03\sec\vartheta$~GeV is the muon critical energy,
$B(M_{\mu2(3)})$ are the branching ratios for the
$\pi_{\mu2}$, $K_{\mu2}$, and $K_{\mu3}$ decays, $F_K^\mu(x)$ is
the muon spectral function for $K_{\mu3}$ decay, and
\[
x_K^\mp = 2m_\mu^2\left[\left(m_K^2-m_\pi^2+m_\mu^2\right)
                      \pm\sqrt{\left(m_K^2-m_\pi^2+m_\mu^2\right)^2
                                      -4m_\mu^2m_K^2}\,\right]^{-1},
\]
The explicit form of $F_K^\mu(x)$ is rather cumbersome but there
is no need to write it out because the $K_{\mu3}$ decay contribution
to the muon flux does not exceed 2.5\,\%~\cite{footnoteKl3effect}.

The solution to Eq.~(\ref{m1}) is given by
\[
{\cal D}_\mu(E,h,\vartheta) = \int_0^h W_\mu(E,h,h',\vartheta)
        G_\mu^{\pi,K}\left({\cal E}(E,h-h'),h',\vartheta\right)dh'.
\]
Here
\begin{equation}\label{W}
W_\mu(E,h,h',\vartheta) =
        \frac{\beta_\mu\left({\cal E}(E,h-h')\right)}{\beta_\mu(E)}
             \exp\left[-\int_{h'}^h\frac{E_\mu^{\rm cr}(\vartheta)}
                         {{\cal E}(E,h-h'')}\frac{dh''}{h''}\right]
\end{equation}
and ${\cal E}(E,h)$ is the root of the integral equation
\[
\int_E^{{\cal E}}\frac{dE}{\beta_\mu(E)} = h,
\]
that is, the energy which a muon must have at the top of the
atmosphere in order to reach depth $h$ with energy $E$. As our
analysis demonstrates, the weak (logarithmic) energy dependence of
the functions $a_\mu$ and $b_\mu$ is only essential for
near-horizontal directions and can be disregarded with an accuracy
better than 3\,\% for the directions close to vertical. In this
approximation
\[
{\cal E}(E,h) = \left(E+\frac{a_\mu}{b_\mu}\right)\exp(b_\mu h)
                        -\frac{a_\mu}{b_\mu} \quad \mbox{and} \quad
                         \frac{\beta_\mu\left({\cal E}(E,h)\right)}
                                     {\beta_\mu(E)} = \exp(b_\mu h).
\]
In numerical calculations we use $a_\mu = 2.0$~MeV$\cdot$cm${}^2$/g
and $b_\mu = 3.5\times10^{-6}$~cm${}^2$/g. Eq.~(\ref{W}) can be
simplified in the two particular cases. At
$E\gg E_\mu^{\rm cr}(\vartheta)$ the muon decay can be neglected, so
\[
W_\mu\left(E,h,h',\vartheta\right)
                     \simeq \exp\left[b_\mu\left(h-h'\right)\right].
\]
At $E \ll a_\mu/b_\mu\approx0.57$~TeV, the radiative and
photonuclear energy loss are inessential and thus
\[
W_\mu\left(E,h,h',\vartheta\right) \simeq
       \left[\left(\frac{h'}{h}\right)\left(\frac{E}{E+a_\mu(h-h')}
  \right)\right]^{E_\mu^{\rm cr}(\vartheta)/\left(E+a_\mu h\right)}.
\]

The combined results of our calculations for the vertical
momentum spectrum of conventional muons at sea level can be
summarized (with a 2\,\% accuracy) by the following fitting formula:
\begin{equation}\label{fit1}
{\cal D}_\mu\left(p,h=1030\;\mbox{g/cm}^2,\vartheta=0^\circ\right) =
Cp^{-(\gamma_0+\gamma_1\log p+\gamma_2\log^2p+\gamma_3\log^3p)},
\end{equation}
with parameters presented in Table \ref{tab:fit1} for a few momentum
ranges [here $p$ is the muon momentum in GeV/c and
${\cal D}_\mu(p,h,\vartheta) = (p/E){\cal D}_\mu(E,h,\vartheta)$].
\begin{table}[htb]
\protect\caption[Parameters of the fitting formula~%
                 (\protect\ref{fit1}) for the vertical energy
                 spectrum of conventional muons at sea level]
                {Parameters of the fitting formula~%
                 (\protect\ref{fit1}) for the vertical energy
                 spectrum of conventional muons at sea level.
\label{tab:fit1}}
\center{\begin{tabular}{cccccc}
Momentum range (GeV/c) &
                 $C$ (cm${}^{-2}$s${}^{-1}$sr${}^{-1}$GeV${}^{-1}$)
       & $\gamma_0$ & $\gamma_1$ & $\gamma_2$ & $\gamma_3$ \\\hline
$1                \div 9.2765\times10^2$  & $2.950\times10^{-3}$
       &   0.3061   &   1.2743   &   -0.2630  &   0.0252  \\
$9.2765\times10^2 \div 1.5878\times10^3$  & $1.781\times10^{-2}$
       &   1.7910   &   0.3040   &    0       &   0        \\
$1.5878\times10^3 \div 4.1625\times10^5$  & $1.435\times10^{ 1}$
       &   3.6720   &   0        &    0       &   0        \\
$ > 4.1625\times10^5$                     &            $10^{ 3}$
       &   4        &   0        &    0       &   0
\end{tabular}}
\end{table} 

\begin{figure}[!t]
\center{\mbox{\epsfig{file=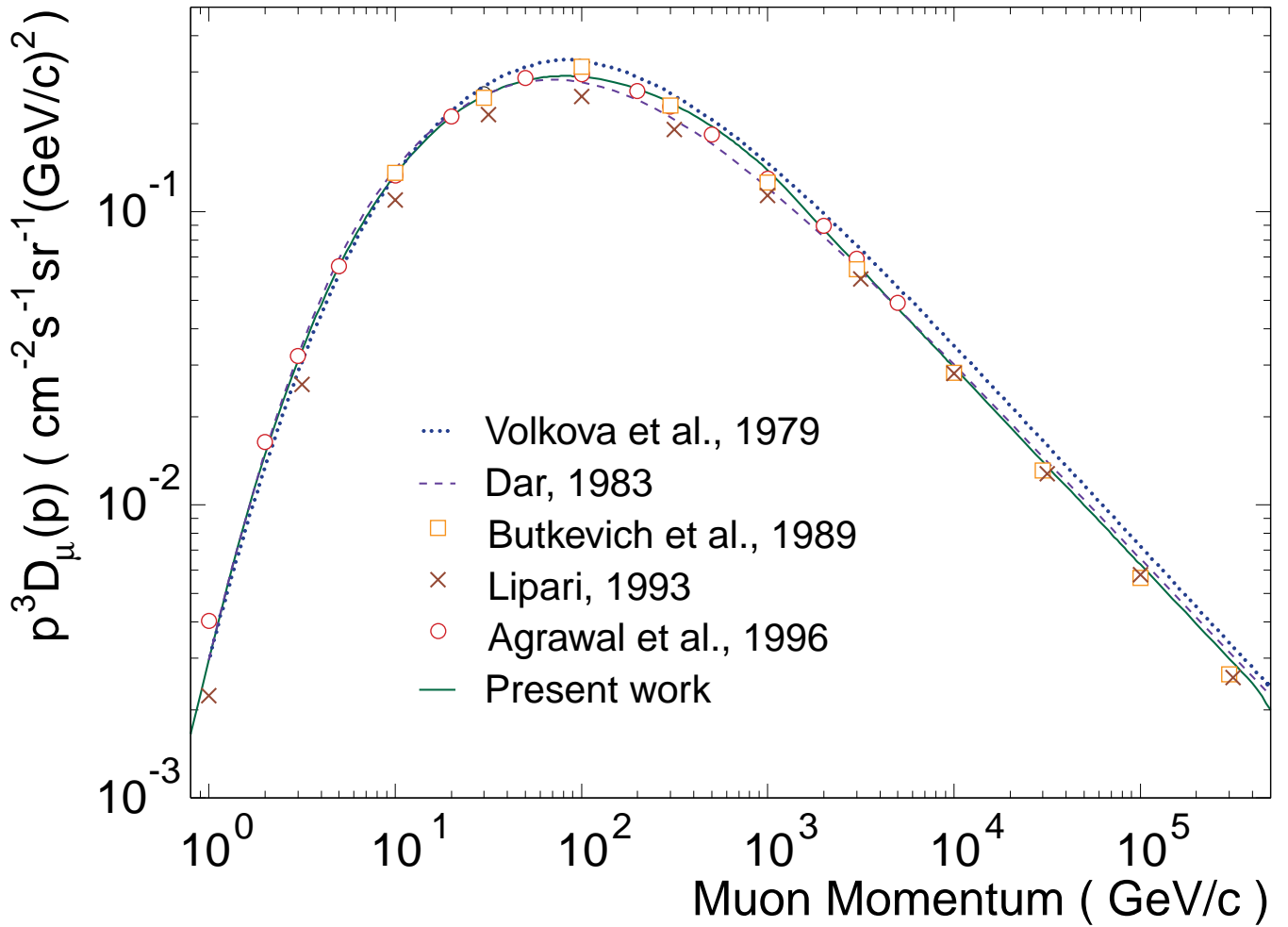,height=13.0cm}}}
\vspace{0.6cm}
\protect\caption[Comparison of vertical differential momentum
                 spectra of conventional muons at sea level
                 calculated by different wokers]
                {Vertical differential momentum spectra of
                 conventional muons at sea level calculated by
                 Volkova {\em et al.}~\protect\cite{Volkova79},
                 Dar~\protect\cite{Dar83},
                 Butkevich {\em et al.}~\protect\cite{Butkevich89},
                 Lipari~\protect\cite{Lipari93},
                 Agrawal {\em et al.}~\protect\cite{Agrawal96},
                 and in present work.
\label{fig:Comp}}
\end{figure}

Figure~\ref{fig:Comp} compares our result for the vertical
differential muon spectrum at sea level with the results of
Volkova {\em et al.}~\cite{Volkova79}, Dar~\cite{Dar83},
Butkevich {\em et al.}~\cite{Butkevich89}, Lipari~\cite{Lipari93},
and Agrawal  {\em et al.}~\cite{Agrawal96}.
In this comparison, we used the fitting formulas from Refs.~%
\cite{Volkova79,Dar83}, and the corresponding tables from
Refs.~\cite{Butkevich89,Lipari93,Agrawal96}.

In Table~\ref{tab:comp}, we show the ratio of each calculated
spectrum from Refs.~\cite{Volkova79,Dar83,Butkevich89,Lipari93,%
Agrawal96} to ours for $E = 1,10,\ldots,10^6$~GeV. The ratios are
inside the wide range $0.75 \div 1.48$. In the
momentum region from $\sim5$ to $5\times10^3$ GeV/c, our
result is in very good agreement with the recent Monte Carlo
calculation by Agrawal {\em et al.}~\cite{Agrawal96}:
the discrepancy is less than 6\,\%. This is consistent with
the uncertainties of both calculations caused by the uncertainties
in the input parameters. At low energies ($1\div10$~GeV), our
calculation agrees closely with the fitting formulas by
Volkova {\em et al.}~\cite{Volkova79} and Dar~\cite{Dar83}.
\begin{table}[!h]
\protect\caption[The ratios of vertical differential spectra of
                 conventional muons calculated by different workers
                 to ours]
                {The ratios of vertical differential spectra of
                 conventional muons calculated by different workers
                 to ours. 
\label{tab:comp}}
\center{\begin{tabular}{cccccccc}
Ref.              &\multicolumn{7}{c}{$E$ (GeV)}       \\\cline{2-8}
                  & $1$ & $10$ &$10^2$&$10^3$&$10^4$&$10^5$&$10^6$\\
                                                              \hline
\cite{Volkova79}  &1.010&0.996 &1.135 &1.056 &1.189 &1.156 &1.483 \\
\cite{Dar83}      &1.001&1.046 &0.958 &0.873 &1.023 &1.047 &1.405 \\
\cite{Butkevich89}& --  &1.015 &1.079 &0.909 &0.958 &0.902 &1.140 \\
\cite{Lipari93}   &0.753&0.820 &0.858 &0.823 &0.955 &0.923 &1.160 \\
\cite{Agrawal96}  &1.355&0.992 &1.017 &0.938 & --   & --   & --   \\
\end{tabular}}
\end{table} 

\protect\section{Charm production and prompt muons}
\label{sec:PM}

The prompt muon and neutrino component of the cosmic ray flux
originates from the decay of short-lived particles (mainly charmed
hadrons $D^\pm$, $D^0$, $\overline{D}{}^0$, $\Lambda_c^+,\ldots$)
produced in interactions of cosmic rays with the atmosphere. For the
last fifteen years, a lot of papers with calculations of prompt
lepton production in the atmosphere have been published
with very different outputs. Suffice it to say that the predicted
energy at which the vertical sea-level PM flux becomes equal to
that of muons from $\pi$ and $K$ decays varies from $\sim20$~TeV to
$\sim10^3$~TeV, depending on an adopted charm production model.
Early works~\cite{Inazawa83,Elbert83,Volkova83-85,Castagnoli84,%
Minorikawa85-86,Inazawa86,Allkofer87,Volkova87} were based on
empirical {\em ad hoc} models for open-charm production and some
extrapolations of the accessible (rather fragmentary) accelerator
data to the orders-of-magnitude higher energies of the primary and
secondary particles participant in cosmic-ray interactions. The
successive works apply more advanced phenomenological approaches
to the charm-production problem~\cite{Bugaev89a,Thunman95-96,%
Bugaev87-88,Bugaev89b,Treichel92,Battistoni96} or a set of
parametrizations for the energy dependence of the inclusive cross
sections those qualitatively describe the main features of some
popular models for charm production~\cite{Zas93-94,Pal94}.
Let us glance off two recent approaches based on the perturbative
QCD and the Dual Parton Model (DPM)~\cite{Thunman95-96,Battistoni96}.

Thunman {\em et al.}~\cite{Thunman95-96} apply a state-of-the-art
model to simulate charm hadroproduction through pQCD processes.
To leading order in the coupling constant, $\alpha_s$, these are
the gluon-gluon fusion ($gg \rightarrow c\overline{c}$) and the
quark-antiquark annihilation ($q\overline{q} \rightarrow
c\overline{c}$). The next-to-leading order, ${\cal O}(\alpha_s)$,
contributions are taken into account by doubling the cross
sections. To simulate the primary and cascade interactions, the
authors use the well-accepted Monte Carlo code {\sl PYTHYA}.
Without going into details of their approach, we emphasize that the
PM flux predicted by Thunman {\em et al.} is one of the lowest ones.
It overcomes the vertical $\pi,K$-muon flux at energy of about
$2\times10^3$ TeV and therefore is undistinguished in present-day
ground-based and underground/water muon experiments.

In the paper by Battistoni {\em et al.}~\cite{Battistoni96}, a new
Monte Carlo calculation of the PM fraction in atmospheric showers
was made using the {\sl DPMJET-II} code based on the two-component
DPM and interfaced to the shower code {\sl HEMAS}. The calculation
does not yield the absolute PM flux but, from the estimated
prompt-to-conventional muon ratio, one can see a leastwise
qualitative agreement with the result of Ref.~\cite{Thunman95-96}.
In particular, according to the DPM, the prompt component overcomes
the conventional one in the region of a thousand TeV (not reachable
with the simulated statistics).

In our previous works~\cite{Bugaev89a,Bugaev87-88,Bugaev89b}, the
two different phenomenological nonperturbative approaches to the
charm-production problem have been applied, the Recombination
Quark-Parton Model (RQPM) and the Quark-Gluon String Model (QGSM).
In the present calculation, we use just these two models. For this
reason, the most salient features of them will be outlined below in
this Section. The RQPM will be discussed at greater length,
considering that the QGSM is well accepted and covered adequately in
the literature~\cite{QGSM} (see also Ref.~\cite{DeRujula93}
and~\cite{Anzivino94} for reviews). As an example of a calculation
giving a particularly high PM flux, we will also sketch a
semiempirical model put forward by Volkova
{\em et al.}~\cite{Volkova87}. The comprehensive reviews of the
current experimental status of the charm production problem can be
found in Ref.~\cite{CharmReview}.

\protect\subsection{Models for charm hadroproduction}
\label{ssec:Charm}

\protect\subsubsection{Recombination quark-parton model (RQPM)}
\label{sssec:RQPM}

The RQPM is one of the models with ``intrinsic charm''. The models
of this class are based on the following key assumptions.
\begin{description}
\item[(i)]  The projectile wave contains an intrinsic-charm Fock
            component (see Refs.~\cite{Brodsky80-81,Brodsky92}). As
            an example, Figure~\ref{fig:Fock} shows the component
            $|uudc\overline{c}\rangle$ generated by the virtual
            subprocess $gg \rightarrow c\overline{c}$ where the
            initial gluons couple to two (or more) valence quarks of
            the projectile.
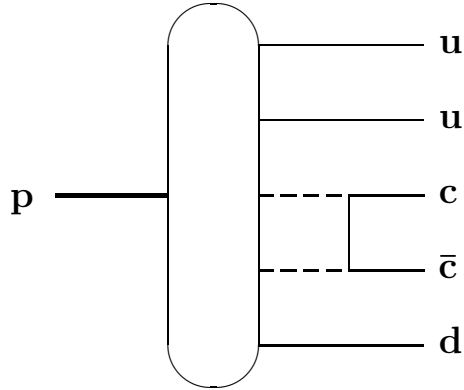
\begin{figure}[!h]
\begin{center}
\begin{picture}(70,60) 
\unitlength=1.0mm
\multiput(10.0,29.8)(0, 0.1){5}{\line(15,0){15}}                    %
\multiput(37.0,20.0)(3, 0  ){4}{\line( 2,0){ 2}}                    %
\multiput(37.0,30.0)(3, 0  ){4}{\line( 2,0){ 2}}                    %
\multiput(49.0,20.0)(0,10  ){2}{\line(10,0){10}}                    %
\multiput(37.0,40.0)(0,10  ){2}{\line(22,0){22}}                    %
\multiput(61.0,39.0)(0,10  ){2}{\Large\bf u}                        %
     \put(37.0,10.0){\line(22,0){22}}                               %
     \put(49.0,20.0){\line(0,10){10}}                               %
     \put(31.0,30.0){\oval(12,51)}                                  %
     \put( 4.0,28.5){\Large\bf  p}                                  %
     \put(61.0, 9.0){\Large\bf  d}                                  %
     \put(61.0,19.0){\Large\bf\=c}                                  %
     \put(61.0,29.0){\Large\bf  c}                                  %
\end{picture} 
\end{center}
\protect\caption[Intrinsic $|uudc\overline{c}\rangle$ Fock component
                 in the wave function of a projectile proton]
                {Intrinsic $|uudc\overline{c}\rangle$ Fock component
                 in the wave function of a projectile proton.
\label{fig:Fock}}
\end{figure}
\item[(ii)] The interaction of partons in the final state leads to a
            recombination (or coalescence) of the charmed quark with
            projectile fragments and to production of leading
            charmed hadrons~\cite{Hwa83,Bugaev85a,Brodsky87}.
\end{description}

An indication in favor of these models was found in muon--nucleon
scattering~\cite{EMC}. It was shown that there exists a visible
excess of the charmed particle yield at $x_F \gtrsim 0.15$ and
$Q^2 \lesssim 40$ GeV${}^2$ over the model expectations based on
the photon-gluon fusion and conventional QCD evolution. The upper
bound for the probability to find an intrinsic-charm Fock component
in the proton wave is about 0.6\,\%.

It has been shown by Brodsky {\em et al.}~\cite{Brodsky92} that the
diagrams with intrinsic charm, in which a $c\overline{c}$ pair is
coupled to more than one constituent of the projectile hadron, are
suppressed by powers of $M_{c\overline{c}}^2(1-x_c)$ (here
$M_{c\overline{c}}$ is the invariant mass of the pair and $x_c$ is
the fraction of hadron momentum carried by a parton), i.e. the
relative contribution of the intrinsic-charm mechanism to the
longitudinal momentum distribution of charmed hadrons is expected
to be especially large in the fragmentation region of a projectile.
In other words, intrinsic-charm models predict relatively hard
inclusive spectra. At the same time, the total inclusive cross
section can be rather large (it depends strongly on the assumptions
about the charm structure function of the projectile hadron). These
features cannot be obtained in perturbation theory (see e.g.
Ref.~\cite{E653'92} where a comparison of 600 GeV $\pi^-$ emulsion
data with the next-to-leading order pQCD predictions was made).

In the RQPM, the process of hadronization occurs by means of
recombination of quarks to hadrons~\cite{Bugaev85a}. It is assumed
that only slow (``wee'') partons of colliding hadrons take part in
the interaction and the distributions of fast partons do not change
during the collision. Therefore the inclusive spectra of produced
particles (those with small $p_T$ and with not too small $x_F$)
are entirely governed by quark distributions inside the projectile
hadron.

\protect\paragraph%
{\underline{Charm production in hadron-nucleon collisions.}}
\label{par:hN}

Inclusive cross section for production of a meson $M=q\overline{q}$
in $pp$ interaction is
\begin{equation} \label{pp}
x_F\frac{d\sigma_{pp \rightarrow MX}}{dx_F} =
\sum_{ij}\int\sigma_{ij}\left(x_{q_i},x_{q_j}\right)
F_{p_1}^{(1)}\left(x_{q_i}\right)
F_{p_2}^{(3)}\left(x_{q_j},x_q,x_{\overline{q}}\right)
R_M\left(x_{q},x_{\overline{q}};x_F\right)
dx_{q_i} dx_{q_j} dx_q dx_{\overline{q}}.
\end{equation}
Here $q_i$ and $q_j$ are the ``wee'' partons from protons $p_1$ and
$p_2$, respectively ($p_2$ is the projectile), $\sigma_{ij}$ is the
total cross section for $q_iq_j$ interaction, $F_{p_k}^{(m)}$ is the
$m$-parton joint distribution inside the proton $p_k$, and $R_M$ is
the function of recombination of the pair $q\overline{q}$ into
meson $M$. The cross section~(\ref{pp}) is written for the
{\em fragmentation region} of the projectile $p_2$. Let us assume
that the distribution of ``wee'' partons is universal and does
not correlate with the distribution of fast partons. Then
\[
F_{p_2}^{(3)}\left(x_{q_j},x_q,x_{\overline{q}}\right) =
F_{p_2}^{(1)}\left(x_{q_j}\right)
F_{p_2}^{(2)}\left(x_q,x_{\overline{q}}\right).
\]
Considering that
\[
\sigma_{pp}^{\rm tot} =
\sum_{ij}\int\sigma_{ij}\left(x_{q_i},x_{q_j}\right)
F_{p_1}^{(1)}\left(x_{q_i}\right)F_{p_2}^{(1)}\left(x_{q_j}\right)
dx_{q_i} dx_{q_j},
\]
yields
\[
x_F\frac{d\sigma_{pp\rightarrow M X}}{dx_F} =
\sigma_{pp}^{\rm tot}\int F_p^{(2)}\left(x_q,x_{\overline{q}}\right)
R_M\left(x_q,x_{\overline{q}};x_F\right)dx_qdx_{\overline{q}}.
\]
In a similar spirit one can derive the inclusive cross
section for the generic reaction $iN \rightarrow fX$:
\begin{equation}\label{iN}
x_F\frac{d\sigma_{iN \rightarrow fX}}{dx_F}=\sigma_{iN}^{\rm tot}(s)
\int F_i\left(\{x_k\}\right)R_f\left(\{x_k\};x_F\right)\prod_k dx_k.
\end{equation}
Here $x_k$ is the fraction of the projectile momentum which belongs
to the parton $q_k$, $F_i\left(\{x_k\}\right)$ is the two- or
three-quark distribution in the projectile hadron $i$ and
$R_f\left(\{x_k\};x_F\right)$ is the function of recombination of
two or three quarks into hadrons.

It would appear reasonable that far away from the threshold of
open-charm production, the parton distributions and recombination
functions do not depend on the projectile particle energy. Then the
$s$-dependence of the $d\sigma_{iN \rightarrow fX}/dx_F$ is
determined by the energy dependence of the total cross section for
the $iN$ interaction, $\sigma_{iN}^{\rm tot}(s)$, and therefore the
scaling violation is fairly small. As in the case of light particle
production, we use for the $\sigma_{iN}^{\rm tot}(s)$ the model of
elastic amplitude from Ref.~\cite{Vall86} which predicts that the
total cross section grows as $\ln s$ at the asymptotic energies.

We assume that the $c$-quark sea in a hadron is essentially
nonperturbative and it is characterized by a flat momentum spectrum
(see e.g. Ref.~\cite{Brodsky80-81}). According to the parton
conception, in the infinite momentum frame, the lifetime of
fluctuations containing heavy quarks is very large; the flatness of
heavy-quark spectra follows from a simple picture of a hadron as an
aggregate of partons with approximately equal velocities and from
calculations of structure functions for strongly coupled states.

To calculate the two- and three-quark distributions,
$F_i\left(\{x_k\}\right)$, we use the statistical approach by
Kuti and Weisskopf~\cite{Kuti71}.
The functions $F_i\left(\{x_k\}\right)$ are constructed through
``uncorrelated'' parton distributions $f_k^{\rm val}(x)$ and
$f_k^{\rm sea}(x)$ for valence and sea quarks ($k = u, d, s, c$) and
through the correlation functions $G^i(1-x)$. For example, the
two-particle distribution of $u$ and $\overline{c}$ quarks in a
proton is of the form
\[
F_p^{(2)}\left(x_u,x_{\overline{c}}\right) =
   \left[2G_u^p(1-x_u-x_{\overline{c}})f_u^{\rm val}(x_u)+
          G_0^p(1-x_u-x_{\overline{c}})f_u^{\rm sea}(x_u)\right]
          f_c^{\rm sea}(x_{\overline{c}}).
\]
The three-particle distribution of $u$, $d$, and $c$ in a proton is
\begin{eqnarray*}
F_p^{(3)}\left(x_u,x_d,x_c\right)                           & = &
  \left[2G_{ud}^p\left(1-\sum x_q\right)f_u^{\rm val}(x_u)+
         G_{ d}^p\left(1-\sum x_q\right)f_u^{\rm sea}(x_u)\right]
                    f_d^{\rm val}(x_d)f_c^{\rm sea}(x_c) \\ & + &
  \left[2G_{u }^p\left(1-\sum x_q\right)f_u^{\rm val}(x_u)+
         G_0   ^p\left(1-\sum x_q\right)f_u^{\rm sea}(x_u)\right]
                    f_d^{\rm sea}(x_d)f_c^{\rm sea}(x_c),
\end{eqnarray*}
\[
\sum x_q = x_u+x_d+x_c.
\]

Both the uncorrelated distributions and correlation functions for
light quarks and gluons in a proton and pion were calculated by
Takasugi~\cite{Takasugi82} in the framework of the statistical model
using all appropriate accelerator data. It can be shown that the
correlation functions are little affected by introducing the sea of
charmed quarks and hence we will use the results of
Ref.~\cite{Takasugi82} without any modifications. In so doing and
using Eq.~(\ref{iN}), the uncorrelated $c$-distributions,
$f_c^{\rm sea}(x)$, could be basically extracted from the data on
charm production. In fact the realization of this program is
somewhat limited because Eq.~(\ref{iN}) only holds at asymptotic
energies (far away from the charm-production threshold) and besides,
the available accelerator data at high energies cover a narrow
range, $0.1 \lesssim x_F \lesssim 0.9$. Within this range, the best
fit of the ISR data on $\Lambda_c$ production in $pp$
interactions~\cite{Basile81}  
and the EMC data on charm production in deep-inelastic muon
scattering~\cite{EMC} is achieved with the following simple
parametrizations~\cite{Bugaev85a}:
\[
f_c^{\rm sea}(x) = \left\{\begin{aligned}
5.5\times10^{-3}x^{-0.5}(1-x)^{-1.83} & \quad\text{for proton}, \\
7.7\times10^{-3}x^{-1  }(1-x)^{-0.85} & \quad\text{for pion}.
\end{aligned}\right.
\]
In our calculations, we do not make distinctions between
pseudoscalar and vector charmed mesons of an identical quark
composition at production. So, by a $D$ meson production cross
section is meant the overall cross section for production of $D$
and $D^*$ mesons.

For the recombination functions of quarks into $D$ and
$\Lambda_c$ we use the formula derived by Hwa in his valon
model~\cite{Hwa80},
\begin{eqnarray*}
R_D(x_1,x_2;x) & = & \frac{x}{B(a,b)}\left(\frac{x_1}{x}\right)^a
   \left(\frac{x_2}{x}\right)^b\delta(x_1+x_2-x), \\
R_{\Lambda_c}(x_1,x_2,x_3;x) & = &
\frac{x}{B(a,b)B(a,a+b)} \left(\frac{x_1x_2}{x}\right)^a
\left(\frac{x_3}{x}\right)^b\delta(x_1+x_2+x_3-x).
\end{eqnarray*}
Here $B(a,b)$ is the beta-function, $a$ and $b$ are the constants
defined by the form of the valon distributions. Regarding the valons
as constituent quarks bound nonrelativistically in a bag, it can be
shown~\cite{Hwa80} that their average momenta, $\langle x_i\rangle$,
are proportional to their masses, $\hat{m}_i$. Then, considering the
two-valon distribution in a $D^0$-meson, we have
\[
\frac{a}{b} =
\frac{\langle x_u\rangle}{\langle x_{\overline{c}}\rangle} =
\frac{\hat{m}_u}{\hat{m}_c} \simeq \frac{1}{6}.
\]
Below, we adopt $a = 1$ in all numerical calculations.

\protect\paragraph{\underline{Nuclear effects.}}
\label{par:NucEff}

In order to take the nuclear effects into account, we use the
additive quark model~\cite{AQM}. Let us assume that passing over the
target nucleus, $A$, a valence quark of the projectile behaves as a
free particle between its collisions with nuclei. If at a collision
with a nucleus the quark loses the bulk of its momentum, that quark
may be thought of as captured by the target and its contribution to
the production (through the recombination) of hadrons with large
$x_F$ can be neglected. On the contrary, the quark which escape
collisions can hadronize by recombining with slow quark(s) as
described above.  Because our prime interest is in the high-energy
range and in the fragmentation region of projectile particles, one
can neglect the interaction of secondary hadrons with the target
nucleus. Indeed, the time of generation of hadrons is proportional
to their momenta and fast hadrons are produced outside the nucleus.
In line with these assumptions, the invariant cross section for
inclusive production of hadrons in hadron-nucleus collisions is
expressed in terms of the ``recombination'' hadron-nucleus cross
sections and the probabilities for capturing valence quarks by the
target nucleus. Using standard ``nuclear optics''
techniques~\cite{Berlad80} and the additive-quark-model relations
for the total cross sections~\cite{AQM},
$2\sigma_{pp} \simeq 3\sigma_{\pi p} \simeq 2\sigma_{qp}$
($q = u, \overline{u}, d, \overline{d}$), one can derive the
following formulas for the inclusive charm-production cross
sections~\cite{Bugaev85a}:
\begin{eqnarray*}
\frac{d\sigma_{pA \rightarrow D^+X}}{dx_F}                 & = &
3\left(\frac{\sigma_{\pi A}-\sigma_{qA}}{\sigma_{pp}}\right)
\frac{d\sigma_{pp \rightarrow D^+X}}{dx_F},             \\
\frac{d\sigma_{pA \rightarrow D^-X}}{dx_F}                 & = &
3\left(\frac{\sigma_{pA}-\sigma_{qA}}{\sigma_{pp}}\right)
\frac{d\sigma_{pp \rightarrow D^-X}
                     ^{[d_{\rm v}\overline{c}]}}{dx_F}+
3\left(\frac{\sigma_{\pi A}-\sigma_{qA}}{\sigma_{pp}}\right)
\frac{d\sigma_{pp \rightarrow D^-X}
                     ^{[d_{\rm s}\overline{c}]}}{dx_F}, \\
\frac{d\sigma_{pA \rightarrow D^0X}}{dx_F}                 & = &
3\left(\frac{\sigma_{\pi A}-\sigma_{qA}}{\sigma_{pp}}\right)
\frac{d\sigma_{pp \rightarrow D^0X}}{dx_F},             \\
\frac{d\sigma_{pA \rightarrow \overline{D}{}^0X}}{dx_F}    & = &
 \left(\frac{\sigma_{pA}+\sigma_{\pi A}-2\sigma_{qA}}
                                               {\sigma_{pp}}\right)
\frac{d\sigma_{pp \rightarrow \overline{D}{}^0X}
                     ^{[u_{\rm v}\overline{c}]}}{dx_F}+
3\left(\frac{\sigma_{\pi A}-\sigma_{qA}}{\sigma_{pp}}\right)
\frac{d\sigma_{pp \rightarrow \overline{D}{}^0X}
                     ^{[u_{\rm s}\overline{c}]}}{dx_F}, \\
\frac{d\sigma_{pA \rightarrow \Lambda^+_cX}}{dx_F}         & = &
3\left(\frac{\sigma_{pA}-\sigma_{\pi A}}{\sigma_{pp}}\right)
\frac{d\sigma_{pp \rightarrow \Lambda^+_cX}
                     ^{[u_{\rm v}d_{\rm v}c]}}{dx_F}+
 \left(\frac{\sigma_{pA}+\sigma_{\pi A}-2\sigma_{qA}}
                                            {\sigma_{pp}}\right)
\frac{d\sigma_{pp \rightarrow \Lambda^+_cX}
                     ^{[u_{\rm v}d_{\rm s}c]}}{dx_F}    \\ & + &
\frac{3}{2}\left(\frac{\sigma_{pA}-\sigma_{qA}}{\sigma_{pp}}\right)
\frac{d\sigma_{pp \rightarrow \Lambda^+_cX}
                     ^{[u_{\rm s}d_{\rm v}c]}}{dx_F}+
3\left(\frac{\sigma_{\pi A}-\sigma_{qA}}{\sigma_{pp}}\right)
\frac{d\sigma_{pp \rightarrow \Lambda^+_cX}
                     ^{[u_{\rm s}d_{\rm s}c]}}{dx_F},   \\
\frac{d\sigma_{\pi^+A \rightarrow DX}}{dx_F}               & = &
2\left(\frac{\sigma_{\pi A}-\sigma_{qA}}{\sigma_{\pi p}}\right)
\frac{d\sigma_{\pi p \rightarrow DX}}{dx_F} \qquad
 (D = D^\pm, D^0, \overline{D}{}^0).
\end{eqnarray*}
Here $d\sigma_{ip \rightarrow fX}^{[\cdots]}/dx_F$ is the
contribution to the $iN$ cross section from a quark diagram with a
final hadron $f$ that contains the leading valence (v) or sea
(s) quarks indexed in the brackets. To sufficient accuracy, the
total cross sections in the foregoing equations are assumed to
be energy-independent. In our numerical evaluations, we set
$\sigma_{iA} = \sigma_{iA}^0$ for the hadron-nucleus cross sections
(see Section~\ref{sec:Cascade}) and $\sigma_{qp} = 13.0$~mb for the
quark-proton cross section~\cite{Berlad80}.
The numerical results are represented in the traditional form,
\[
\frac{d\sigma_{i A \rightarrow f X}}{d x_F} =
A^{\alpha(x_F)}\frac{d\sigma_{i N \rightarrow f X}}{d x_F}.
\]
For the reactions $pA \rightarrow D^+X$, $pA \rightarrow D^0X$, and
$\pi A \rightarrow DX$ ($D = D^\pm, D^0, \overline{D}{}^0$),
$\alpha = 0.765$, independently of $x_F$. It should be pointed out
that accelerator data at low energies show a higher value of
$\alpha$. For example, in the WA82 experiment~\cite{WA82'92}
(a 340~GeV $\pi^-$ beam) the value $\alpha = 0.92 \pm 0.06$ was
obtained for $D$ mesons with $\langle x_F \rangle = 0.24$. However,
it seems plausible that this is a reflection of the
``near-threshold effect'' and the $\alpha$ will decrease with a rise
 of the projectile energy. In any event, the non-perturbative effects
should become more important as $\sqrt{s}$ and $x_F$ increase and
therefore the shadowing is expected to become more essential at
higher center-of-mass energies and at large $x_F$~\cite{Vogt92}.

For the rest reactions and within the range
$0.10 \leq x_F \leq 0.95$, the functions $\alpha(x_F)$ may be
parametrized as follows:
\begin{align*}
\alpha_{pA \rightarrow \overline{D}{}^0X}(x) & =
                              0.754-0.034x-0.008x^2+0.020x^3, \\
\alpha_{pA \rightarrow D^-X}(x)              & =
                              0.769-0.158x+0.272x^2-0.174x^3, \\
\alpha_{pA \rightarrow \Lambda^+_cX}(x)      & =
                              0.780-0.367x+0.672x^2-0.456x^3.
\end{align*}
These results do not contradict the accelerator data even at very
low energies, although the data are rather uncertain yet.
For example, the BIS-2 experiment~\cite{BIS2'93} (a 37.5--70~GeV
neutron beam) gives $\langle\alpha\rangle=0.73\pm0.23$ for
$\overline{D}{}^0$ production.

As discussed above, we assume that the captured quarks take no part
in the recombination. This leads to a small underestimation of the
cross sections, because some portion of wounded quarks actually will
recombine. Let us estimate the upper limit of the $\alpha$ assuming
that {\em all} the valence quarks of the projectile can recombine.
This assumption yields
\[
\frac{d\sigma_{iA \rightarrow fX}}{dx_F} =
                      \left(\frac{\sigma_{iA}}{\sigma_{iN}}\right)
                          \frac{d\sigma_{iN \rightarrow fX}}{dx_F}
\]
and thus $\alpha\leq0.85$ for $\pi A \rightarrow D X$ and
$\alpha\leq0.79$ for $p A \rightarrow D\,(\Lambda_c)X$. This
estimate demonstrates that the uncertainty in the $A$-dependence
within our simplified approach does not exceed $\sim15$\,\% for
the air nuclei.

\protect\paragraph{\underline{$Z$-factors.}}
\label{par:ZRQPM}
Owing to the mentioned small scaling violation, the fractional
moments $Z_{fi}$ calculated with the RQPM from Eq.~(\ref{Z}) are
energy dependent. They can be approximated with an accuracy of
(2--3)\,\% by the following expression:
\begin{equation} \label{ZRQPM}
Z_{fi}(\gamma,E) = Z_{fi}(\gamma,E_\gamma)
                   \left(\frac{E}{E_\gamma}\right)^{\xi_\gamma},
\end{equation}
\begin{table*}[!h]
\protect\caption[Parameters of fitting formula~(\protect\ref{ZRQPM})
                 for the fractional moments calculated with the
                 RQPM]
                {Parameters $Z_{fi}(\gamma,E_\gamma)$ of fitting
                 formula (\protect\ref{ZRQPM}) for the fractional
                 moments $Z_{fi}(\gamma,E)$ calculated with the RQPM
                 for the two values of $\gamma$.
\label{tab:ZRQPM}}
\center{\begin{tabular}{lccccc}
       &                  &                  &  $f$             &
                          &              \\\cline{2-6} & & & & & \\
  $i$  & $D^+$            &  $D^-$           &  $D^0$           &
                      $\overline{D}{}^0$&  $\Lambda_c^+$    \\\hline
       &                  & &$\gamma = 1.62$&&                   \\
  $p$  &$4.6\times10^{-4}$&$6.5\times10^{-4}$&$3.8\times10^{-4}$&
                           $6.9\times10^{-4}$&$4.9\times10^{-4}$ \\
$\pi^+$&$1.3\times10^{-3}$&$9.0\times10^{-4}$&$9.0\times10^{-4}$&
                      $1.3\times10^{-3}$&$6.0\times10^{-4}$ \\\hline
       &                  &&$\gamma = 2.02$&&                    \\
  $p$  &$5.4\times10^{-4}$&$7.9\times10^{-4}$&$4.5\times10^{-4}$&
                           $8.6\times10^{-4}$&$6.2\times10^{-4}$ \\
$\pi^+$&$1.8\times10^{-3}$&$1.2\times10^{-3}$&$1.2\times10^{-3}$&
        $1.8\times10^{-3}$&$7.9\times10^{-4}$
\end{tabular}}
\end{table*} 
where $E$ is the energy of secondary particle $f$
($f = D^\pm,\,D^0,\,\overline{D}{}^0,\,\Lambda_c^+$), $E_\gamma$
and $\xi_\gamma$ are the constants dependent on the primary
cosmic-ray spectrum. In particular,
\begin{eqnarray*}
E_\gamma & = & 10^3\;\mbox{GeV},\quad
            \xi_\gamma = 0.096\quad\mbox{for}\;\gamma = 1.62, \\
E_\gamma & = & 10^6\;\mbox{GeV},\quad
            \xi_\gamma = 0.076\quad\mbox{for}\;\gamma = 2.02.
\end{eqnarray*}

Parameters $Z_{fi}(\gamma,E_\gamma)$ for $i = p, \pi^+$ are
presented in Table \ref{tab:ZRQPM}.
For $i = n, \pi^-$ one can use the relations
\begin{gather*}
Z_{D^+n} = Z_{D^0p},                                   \quad
Z_{D^-n} = Z_{\overline{D}^0p},                        \quad
Z_{D^0n} = Z_{D^+p},                                   \\
Z_{\overline{D}^0n} = Z_{D^-p},                        \quad
Z_{\Lambda_c^+n} = Z_{\Lambda_c^+p},                   \quad
Z_{D^+\pi^-} = Z_{\overline{D}^0\pi^-} = Z_{D^0\pi^+}, \\
Z_{D^-\pi^-} = Z_{D^0\pi^-} = Z_{\overline{D}^0\pi^+}, \quad
Z_{\Lambda_c^+\pi^-} = Z_{\Lambda_c^+\pi^+},
\end{gather*}
which follow from considerations of the isotopic symmetry.

\protect\subsubsection{Quark-gluon string model (QGSM)}
\label{ssec:QGSM}

The QGSM~\cite{QGSM} is a non-perturbative approach to the
description of hadron collisions. It is based on the topological
$1/N_f$ expansion of QCD diagrams for elastic
scattering~\cite{t'Hooft74} (associated with the multiple pomeron
exchange expansion) and the string model of hadrons and hadronic
interactions. The particles are produced in this model by breaking
the strings connecting the incident hadron's constituents (quarks and
diquarks).

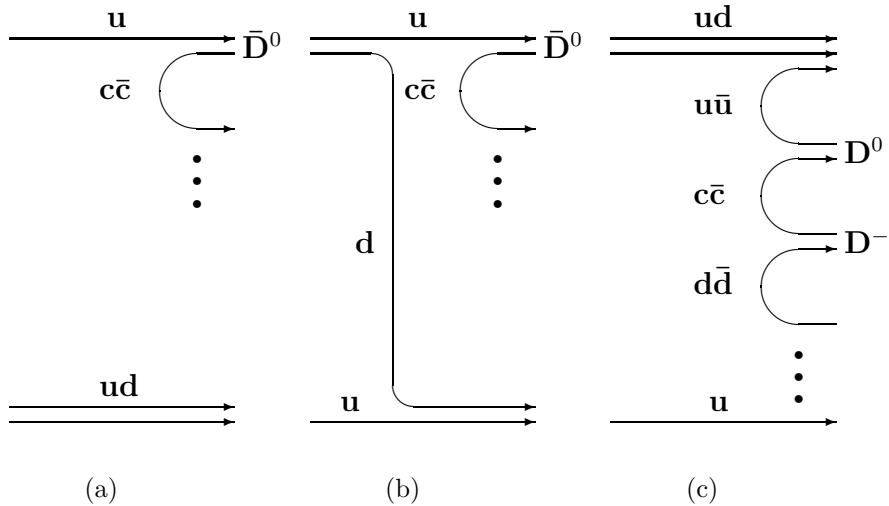
\begin{figure}[!h]
\centering
\begin{picture}(120,70) 
\unitlength=1.0mm                                                   %
\multiput(  0.0,65.0 )(40.0, 0.0){3}{\vector(1,0){30}}              %
\multiput( 25.0,53.0 )(40.0, 0.0){2}{\vector(1,0){ 5}}              %
\multiput(105.0,37.0 )( 0.0,12.0){3}{\vector(1,0){ 5}}              %
\multiput(  0.0,14.0 )(40.0, 0.0){3}{\vector(1,0){30}}              %
\multiput(  0.0,16.0 )(80.0,47.0){2}{\vector(1,0){30}}              %
\multiput( 25.0,63.0 )(40.0, 0.0){2}{\line(5,0){5}}                 %
\multiput(105.0,27.0 )( 0.0,12.0){3}{\line(5,0){5}}                 %
\multiput( 25.0,58.04)(40.0, 0.0){2}{\oval(10,10)[tl]}              %
\multiput( 25.0,58.04)(40.0, 0.0){2}{\oval(10,10)[bl]}              %
\multiput(105.0,32.04)( 0.0,12.0){3}{\oval(10,10)[tl]}              %
\multiput(105.0,32.04)( 0.0,12.0){3}{\oval(10,10)[bl]}              %
\multiput( 25.0,43.0 )( 0.0, 3.0){3}{\circle*{1.0}}                 %
\multiput( 65.0,43.0 )( 0.0, 3.0){3}{\circle*{1.0}}                 %
\multiput(105.0,23.0 )( 0.0,-3.0){3}{\circle*{1.0}}                 %
\multiput( 13.0,66.5 )(40.0, 0.0){2}{\large\bf u}                   %
\multiput( 44.0,15.5 )(49.0, 0.0){2}{\large\bf u}                   %
\multiput( 31.0,62.5 )(40.0, 0.0){2}{\large\bf\=D$^0$}              %
\multiput( 12.0,17.5 )(79.0,49.0){2}{\large\bf ud}                  %
     \put( 12.0,57.0 ){\large\bf c\=c}                              %
     \put( 52.5,57.0 ){\large\bf c\=c}                              %
     \put( 51.0,60.0 ){\line(0,-41){41}}                            %
     \put( 40.0,63.0 ){\line(8,0){8}}                               %
     \put( 48.0,60.04){\oval(6,6)[tr]}                              %
     \put( 54.0,19.04){\oval(6,6)[bl]}                              %
     \put( 54.0,16.0 ){\vector(1,0){16}}                            %
     \put( 46.0,36.5 ){\large\bf d}                                 %
     \put(111.0,48.5 ){\large\bf D$^0$}                             %
     \put(111.0,36.5 ){\large\bf D$^-$}                             %
     \put( 91.0,55.0 ){\large\bf u\=u}                              %
     \put( 91.0,43.0 ){\large\bf c\=c}                              %
     \put( 91.0,31.0 ){\large\bf d\=d}                              %
     \put( 10.0, 4.0 ){\rm (a)}                                     %
     \put( 50.0, 4.0 ){\rm (b)}                                     %
     \put( 90.0, 4.0 ){\rm (c)}                                     %
\end{picture} 
\protect\caption[Fragmentation of quark chains into $D$ mesons in
                 the QGSM]
                {Fragmentation of quark chains into $D$ mesons in
                 the QGSM: (a,b) favored fragmentation into
                 \protect$\overline{D}{}^0$; (c) unfavoured
                 fragmentation into $D^-$ and $D^0$.
\label{fig:frag}}
\end{figure}

The QGSM is considered to be one of the most satisfactory of
the tools available to represent open-charm production. It describes
a great body of data on hadronic interactions at all available
energies. However, the model is not free from difficulties. For
instance, the QGSM predicts clear-cut flavor correlations. In
particular, there must be preferential production of
$\overline{D}{}^0$ mesons in $pp$ collisions (``favored
fragmentation'') owing to $(u\!-\!ud)$ composition of the proton
and $(\overline{c}u)$ composition of $\overline{D}{}^0$
(Figure~\ref{fig:frag}). This prediction is not supported by
experiment~\cite{LEBC-EHS'88}, although this disagreement can be
caused in part by bad flavor identification in the experiment (see
Ref.~\cite{DeRujula93} for a discussion).

To calculate the inclusive cross sections one must know the
distribution functions of the dressed quarks (constituents) of the
colliding hadrons and the fragmentation functions of these
constituents into charmed particles. These functions can be
approximately determined by the use of Regge model
arguments~\cite{Kaidalov81}, in terms of intercepts
$\alpha_R\sim-\alpha_N \sim 0.5$, of known Regge poles and the
intercept of the $c\overline{c}$ Regge trajectory, $\alpha_\psi$,
on which there is no direct experimental information. Hence
$\alpha_\psi$ is a free parameter of the model. It governs, in
particular, the steepness of the inclusive spectra of charmed
particles. If the $c\overline{c}$ trajectories are linear (as it
is in the case of light quarks and generally in the string models
of hadrons), the intercept of the $\psi$ trajectory is fairly large
($\simeq -2.2$) and the longitudinal momentum distributions of
charmed hadrons are rather steep. A complete list of the
distribution and fragmentation functions as well as the values of
their various parameters are given in Ref.~\cite{QGSM}.

Our calculations of the inclusive cross sections within the
framework of the QGSM have been done without attempts to optimize
the set of parameters of the model. In particular, we do not
include the intrinsic charm component as it was suggested
recently~\cite{Anzivino94}. Below, we are dealing with a qualitative
analysis of the QGSM prediction for charm production at cosmic-ray
energies rather than with a close examination of the model. For this
reason, in evaluating the nuclear effect within the QGSM, we adopt
$\alpha = 0.72$ for all processes under consideration. This
simplification can lead to a small ($<15$\,\%) error in the
$Z$-factors, compared to the exact calculation within the additive
quark model.
The energy dependence of the factors $Z_{fi}(\gamma,E)$ calculated
with the QGSM is somewhat different as compared with the RQPM
prediction. The parametrization~(\ref{ZRQPM}) is valid for the QGSM
only at very high energies ($\gtrsim 10^3$ TeV) and the parameters
$\xi_\gamma$ are in general different for different reactions
$i A\rightarrow f X$.  The parameters $Z_{fi}(\gamma,E_\gamma)$ and
$\xi_\gamma$ for $i = p,n,\pi^+$ and $\pi^-$ are presented in Table
\ref{tab:ZQGSM} at $\gamma = 2.02$ (above the knee energy region).
The energy dependence of the $Z$-factors at $E < 10^3$ TeV
can be found in Ref.~\cite{Bugaev89a}.
\begin{table}[h]
\protect\caption[Parameters of fitting formula~(\protect\ref{ZRQPM})
                 for the fractional moments calculated with the
                 QGSM]
                {Parameters $Z_{fi}(\gamma,E_\gamma)$ and
                 $\xi_\gamma$ (in parentheses) of fitting
                 formula~(\protect\ref{ZRQPM}) for the fractional
                 moments $Z_{fi}(\gamma,E)$ calculated with the QGSM
                 for $\gamma=2.02$ at $E\gtrsim10^3$~TeV.
\label{tab:ZQGSM}}
\center{\begin{tabular}{cccccc}
                         &  &  & $f$ & & \\\cline{2-6} & & & & & \\
  $i$  &$D^+$& $D^-$ & $D^0$  & $\overline{D}{}^0$ &
                                             $\Lambda_c^+$ \\\hline
  $p$  &$6.5\times10^{-5}$&$9.9\times10^{-5}$ & $7.1\times10^{-5}$ &
                           $2.1\times10^{-4}$ & $9.5\times10^{-4}$\\
       &(0.050) & (0.046) &(0.050) &       (0.044)      &  (0.041)\\
  $n$  &$7.1\times10^{-5}$&$1.9\times10^{-4}$ & $6.5\times10^{-5}$ &
                           $1.2\times10^{-4}$ & $9.5\times10^{-4}$\\
       &(0.050) & (0.045) &(0.050) &       (0.045)      &  (0.041)\\
$\pi^+$&$5.5\times10^{-4}$&$1.4\times10^{-4}$ & $1.4\times10^{-4}$ &
                           $5.5\times10^{-4}$ & $1.5\times10^{-5}$\\
       &(0.041) & (0.048) &(0.048) &       (0.041)      &  (0.035)\\
$\pi^-$&$1.4\times10^{-4}$&$5.5\times10^{-4}$ & $5.5\times10^{-4}$ &
                           $1.4\times10^{-4}$ & $1.5\times10^{-5}$\\
       &(0.041) & (0.048)& (0.048) &       (0.041)      &  (0.035)
\end{tabular}}
\end{table} 

\protect\subsubsection{Semiempirical model (VFGS)}
\label{sssec:VFGS}

The model of Volkova {\em et al.}~\cite{Volkova87} (let us call it
the VFGS model) is a typical example of an approach which proceeds
from a parametrization of available accelerator data for inclusive
spectra of charmed particles together with some additional
assumptions to extrapolate the parametrization to the kinematic
regions, where the data on the inclusive charm production cross
sections are absent.

Volkova {\em et al.} make use a very steep inclusive spectrum of
produced $D$-mesons ($\propto(1-x_D)^5/x_D$, where $x_D$ is the
ratio of the $D$-meson energy to the nucleon energy in the lab{}.
frame) with a sharp cut-off in the central region ($d\sigma/d x_D=0$
at $x_D \leq 0.05$). In spite of such cut-off the integral
$\int(d\sigma/dx_D)dx_D$ was normalized to the total $D\overline{D}$
cross section, $\sigma_{pp}^{D\overline{D}}(E_N)$. Considering the
accelerator data at $E_N \gtrsim 1$ TeV together with some
implications of the QGSM, it has been adopted that
\[
\sigma_{pp}^{D\overline{D}}(E_N) = \Big\{\begin{aligned}
   0.48(\log E_N-3.075)\;\text{mb}\quad
                & \text{for 1 TeV $\leq E_N < 500$ TeV}, \\
   1.26\;\text{mb} \quad
                & \text{for $E_N \geq$ 500 TeV}. \end{aligned}
\]
A consequence of this assumption is a relatively strong scaling
violation in the fragmentation region.

The VFGS model predicts comparatively large PM flux (see below)
since, owing to the cut-off, all produced particles are in the
fragmentation region of a projectile (i.e. there is no the central
part of the inclusive spectrum). It was also assumed that
(independently of $x_F$) $\alpha = 1$ and 2/3 for reactions with
$D$ mesons and $\Lambda_c^+$ hyperons in the final state,
respectively.

The approach of Ref.~\cite{Volkova87} includes some other assumptions
which also tend to increase the PM fraction in comparison with our
result. The most important ones are concerned with the primary
spectrum, semileptonic decays of charmed particles and with certain
elements of the nuclear-cascade model.  A more detailed comparison of
the approach under consideration against the RQPM and the QGSM, in
connection with the PM problem, has been done in
Ref.~\cite{Bugaev89b}.

\protect\subsection{Prompt muon flux at sea level}
\label{ssec:PMF}

\protect\subsubsection{Interactions and decay of charmed particles}
\label{sssec:ID}

As we neglect the production of nucleons, pions, and kaons
by charmed particles and charm regeneration, the transport
equations for $D$ and $\Lambda_c$ spectra are identical in form to
Eq.~(\ref{Ktrans}) for kaons.  Notice that the PM flux weakly depends
on the specific values of the inelastic cross sections for $D$ and
$\Lambda_c$ up to about $10^4$ TeV of muon energy, due to very short
lifetimes of these particles. Thus a rough estimation of
$\sigma_{DA}^{\rm inel}$ and $\sigma_{\Lambda_c^\pm A}^{\rm inel}$
will suffice for our purposes. We use the same formula~(\ref{sigma})
as for the light hadrons with $\sigma_{DA}^0 = 100$~mb
($D = D^\pm,D^0,\overline{D}{}^0$) and
$\sigma_{\Lambda_c^\pm A}^0 = 200$~mb.

Calculation of the PM flux can be performed in almost perfect
analogy to the conventional muon fluxe with the only one essential
difference: the PM generation function includes a rich variety of
multiparticle semileptonic decay modes. Thus the inclusive approach
is best suited to the problem. The corresponding muon generation
function may be written as
\begin{equation}\label{GPM}
G_\mu^{D,\Lambda_c}(E,h,\vartheta) =
\sum_{i=D^\pm,D^0,\overline{D}{}^0,\Lambda_c}\!\!\!\!
B(i\rightarrow\mu\nu X)\;\frac{E_i^{\rm cr}(\vartheta)}{hE}
                             \int_{x_i^-}^{x_i^+}F_i^\mu(x)
                  {\cal D}_i\left(\frac{E}{x},h,\vartheta\right)dx.
\end{equation}
Here $F_i^\mu(x)$ is the normalized spectrum of muons in the
inclusive decay $i\rightarrow\mu\nu_\mu X$ ($x = E/E_i$) and
\[
x_i^\mp = 2m_\mu^2\left[\left(m_i^2+m_\mu^2-s_X\right)
                          \pm\sqrt{\left(m_i^2+m_\mu^2-s_X\right)^2
                                      -4m_\mu^2m_i^2}\,\right]^{-1},
\]
with $s_X$ the minimal invariant mass square for the hadron system
$X$. The other designations are completely similar to the ones
previously used.

To simplify matters we consider the inclusive decay
$i\rightarrow\mu\nu X$ as a 3-particle one.
We assume the simplest form of matrix elements according to
Ref.~\cite{CharmDecay}. The form factors involved
(one for $D\rightarrow\mu\nu_\mu X$ and three for
$\Lambda_c\rightarrow\mu\nu_\mu X$) are replaced with their
averaged values. In so doing the mass square of the ``$X$-particle'',
$s_X^{\rm eff}$, may be fitted in such a way as to correlate
the calculated and experimental values for the differential and total
decay rates. Omitting rather tedious details of the calculation, we
present the final formulas for the muon spectral functions
$F_D^\mu(x)$ and $F_{\Lambda_c}^\mu(x)$ in Appendix~\ref{app:Decay}.

\protect\subsubsection{Parametrization of the calculated PM flux}
\label{sssec:Par}

In the energy region 5 TeV~$\lesssim E \lesssim 5\times10^3$~TeV the
differential spectra of PM in the vertical direction at sea level,
${\cal D}_\mu^{\rm pr}(E)$, calculated in Ref.~\cite{Bugaev89a} with
the RQPM and the QGSM, can be approximated by
\begin{equation}\label{fitPMdif}
{\cal D}_\mu^{\rm pr}\left(E,h=1030\;\mbox{g/cm}^2,\vartheta=
         0^\circ\right) = C'\left(\frac{E_b}{E}\right)^{\gamma'}
        \left[1+\left(\frac{E_b}{E}\right)^{\gamma'-1}\right]^{-a}.
\end{equation}
Here
\begin{align*}
C' & = 4.53 \times 10^{-18}\;{\rm cm}^{-2}{\rm s}^{-1}
                           {\rm sr}^{-1}{\rm GeV}^{-1}, \quad
\gamma' = 2.96, \quad a = 0.152 \quad \mbox{(RQPM)}, \\
C' & = 1.09 \times 10^{-18}\;{\rm cm}^{-2}{\rm s}^{-1}
                           {\rm sr}^{-1}{\rm GeV}^{-1}, \quad
\gamma' = 3.02, \quad a = 0.165 \quad \mbox{(QGSM)},
\end{align*}
and $E_b = 10^5$ GeV in both cases.  Eq.~(\ref{fitPMdif}) fits the
numerical results with accuracy better than 4\,\%. With the same
accuracy it is also valid for zenith angles
$\vartheta\lesssim80^\circ$ in the energy interval
$(10\div10^3)$~TeV, i.e.  within the ``region of isotropy'' of the
PM flux (see Ref.~\cite{Bugaev89a} for more details). Beyond the
interval $(5\div5\times10^3)$~TeV, Eq.~({\ref{fitPMdif}) can be
used as an extrapolation of our result which would suffice for
calculating the muon DIR.

It is interesting to note that the RQPM and QGSM predict very
different values for the muon charge
ratio~\cite{footnoteChargeRatio}.
The energy dependencies of the charge ratios may be approximated by
\[
\frac{{\cal D}_{\mu^+}^{\rm pr}}{{\cal D}_{\mu^-}^{\rm pr}} =
\left\{\begin{aligned}
0.864-0.006\log^2\left(E/E_R\right)  & \quad \text{for RQPM}, \\
1.250+0.008\left(E/E_R\right)^{0.73} & \quad \text{for QGSM},
\end{aligned}\right.
\]
with $E_R=10$~TeV. These approximations are valid in the energy
range $3 \div 10^3$~TeV at all zenith angles with an accuracy better
than 2\,\%.

From Eq.~(\ref{fitPMdif}) we find the following expression for the
integral PM spectrum:
\[
{\cal I}_\mu^{\rm pr}\left(E,h=1030\;\mbox{g/cm}^2,\vartheta =
      0^\circ\right) = \frac{C'E_b}{(\gamma'-1)(1-a)}\left\{\left[
    1+\left(\frac{E_b}{E}\right)^{\gamma'-1}\right]^{1-a}-1\right\}.
\]
A comparison of our calculation of the PM flux with the results of
other authors can be found in
Refs.~\cite{Bugaev89a,Bugaev87-88,Bugaev89b} (see also~\cite{Pal94}).

According to Ref.~\cite{BNO90}, the differential and integral PM
spectra calculated in the VFGS model can be approximated (at all
zenith angles) by
\begin{align*}
{\cal D}_\mu^{\rm pr}\left(E,h=1030\;\text{g/cm}^2,\vartheta\right)
              & = 2.92\times10^{-5}E^{-2.48}\;
          {\rm cm}^{-2}{\rm s}^{-1}{\rm sr}^{-1}{\rm GeV}^{-1}, \\
{\cal I}_\mu^{\rm pr}\left(E,h=1030\;\text{g/cm}^2,\vartheta\right)
              & = 1.97\times10^{-5}E^{-1.48}\;
          {\rm cm}^{-2}{\rm s}^{-1}{\rm sr}^{-1}.
\end{align*}
($E$ in GeV.) This approximation holds true to about $10^3$ TeV.

\protect\section{Calculated sea-level muon spectra vs experiment}
\label{sec:SLM}

Comparison of the calculated differential and integral muon spectra
with direct data from spectrometers and indirect data extracted from
underground measurements is shown in Figures~\ref{fig:Dif} (a,b)
and \ref{fig:Int} (a,b). The ground-based measurements can be
classified as absolute and non-absolute (normalized). In line with
this arrangement we present here the following three groups of
experiments.

\vspace{3mm}

\begin{quote}
\begin{itemize}
\item{\sf \underline{Absolute ground-based measurements}}
\vspace{2mm}\\
     with MARS apparatus in Durham (Aurela {\em et al.}~%
     \cite{MARS63}, Ayre {\em et al.}~\cite{MARS75});
     Nottingham spectrograph (Baber {\em et al.}~\cite{Baber68},
     Rastin~\cite{Rastin84});
     spectrometer near College Station, Texas (Bateman
     {\em et al.}~\cite{Bateman71});
     Kiel spectrographs (Allkofer {\em et al.}~\cite{Allkofer71}),
     MASS apparatus at Prince Albert, Saskatchewan (De Pascale
     {\em et al.}~\cite{MASS93});
     EAS-TOP array at Campo Imperatore, Gran Sasso (Aglietta
     {\em et al.}~\cite{EAS-TOP95}).

\item{\sf \underline{Non-absolute ground-based measurements}}
\vspace{2mm}\\
     with Durgapur spectrograph (Nandi and Sinha~\cite{Nandi72}, the
     data were normalized to the Nottingham spectrum~\cite{Rastin84}
     at $p = 20$ GeV/c);
     Durham spectrograph MARS (Thompson {\em et al.}~\cite{MARS77},
     the data were normalized to the previous MARS
     results~\cite{MARS75} at 261 GeV/c);
     L3 detector at CERN, (Bruscoli and Pieri~\cite{L3Cosmic93},
     the absolute intensity in the momentum range 40--70 GeV/c and
     its error were taken from the Kiel result~\cite{Allkofer71}).

\item{\sf \underline{Indirect data}}
\vspace{2mm}\\
     from several detectors in the Kolar Gold Fields
     (Ito~\cite{KGF90a}, Miyake {\em et al.}~\cite{KGF64b}, Adarkar
     {\em et al.}~\cite{KGF90b});
     unimodular scintillation detector ``Collapse'' of the Institute
     for Nuclear Research (INR) at the Artyomovsk Scientific
     Station (Khalchukov {\em et al.}~\cite{Collapse85});
     Baksan underground scintillation telescope of INR situated in
     North Caucasus (Andreyev {\em et al.}~\cite{BNO87,BNO90},
     Bakatanov {\em et al.}~\cite{BNO92});
     X-ray emulsion chambers of Moscow State University situated in
     the Moscow metro (Zatsepin {\em et al.}~\cite{MSU94});
     proton decay detector Fr\'{e}jus under the Alps
     (Rhode~\cite{Frejus94}),
     detector MACRO at the Gran Sasso National Laboratory
     (Ambrosio {\em et al.}~\cite{MACRO95}).
\end{itemize}
\end{quote}

\vspace{3mm}

The marked curves in Figures~\ref{fig:Dif} and \ref{fig:Int} refer to
the differential and integral muon spectra, respectively, calculated
without the PM contribution (``$\pi,K$''-muons) and with the PM
contribution according to the three charm production models (QGSM,
RQPM, and VFGS) under consideration. As seen from the Figures, the
PM contribution to the sea-level muon flux calculated with the QGSM
is very small: up to $p = 100$ TeV/c it does not exceed 16\,\% for
the differential spectrum and 22\,\% for the integral spectrum.

Unfortunately, it is difficult to extract some quantitative
assessment for the validity of our nuclear cascade model from the
presented set of data even at $p \lesssim 1$~TeV/c. As is seen
from Figures~\ref{fig:Dif}\,(a) and~\ref{fig:Int}\,(a), a wide
disagreement between the results of different experiments takes place
despite the fact that the quoted errors are relatively small in the
majority of the experiments. It indicates the existence of
significant systematic errors in some experiments which may be as
much as (30--35)\,\% at momenta 10 to 1000~GeV/c.

It should be noted in this connection that only statistical errors
are indicated in the data points of the MASS experiment. According
to Ref.~\cite{MASS93}, the systematic errors in the MASS experiment
may be as much as 15\,\% at $p \gtrsim 40$~GeV/c. The systematics in
the non-absolute measurements is, as a general rule, unknown. For
example, no attempt was made to estimate the systematic errors in
the CERN L3 experiment~\cite{L3Cosmic93}. In our opinion, the L3
spectrum was underestimated owing to incompletely correct
normalization.

At $p \lesssim 2$\,TeV/c our prediction, regardless of the
charm production model, is in very good agreement with the
Nottingham direct and absolute measurements~\cite{Rastin84}.

\clearpage
\begin{figure}[t]
\center{\mbox{\epsfig{file=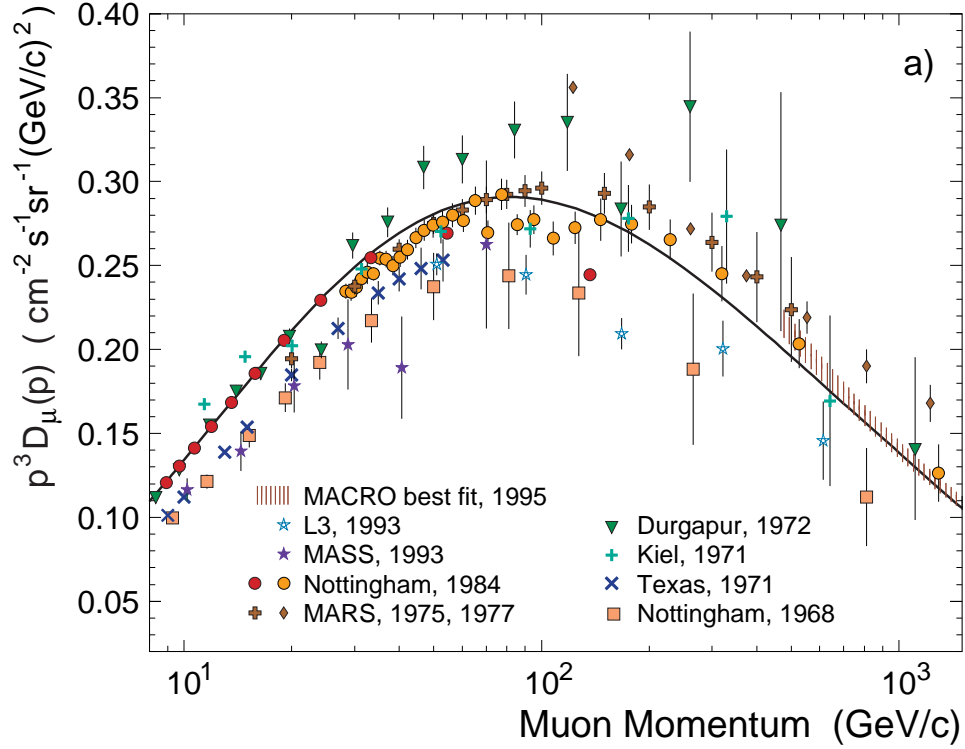,height=10.0cm}}
\vskip 5mm
        \mbox{\epsfig{file=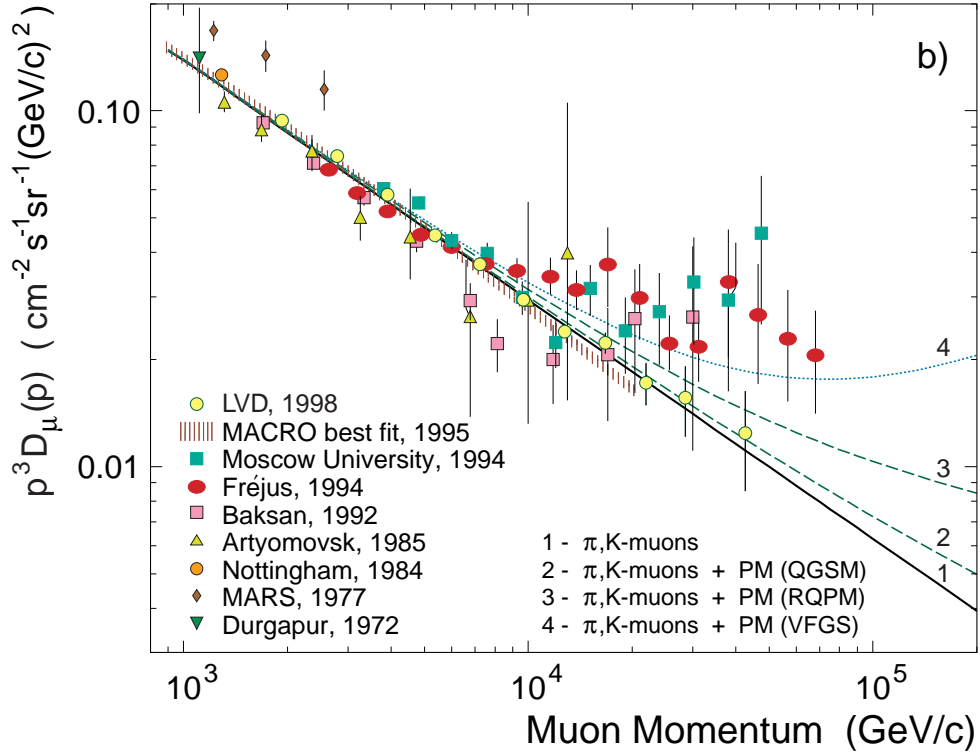,height=10.0cm}}}
\vspace{1.0cm}
\protect\caption[Differential muon momentum spectrum at sea level]
                {Vertical differential momentum spectrum of muons
                 at sea level. The direct data are taken from
                 Refs.~\protect\cite{Baber68,Bateman71,Allkofer71,%
                 Nandi72,MARS75,MARS77,Rastin84,MASS93,L3Cosmic93}
                 and indirect (underground) data are from
                 Refs.~\protect\cite{MACRO95,LVD98,Collapse85,%
                 BNO92,Frejus94,MSU94}. The shaded areas are for the
                 MACRO fit~\protect\cite{MACRO95}.
                 The solid curves represent the results of this
                 work for the conventional ($\pi,K$) differential
                 muon spectrum and for the $\pi,K$ muon spectrum
                 plus the PM contribution calculated according to
                 QGSM, RQPM, and VFGS.
\label{fig:Dif}}
\end{figure}

\clearpage
\begin{figure}[t]
\center{\mbox{\epsfig{file=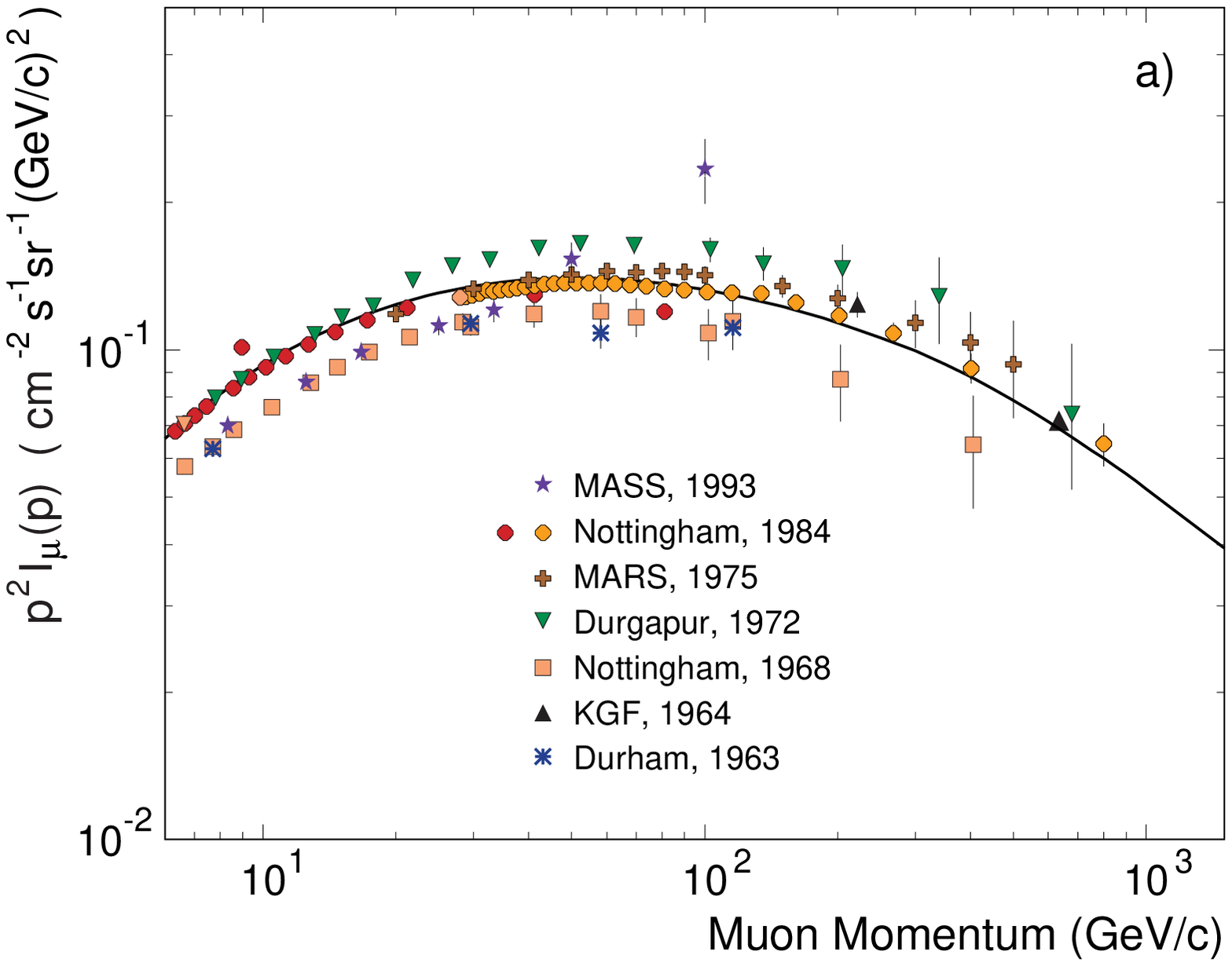,height=10.0cm}}
\vskip 5mm
        \mbox{\epsfig{file=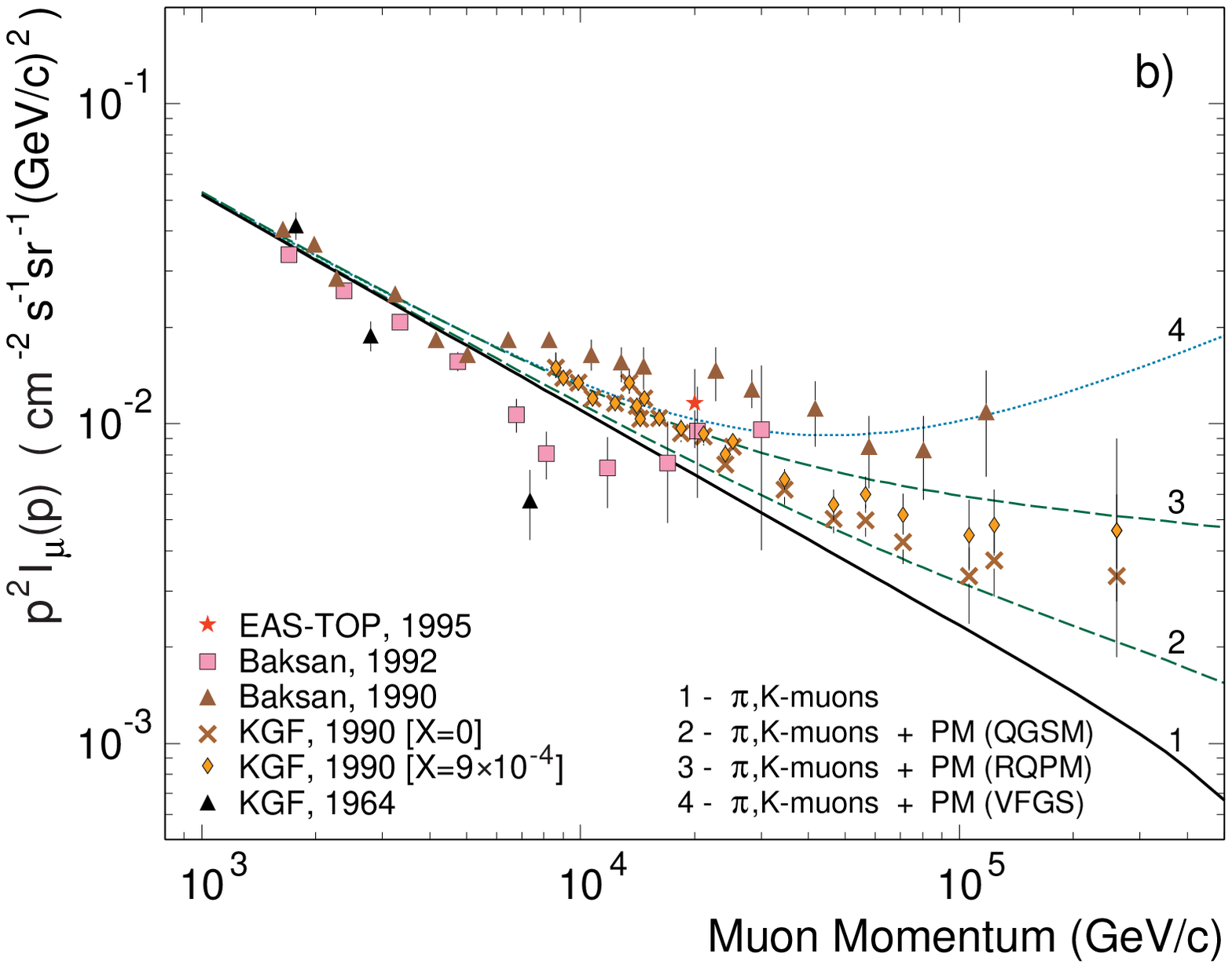,height=10.0cm}}}
\vspace{1.0cm}
\protect\caption[Integral muon momentum spectrum at sea level]
                {Vertical integral momentum spectrum of muons at
                 sea level. The direct data are taken from
                 Refs.~\protect\cite{MARS63,Baber68,Nandi72,%
                 MARS75,Rastin84,MASS93,EAS-TOP95} and indirect
                 (underground) data are from Refs.~\protect\cite%
                 {BNO90,KGF90a,KGF64b,BNO92}.
                 The solid curves represent the results of this
                 work for the conventional ($\pi,K$) integral muon
                 spectrum and for the $\pi,K$ muon spectrum plus the
                 PM contribution calculated according to QGSM, RQPM,
                 and VFGS.
\label{fig:Int}}
\end{figure}

\clearpage

At energies above a few TeV we only have indirect data at our
disposal~\cite{footnoteHorFlux} and the uncertainties (both
statistical and systematic) are vastly greater here.
The data of Refs.~\cite{BNO90,KGF90a,MACRO95,KGF64b,KGF90b,Frejus94}
have been deduced from the muon DIR measured in different rocks
(Baksan, Kolar, Alpine, Gran Sasso). We will dwell on the initial
underground data in Section~\ref{sec:DIR}. Here,
it should be pointed out that in all underground experiments, among
the systematic uncertainties related to inhomogeneities in density
and chemical composition of the matter overburden, topographical map
resolution, muon range-energy relation, muon range fluctuations,
effective differential aperture of the array, etc{}., another
uncertainty is essential. It results from the necessity to
assign some model for the energy spectrum and zenith-angle
distribution of muons at sea level which are functions of the PM
fraction in the muon flux or, to be more specific, the ratio $X$ of
prompt muon spectrum to the $\pi+K$ production one.
Hence one is forced to assume some value of the ratio $X$ (as a
function of energy) when reconstructing the vertical muon spectrum
on surface.  But the greater the adopted value of $X$, the harder
the resultant spectrum. For this reason alone the conversion
procedure is fairly ambiguous.

As an illustration we consider the KGF results. The KGF muon
spectrum in the energy range $(200\div7500)$ GeV was
deduced~\cite{KGF64b} using the underground data from
Ref.~\cite{KGF64a} and assuming $X = 0$, what is quite reasonable
for this range. But the data at higher energies~\cite{KGF90a} (see
also Ref.~\cite{KGF90b}) demand a nonzero $X$. To estimate the ratio
$X$, the authors have assumed a pion production spectrum of the form
$F(E_\pi) \propto E_\pi^{-\gamma}$ and a $K/\pi$ ratio of 0.15.  The
$X$ ratio was assumed to be a constant. Then a $\chi^2$ analysis
indicated that with $\gamma = 2.7$ for muon energy of 8 to 250 TeV,
there is PM production at the level of $X = (9\pm2)\times10^{-4}$.
In Figure~\ref{fig:Int}\,(b), we show this result (the corresponding
data points are represented by diamonds) together with the spectrum
deduced on the assumption that $X = 0$ (the data points are
represented by symbols $\times$). As would be expected, the spectrum
reconstructed with $X = 0$ is softer. It is not difficult to
understand that the final result is subject to variation also in
response to variation of the adopted $K/\pi$ ratio and
$\gamma$~\cite{footnoteKGF1}. It should also be recognized that the
real spectra of muons and mesons are far short of being power-law
ones.

Let us touch briefly on some essential points of the rest of the
underground data presented in Figures~\ref{fig:Dif}\,(b) and
\ref{fig:Int}\,(b).

In the Baksan experiment~\cite{BNO90}, $X=(1.5\pm0.5)\times10^{-3}$
was found as the best fit of the calculated total intensity of
conventional and prompt muons to the experimental data, assuming a
power-law primary nucleon flux with the spectral index
$\gamma_N = 1.65$.

In Ref.~\cite{Frejus94} the complete data set of downgoing muons
recorded with the Fr\'{e}jus detector~\cite{Frejus89-90} has
been reanalyzed. However in this analysis, the sea-level spectrum
was derived using in essence the continuous loss approximation with
some effective and energy-independent energy loss coefficients. The
muon range fluctuations are discussed in Ref.~\cite{Frejus94}
exclusively to estimate the uncertainty of the analysis. But it is
a matter of common knowledge that, on calculation of the muon DIR,
the continuous loss approximation results in downward bias and the
corresponding error increases fast with depth~\cite{Zatsepin62,%
Takahashi84,Bugaev85b,Naumov93}. It is our opinion that the muon
spectrum obtained in Ref.~\cite{Frejus94} was significantly
overestimated while the systematic errors were underestimated for
$E \gtrsim 10$~TeV in consequence of the oversimplified analysis.

The MACRO fit~\cite{MACRO95} presented in
Figs.~\ref{fig:Dif}\,(a,b) by shaded areas has the following form:
\begin{equation}\label{MACROfit}
{\cal D}_\mu^{\rm MACRO}
\left(E,h=10^3\;\mbox{g/cm}^2,\vartheta\right)
        = C_0\left(\frac{E}{1\;\mbox{GeV}}\right)^{-\gamma_\mu}
                           \left(\frac{1}{1+\frac{1.1E\cos\vartheta}
{115\,\mbox{\small\rm GeV}}}+\frac{0.054}{1+\frac{1.1E\cos\vartheta}
{850\,\mbox{\small\rm GeV}}}\right),
\end{equation}
with $C_0 = (0.26 \pm 0.01)$~cm$^{-2}$s$^{-1}$sr$^{-1}$GeV$^{-1}$
and $\gamma_\mu=2.78\pm0.01$. 
The quoted errors are due to statistics and the topographical map
resolution. According to Ref.~\cite{MACRO95}, the overall systematic
error resulting from rock density uncertainties and hard energy loss
cross sections is about 5\,\% in $C_0$ and, what is much more
important, 3\,\% in $\gamma_\mu$. But a 3\,\% variation in
$\gamma_\mu$ corresponds to uncertainties of 47\,\%, 78\,\% and more
than 100\,\% in the surface muon flux at energies of $10^2$, $10^3$
and $10^4$~GeV, respectively. Therefore, the result of MACRO is
greatly uncertain and {\em pro forma} it is not in contradiction with
all the rest of indirect data shown in Figure~\ref{fig:Dif}.

The results of the rest of the underground experiments, were obtained
with quite different methods.  The experiment with the Artyomovsk
100-ton installation ``Collapse''~\cite{Collapse85} (situated in a
salt mine at the depth of 570 m w.e.) detects the energy release of
the showers produced by cosmic-ray muons in the salt and scintillator
($C_{10}H_{22}$).  In the Baksan ``calorimetric''
experiment~\cite{BNO92}, the integral muon intensity at the position
of the scintillation telescope (8.5 km w.e.) was evaluated from the
spectrum of electromagnetic cascades generated by muons in the
telescope. To find the muon intensity at the surface, the authors
used a conversion procedure similar to that which was used in
Refs.~\cite{BNO87,BNO90}. Due to a 10\,\% error in the calibration of
the energy evolution in the detector, the systematic error in the
determination of the absolute muon intensity can reach 25\,\% in this
experiment. One might expect a supplement systematic uncertainty due
to the conversion procedure. Comparing with the results of
other experiments, the authors moved up their data by
12\,\%.  We use the same normalization in Figs.~\ref{fig:Dif}\,(b)
and~\ref{fig:Int}\,(b).  The data of Moscow State University
(MSU)~\cite{MSU94} were extracted from a multidimensional analysis of
the measured energy and angular distributions of electron-photon
cascades generated by muons in X-ray emulsion chambers. However, the
output of this method is also very sensitive to the adopted models
for the primary spectrum and charm production.  According to
Refs.~\cite{MSU94}, the estimated primary spectrum index is
$\gamma_N=1.64\pm0.03$ at nucleon energies $(20\div400)$~TeV and the
best-fit $X$ ratio changes from $(2.6\pm0.8)\times10^{-3}$ at
$E=5$~TeV to $(3.3\pm1.0)\times10^{-3}$ at $E=40$~TeV.

Up to about 5 TeV/c, our prediction for the differential and
integral spectra (irrespective of a charm production model) does
not contradict to the results of the Artyomovsk detector and the
X-ray emulsion chambers of MSU. Below a few TeV the predicted
spectrum agrees well with the data from Baksan, Fr\'{e}jus and
MACRO extracted from the muon DIR, as well as with the Baksan data
obtained from the spectrum of electromagnetic
cascades~\cite{BNO92}.

The region from 5 to 15--20~TeV/c is rather oracular: the data of
KGF~\cite{KGF64b},Artyomovsk~\cite{Collapse85}, Baksan~\cite{BNO92}
and one data point of MSU~\cite{MSU94} show a broad dip in the
differential and/or integral spectra, 
whereas the rest of the data indicates some flattening or even a
bulge~\cite{BNO90}.

Above $\sim 20$ TeV/c, the data of Baksan~\cite{BNO90},
MSU~\cite{MSU94} and Fr\'{e}jus~\cite{Frejus94} clearly indicate a
significant flattening of the muon spectrum. Neither the QGSM nor the
RQPM can explain this effect; even the maximum VFGS flux is not
sufficiently large to this end, although the VFGS flux is not in
contradiction with these data. It will be demonstrated in
Section~\ref{sec:DIR} that this flattening is not confirmed by the
body of direct underground data, while the late result of
KGF~\cite{KGF90a} seems to be (somewhat) more credible. It is also of
interest that, irrespective of a charm production model, our
prediction for the horizontal muon spectrum is in agreement with the
corresponding MSU data~\cite{MSU94} up to about 40~TeV/c. The KGF
spectrum~\cite{KGF90a} obtained at $X=(9\pm2)\times10^{-4}$ is in
qualitative agreement with the RQPM prediction. Apparently the
inconsistency of the data from different experiments gives no way of
deducing a definite conclusion about the PM fraction in the
sea-level muon flux.

\protect\section{Muon propagation through matter}
\label{sec:Prop}

To calculate the muon depth-intensity relation we apply the
semianalytical method proposed in Ref.~\cite{Naumov93}. The method
allows us to avoid any simplifying assumptions about the scale
invariance of the cross sections for radiative (direct $e^{+}e^{-}$
pair production, bremsstrahlung) and photonuclear interactions of
muons with matter, and to take into account the real non-power-law
behavior of the muon boundary spectrum. The solution to the transport
equation for the differential muon intensity, ${\cal D}_\mu(E,h)$, is
constructed by iterations, starting from an initial approximation
with the correct high-energy asymptotic behavior. Let us sketch the
basic ideas and formulas.

The equation describing the high-energy muons propagation through a
homogeneous medium may be written
\begin{equation} \label{TE}
\frac{\partial}{\partial h}{\cal D}_\mu(E,h)
                                 -\frac{\partial}{\partial E}
    \left[{\cal B}(E){\cal D}_\mu(E,h)\right] =
         \sum_{k=p,b,n}\int_0^1\left[(1-v)^{-1}\Phi_k(v,E_v)
         {\cal D}_\mu(E_v,h)-\Phi_k(v,E){\cal D}_\mu(E,h)\right]dv,
\end{equation}
with the boundary condition ${\cal D}_\mu(E,0) = {\cal D}_0(E)$.
Here ${\cal D}_0(E)$ is the ground-level muon spectrum,
${\cal B}$ is the rate of the muon energy loss those are
treated as continuous. In the present calculation, ${\cal B}$
includes the ionization energy loss and the part of the loss due to
$e^+e^-$ pair production with $v < v_0 = 2\times10^{-4}$, where $v$
is the fraction of the energy lost by muon
(see Appendix~\ref{app:MuInt} and Ref.~\cite{Bugaev93}).
However the method is independent of the specific choice of
${\cal B}(E)$. The right-hand side of Eq.~(\ref{TE}) describes the
``discrete'' muon energy loss resulting from direct $e^+e^-$ pair
production with $v > v_0$ ($k=p$), bremsstrahlung ($k=b$), and
inelastic nuclear scattering ($k=n$). The corresponding macroscopic
cross sections, $\Phi_k(v,E)$, are defined by
\[
\Phi_k(v,E) = N_0\frac{ d\sigma_k(v,E)}{dv} =
            N_0E\frac{d\sigma_k(E,E')}{dE'}\Big|_{E'=(1-v)E},
\]
where $N_0$ is the number of atoms per 1~g of the matter and $E$
($E'$) is the initial (final) muon energy. It is implied that the
differential cross sections are averaged over the atomic number and
weight of the target nuclei and $d\sigma_k(v,E)/dv = 0$ outside the
ranges $0 \leq v_k^{\min}(E) < v_k^{\max}(E) \leq 1$ allowed by
kinematics. Lastly, $E_v\equiv E/(1-v)$. The summary of the explicit
formulas for the cross sections used in our calculation is presented
in Appendix~\ref{app:MuInt}.

Let us seek the solution of the transport equation~(\ref{TE})
in the form
\begin{equation} \label{Ansatz}
{\cal D}_\mu (E,h) = {\cal D}_0({\cal E}_\varrho(E,h))
\exp\left[-{\cal K}(E,h)\right][1+\delta(E,h)].
\end{equation}
The functions involved are defined by the following chain of
equations:
\[
{\cal K}(E,h) = \int_E^{{\cal E}_\varrho(E,h)}
                  \frac{\xi(E')-\zeta(E')\varrho(E')-{\cal B}'(E')}
                                 {{\cal B}(E')+\varrho(E')E'}\,dE',
\]
\[
\varrho(E) = \sum_k\int_0^1\Phi_k(v,E_v)\eta(v,E)vdv, \qquad
    \xi(E) = \sum_k\int_0^1[\Phi_k(v,E)-\eta(v,E)\Phi_k(v,E_v)]dv,
\]
\[
  \zeta(E) = -\frac{E{\cal D}'_0(E)}{{\cal D}_0(E)}, \qquad
 \eta(v,E) =  \frac{{\cal D}_0(E_v)}{(1-v){\cal D}_0(E)}.
\]
[As is easy to see, $\zeta=\gamma+1$ and $\eta(v,E)=(1-v)^\gamma$
in the special case of a power-law boundary spectrum,
${\cal D}_0(E) \propto E^{-(\gamma+1)}$.] The function
${\cal E}_\varrho(E,h)$ is the only root of the equation
\begin{equation} \label{epsi}
\int_E^{{\cal E}_\varrho}\frac{dE'}{{\cal B}(E')+\varrho(E')E'} = h;
\end{equation}
it can be treated as the effective energy, which a muon must have
at the boundary of the medium in order to reach the depth $h$ having
energy $E$ with a nonzero probability. Lastly, the function
$\delta(E,h)$ satisfies the equation
\begin{eqnarray} \label{corr}
\left[\frac{\partial}{\partial h}-{\cal B}(E)
\frac{\partial}{\partial E}\right]\delta(E,h)           & = &
\sum_k\int_0^1\Phi_k(v,E_v)\left\{\Omega\left(E,E_v,h\right)
                  [1+\delta(E_v,h)]\right. \nonumber \\ &   &
-\left.[1+\omega(E,h)v][1+\delta(E,h)]\right\}\eta(v,E)dv,
\end{eqnarray}
with
\[
\Omega(E,E_v,h) = \frac{{\cal D}_0(E)}{{\cal D}_0(E_v)}
\frac{{\cal D}_0\left({\cal E}_\varrho(E_v,h)\right)}
     {{\cal D}_0\left({\cal E}_\varrho(E,h)\right)}
               \exp\left[{\cal K}(E,h)-{\cal K}(E_v,h)\right],
\]
\[
\omega(E,h) = \frac{Q(E)-Q\left({\cal E}_\varrho(E,h)\right)}
              {{\cal B}(E)+\varrho(E)E}, \qquad
       Q(E) = [\xi(E)-{\cal B}'(E)]E+\zeta(E){\cal B}(E).
\]
Clearly $\delta (E,0)=0$. We shall seek the solution to
Eq.~(\ref{corr}) using an iteration procedure. It is based on the
following consideration.

Let us suppose that the functions $\Phi_k(v,E)$ and $\zeta(E)$
become energy-independent as $E\rightarrow\infty$. If so, it is a
matter of direct verification to prove that the asymptotic behavior
of the function $\delta(E,h)$ is $c_2(h)/E^2$ with $c_2(h)$ an
$E$-independent function. Hence it follows that
$\delta(E_v,h)-(1-v)^2\delta(E,h)\propto(1-v)^2vE^{-3}$ as
$E\rightarrow\infty$. Thus, putting $\delta^{(1)}(E,h) = 0$ as a
first approximation for the function $\delta(E,h)$, the second one
can be found from the equation
\[
\left[\frac{\partial}{\partial h}-{\cal B}(E)\frac{\partial}
{\partial E}-R_2(E,h)\right]\delta^{(2)}(E,h) = \Re_1(E,h),
\]
where we introduced
\[
R_l(E,h) = \sum_k\int_0^1\Phi_k(v,E_v)
  \left\{\Omega\left(E,E_v,h\right)(1-v)^l-[1+\omega(E,h)v]\right\}
                                         \eta(v,E)dv, \quad l\geq 0
\]
and $\Re_1(E,h) \equiv R_0(E,h)$.
Repeating the consideration, one can proof by induction
that $\delta(E,h)-\delta^{(l)}(E,h) \propto c_l(h)/E^l$ as
$E\rightarrow\infty$. Let us define
\[
\Theta_l(E,h)=\delta^{(l)}(E,h)-\delta^{(l-1)}(E,h), \quad l\geq 2,
\]
\[
\Re_l(E,h) = \sum_k\int_0^1\Phi_k(v,E_v)\Omega\left(E,E_v,h\right)
[\Theta_l(E_v,h)-(1-v)^l\Theta_l(E,h)]\eta(v,E)dv, \quad l \geq 2.
\]
Then the following recursion chain of equations for the functions
$\Theta_l(E,h)$ is derivable from the above reasoning:
\begin{equation} \label{Theta}
\left[\frac{\partial}{\partial h}-{\cal B}(E)\frac{\partial}
{\partial E}-R_l(E,h)\right]\Theta_l(E,h) = \Re_{l-1}(E,h),
\quad l \geq 2,
\end{equation}
The solution to Eq.~(\ref{Theta}) is given by
\[
\Theta_l(E,h) = \int_0^h\exp\left[\int_{h'}^h
                 R_l\left({\cal E}_0(E,h-h''),h''\right)dh''\right]
             \Re_{l-1}\left({\cal E}_0(E,h-h'),h'\right)dh',
\]
where ${\cal E}_0(E,h)$ is the root of Eq.~(\ref{epsi}) with
$\varrho \equiv 0$ in its left-hand side.

The formal convergence of this procedure can be proved under quite
general assumptions on the energy dependence of the functions
involved; specifically if the functions ${\cal B}(E)$ and
$\Phi_k(v,E)$ increase monotonically and sufficiently slowly, while
${\cal D}_0(E)$ decreases with energy so that $\zeta(E)$ is a
slightly varying function of energy. It follows from our numerical
analysis that in ``real environment'' the rate of convergence is very
high:  the first approximation ($\delta(E,h) \equiv 0$) works with a
reasonable exactness up to about 6 km w.e. and 3--4 iterations are
suffice to obtain a few-percent accuracy for the differential muon
spectrum at $h \lesssim 18$~km w.e. and $E \gtrsim 1$~GeV.

The results obtained by the method being discussed agree well with
our previous calculations~\cite{Bugaev85b}. At $h \leq 16$~km w.e.
of standard rock, the method was verified by the direct Monte Carlo
calculation, using an updated version of the code by Takahashi
{\em et al.}~\cite{Takahashi84}. The accuracy of our calculation
for the muon depth-intensity relation (DIR),
\[
{\cal I}_\mu(h) = \int_{E_{\rm th}}^\infty{\cal D}_\mu(E,h)dE \qquad
(E_{\rm th} \sim 1\;\mbox{GeV}),
\]
is estimated to be under (2--3)\,\% at all depths of interest. The
systematic errors can only be caused by the uncertainties in the
input parameters, namely, the boundary muon spectrum and the
muon--matter interaction cross sections. An additional error arising
on the comparison with the data of a particular experiment, is
related to the uncertainties in the averaged density and chemical
composition of the matter overburden ($\langle \rho\rangle$,
$\langle Z \rangle$, $\langle A \rangle$, $\langle Z/A \rangle$,
$\langle Z^2/A \rangle$)~\cite{Uncertainties}. Strictly speaking,
the approximation of homogeneous medium may also introduce a
systematic error into the calculation for the real inhomogeneous
media~\cite{Unhomogeneities}.

\protect%
\section{Calculated muon DIR vs underground and underwater data}
\label{sec:DIR}

\protect\subsection{Early underground experiments}

In Figure~\ref{fig:Crouch}, we present a comparison between the
calculated vertical intensity (vs. depth underground) of conventional
muons and the data obtained in early underground experiments performed
with relatively small detectors~\cite{Wilson38,Clay39,Bollinger50,%
Randall51,Avan55,Castagnoli65,Stockel69,Bergamasco71,Crookes73} as
well as the Crouch's 1987 ``World Survey'' data~\cite{Crouch87}.
To expand the comparison, we represent in Figure~\ref{fig:Archive} a
fragment of the same information relevant to shallow depths.

The data obtained by Wilson~\cite{Wilson38} and by Clay and Van
Gemert~\cite{Clay39} in the late 1930s are rather uncertain since
the techniques used were unable to estimate the effects of showers,
scattering and $\delta$-electrons. We have normalized these data to
our curve. All the other data points in Figs.~\ref{fig:Crouch} and
\ref{fig:Archive} are absolute. The Crouch World Survey comprises
the data of different experiments, in particular, the early KGF
data~\cite{KGF64a,KGF65-71} and extensive data from East Rand
Proprietary Mine (ERPM) near Johannesburg~\cite{ERPM}) at great
depths (all the points at $h \gtrsim 7.5$~km w.e.). All the data
were converted by Crouch to standard rock ($Z = 11$, $A = 22$,
$\rho = 2.65$~g/cm${}^3$) with some correction for the depths.
Crouch's original compilation also includes the data from the
depths well beyond 18 km w.e., where the atmospheric muon
contribution is entirely negligible compared to the neutrino
induced muon flux (see below), as well as three data points from
an underwater experiment~\cite{Fyodorov85} (we dropped these
three points intending to discuss the complete set of underwater
data below).

At $h \gtrsim 11$ km w.e., the flux ${\cal I}_\mu^\nu$ of
muons produced by atmospheric neutrino interactions in the
surrounding rock becomes important. The value of ${\cal I}_\mu^\nu$
can significantly vary from one experiment to another due to
different registration thresholds, the topology of the matter
overburden, and so on. So, to account for the neutrino-induced
background, we shall use the specific experimental data
rather than some theoretical predictions. In Figure~\ref{fig:Crouch}
we use the result of Ref.~\cite{Crouch87}:
${\cal I}_\mu^\nu = (2.17\pm0.21)\times10^{-13}\;
    {\rm cm}^{-2}\,{\rm s}^{-1}\,{\rm sr}^{-1}$.

According to Crouch, the presented data at $h \gtrsim 1$~km w.e. can
be approximated by the following empirical function:
\begin{equation}\label{CrFit}
{\cal I}_\mu(h) = \exp(A_1+A_2 h)+\exp(A_3+A_4 h)+{\cal I}_\mu^\nu
\end{equation}
with $A_1 = -11.22 \pm 0.17$, $A_2 = -0.00262 \pm 0.00013$, $A_3 =
-14.10 \pm 0.14$, $A_4 = -0.001213 \pm 0.000021$ (the result of a
least square fit). Here ${\cal I}_\mu(h)$ is in
cm$^{-2}$s$^{-1}$sr$^{-1}$ and $h$ (the depth in standard rock) is in
hg/cm${}^2$ (1 hg/cm${}^2 = 1$ m w.e.). The fit~(\ref{CrFit}) is in
good agreement with the result of the Utah group~\cite{Utah}.

\clearpage

\begin{figure}[H]
\center{\mbox{\epsfig{file=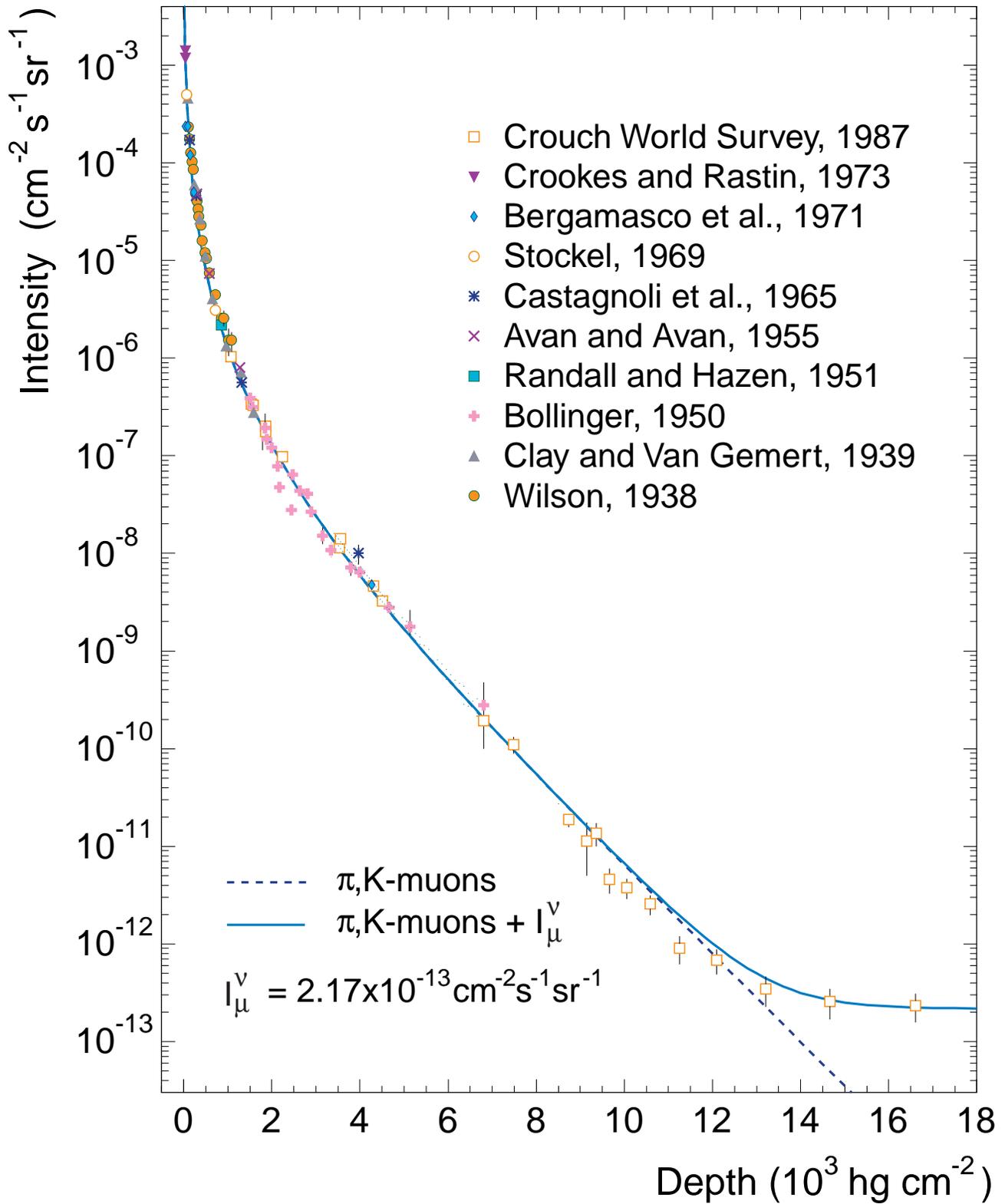,height=20.8cm}}}
\vspace{1.0cm}
\protect\caption[Muon DIR from the early underground experiments]
                {Muon intensity vs standard rock thickness. The
                 data are from Refs.~\protect\cite{Wilson38,%
                 Clay39,Bollinger50,Randall51,Avan55,Castagnoli65,%
                 Stockel69,Bergamasco71,Crookes73,Crouch87}. The
                 dashed curve represents our $\pi,K$-muon DIR, the
                 solid curve represents the same plus the
                 neutrino-induced muon background after
                 Crouch~\protect\cite{Crouch87}.
\label{fig:Crouch}}
\end{figure}

\clearpage

\begin{figure}[H]
\center{\mbox{\epsfig{file=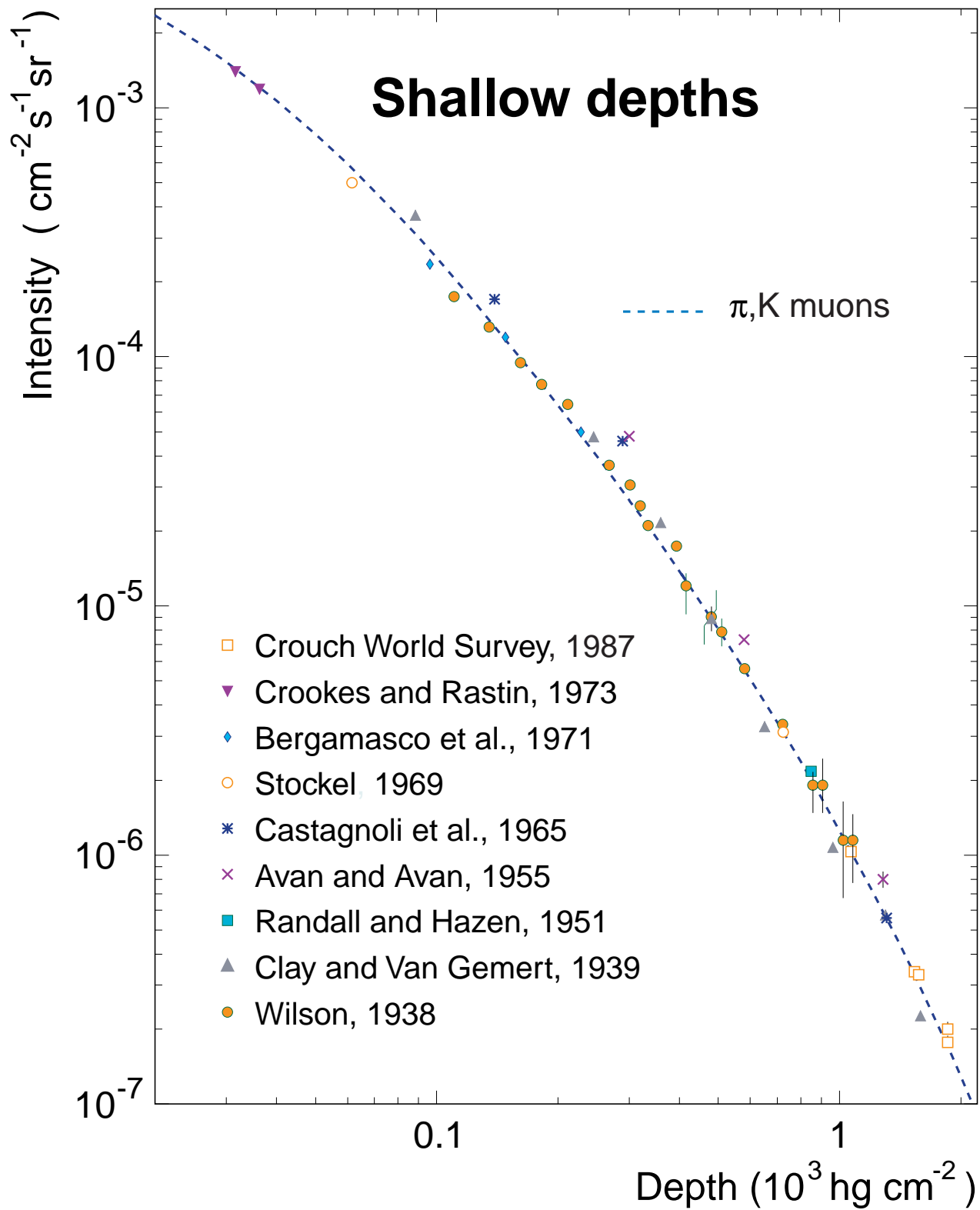,height=20.8cm}}}
\vspace{1.0cm}
\protect\caption[Muon DIR from the early underground experiments
                 (shallow depths)]
                {The same as in Figure~\protect\ref{fig:Crouch} but
                 for shallow depths.
\label{fig:Archive}}
\end{figure}

\clearpage

As Figs.~\ref{fig:Crouch},\,\ref{fig:Archive} suggest, our DIR for
conventional muons agrees well with most of the data within a wide
depth range from about 30 to 9000 m w.e. (the exceptions are the
points from Refs.~\cite{Avan55} and~\cite{Castagnoli65} at
$h \approx 300$ m w.e. and also the points from
Ref.~\cite{Bollinger50} lying in the range 2.1 to 2.5 km w.e.). The
maximum disagreement with the best fit~(\ref{CrFit}) at
$h = (1 \div 7.5)$ km w.e. is about 10\,\%. However, at
$h = (9.5 \div 12)$ km w.e., our intensity noticeably {\em exceeds}
the ERPM data. At $h = 11.5$ km w.e., the disagreement ranges up to
about 77\,\% (or about 30\,\%, if one take the experimental errors
into account). Such an error goes far beyond the expected accuracy
of our calculations and thus are attributable either to
uncertainties in the input parameters (primary spectrum?) or to some
systematics in the ERPM data.  It is clear that the data give no
indication of some PM fraction in the muon flux so we do not show the
corresponding curves here. As an illustration, let us note that, at
$h = 12$ km w.e., the calculated muon intensities with the PM
contribution which result from the QGSM, RQPM, and VFGS are
respectively 1.7, 2.3, and 3.3 times larger than the Crouch best fit.

\vspace{5mm}

\protect\subsection{Kolar Gold Fields}

Figure~\ref{fig:KGF} shows a comparison with the data obtained from
several detectors located at different levels in the deep mine of the
Kolar Gold Fields, Mysore State, South India~\cite{KGF90a} (vertical
telescopes at 745, 1500 and 3375 m w.e., a horizontal telescope at
3375 m w.e. and proton decay detectors at 6045 and 7000 m w.e.).  In
Ref.~\cite{KGF90a}, the neutrino-induced background has been
subtracted from the data using the measured angular distribution
of muons. The four curves in Figure~\ref{fig:KGF} represent our
predictions for the muon intensities without and with adding the PM
contribution from the QGSM, RQPM, and VFGS. Our calculations are done
for the Kolar rock with $\langle Z\rangle=12.9$,
$\langle A\rangle=26.9$, $\langle Z/A\rangle=0.495$,
$\langle Z^2/A\rangle=6.31$, and
$\langle\rho\rangle=3.05$~g/cm${}^3$.

Up to 6--7 km w.e., one can see an excellent agreement between our
predictions and the KGF data, irrespective of the PM flux model.
Contrary to the data presented in Figure~\ref{fig:Crouch}, the KGF
muon DIR visibly exceed the calculated $\pi,K$-muon intensity at
$h \gtrsim 7$~km w.e., hinting at some PM contribution.
Both the RQPM and the VFGS model are in agreement with the
KGF data up to about 10 km w.e., but the VFGS model better fits
the deeper data. This is not in contradiction with the
situation presented in Figure~\ref{fig:Int}\,(b) for the sea-level
integral spectrum when the ambiguities of the conversion procedure
mentioned in Section~\ref{sec:SLM} are taken into account.


\protect\subsection{Baksan}

In Figure~\ref{fig:Baksan} we show a comparison of the calculation
and the data obtained with the Baksan underground scintillation
telescope (North Caucasus, Russia).
The data obtained at zenith angles $50^\circ-70^\circ$
(Ref.~\cite{BNO87}) and $70^\circ-85^\circ$ (Ref.~\cite{BNO90})
were converted by the authors to vertical direction and to
standard rock and the neutrino-induced muon background was
subtracted from the data at high depths.
A systematic difference between the two sets of data takes
place in the depth interval from 6 to 9 km w.e.: in the first set
($50^\circ-70^\circ$) a bump of intensity is clearly visible,
while there is no such bump in the second set of data.

The authors of Ref.~\cite{BNO87} argue that the observed bump can
be interpreted in terms of prompt muons. In our view this is not
the case. The odds are that the bump is caused by errors in the
determination of the oblique depths. Beyond the interval 6--9 km
w.e., the data of both sets fall on a smooth curve. At the same
time, the data at $h \gtrsim 10$ km w.e. may be attributed to the
presence of some PM fraction in the measured underground muon
intensity. Because of rather large experimental errors any model
of charm production under consideration cannot be excluded by the
Baksan data (including the case with the zero PM contribution), but
it seems the data are more favorable for the RQPM. A collation of
Figures~\ref{fig:Baksan} and \ref{fig:Int}\,(b) suggests that
the sea-level integral spectrum reconstructed in Ref.~\cite{BNO90}
from the Baksan muon DIR was distinctly overestimated.


\protect\subsection{Mont Blanc Lab}

Figure~\ref{fig:NUSEX} shows the comparison of the predicted DIR for
the conventional muons with the {\em single muon} intensity measured
with the detectors SCE and NUSEX~\cite{Castagnoli86,NUSEX90} located
in the Mont Blanc Laboratory. Our calculation represents the muon
intensity averaged upon the muon multiplicities and therefore we can
make nothing more than qualitative conclusions from the
comparison.

In the overlapping region ($h \lesssim 7$ km w.e.) the data of both
detectors superimpose and (with allowance for the multi-muon events)
agree with our result. \hspace{1mm} However, as is clear from the
figure, the NUSEX DIR has a much greater
\clearpage

\begin{figure}[!h]
\center{\mbox{\epsfig{file=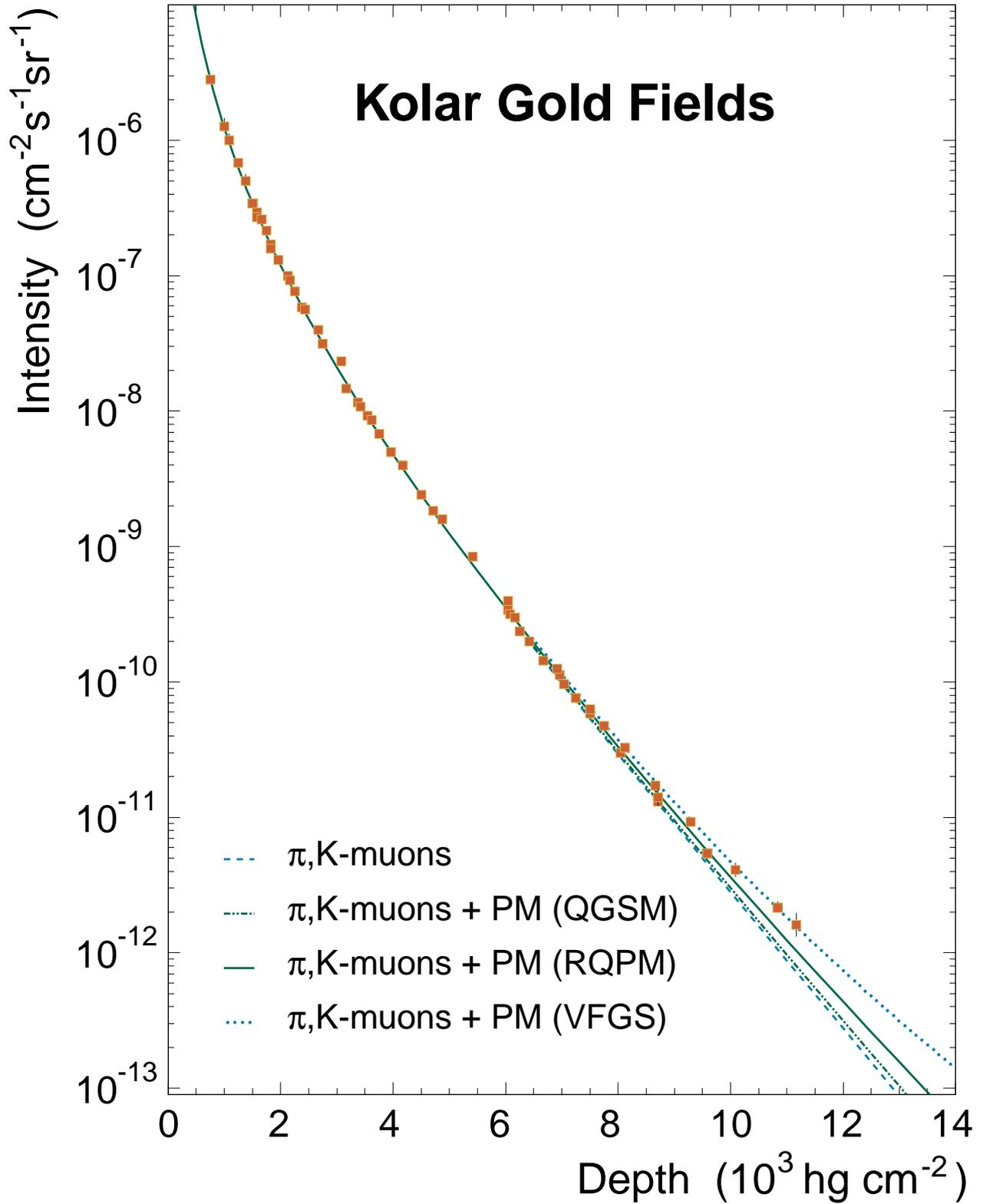,height=20.8cm}}}
\vspace{1.0cm}
\protect\caption[Muon DIR from the KGF underground experiment]
                {Muon intensity vs Kolar rock thickness. The KGF
                 data are from Ref.~\protect\cite{KGF90a}. The
                 curves are for the $\pi,K$-muon DIR and for the
                 DIR with the PM contributions calculated according
                 to the RQPM and VFGS.
\label{fig:KGF}}
\end{figure}

\clearpage

\begin{figure}[!h]
\center{\mbox{\epsfig{file=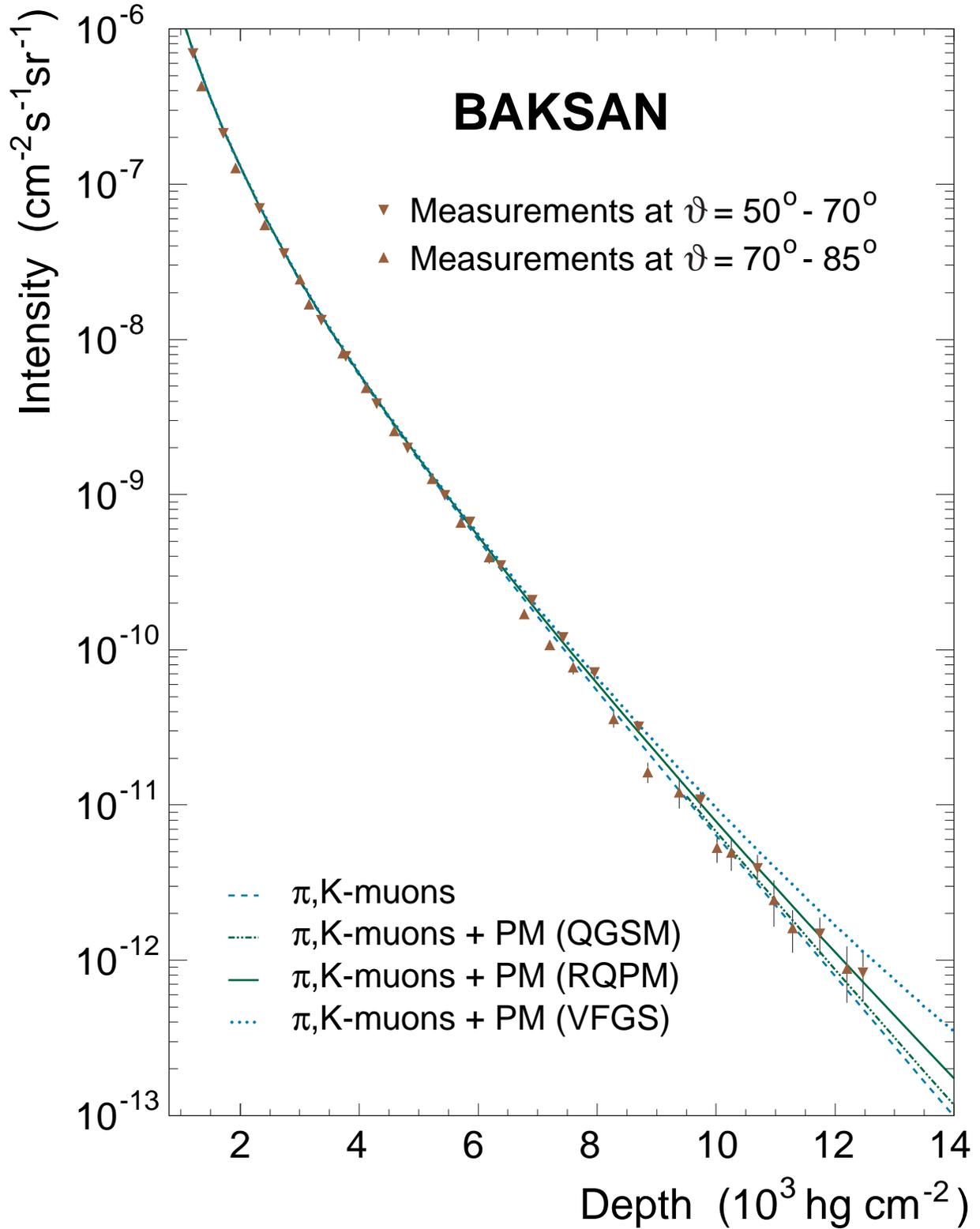,height=20.8cm}}}
\vspace{1.0cm}
\protect\caption[Muon DIR from the Baksan underground experiment]
                {Muon intensity vs standard rock thickness measured
                 with the Baksan scintillation telescope~\protect%
                 \cite{BNO87,BNO90}. The notation for the curves
                 are the same as in Figure~\protect\ref{fig:KGF}.
\label{fig:Baksan}}
\end{figure}

\clearpage

\begin{figure}[!h]
\center{\mbox{\epsfig{file=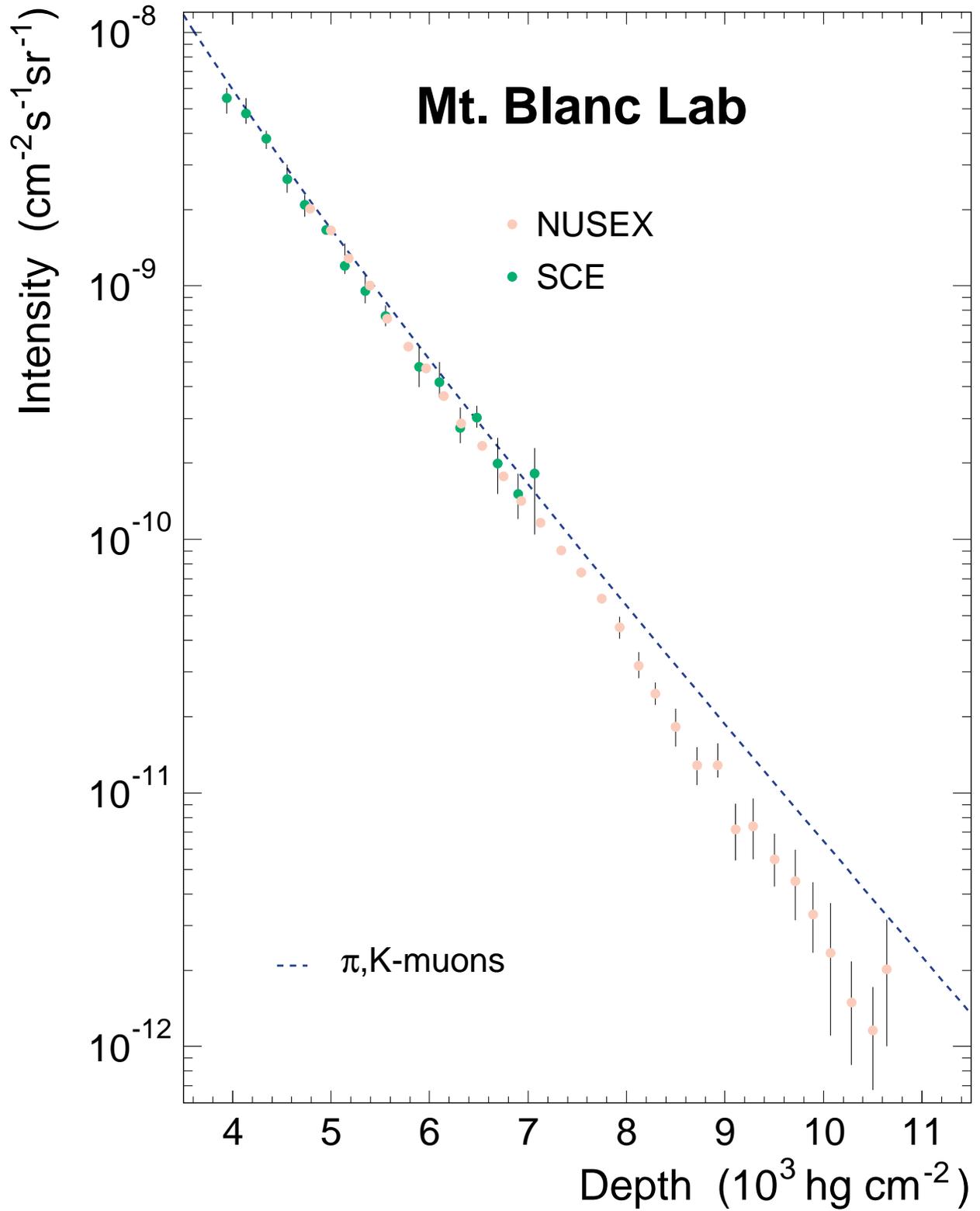,height=20.8cm}}}
\vspace{1.0cm}
\protect\caption[Muon DIR from the SCE and NUSEX underground
                 experiments]
                {Single muon intensity vs standard rock thickness
                 measured in two experiments under Mont Blanc, SCE~%
                 \protect\cite{Castagnoli86} and NUSEX\protect\cite%
                 {NUSEX90}. The curve is for the predicted
                 $\pi,K$-muon DIR.
\label{fig:NUSEX}}
\end{figure}

\clearpage

\noindent 
abrupt grade compared with the predicted one for the $\pi,K$-muon DIR
and, at $(9 \div 11)$ km w.e., the predicted intensity (without any
PM contribution) is 2--3 times higher than the NUSEX data. So large a
discrepancy has no relation to the multi-muon events, whose intensity
decreases with depth quicker than the single-muon one and is
negligible at $h > 6-7$ km w.e.  within a few percent accuracy.

It is our opinion that the NUSEX result at large depths is incorrect.
Notice that the muon DIR measured in the NUSEX experiment has been
converted to standard rock. Although the averaged values of $\rho$,
$Z$, $A$, $Z/A$ and $Z^2/A$ in the Mont Blanc rock are rather close
to the ``standard'' ones, this conversion might be a serious source
of a systematic error because of very complicated and heterogeneous
(layered) chemical composition of the rock (see
Ref.~\cite{Castagnoli86}).

We note here that our calculations are in good agreement with the
result of the French--American muon experiment~\cite{MtBlanc78} also
carried out in the Mt. Blanc tunnel with a GM telescope. The depth
range explored in that experiment was 0.5 to 5 km w.e. and therefore
overlaps in part the SCE--NUSEX depth range. This suggests that the
SCE and NUSEX experiments may have an added source of systematics
related to their experimental procedures. At the same time, the
recent NUSEX measurements of the averaged muon energy
underground~\cite{Castagnoli97} are in good agreement with our
predictions.

\protect\subsection{SOUDAN\,1/2}

Figure~\ref{fig:SOUDAN} represents the comparison of our
prediction with the data from SOUDAN\,1 and SOUDAN\,2 underground
experiments~\cite{SOUDAN90,Kasahara95} (the data points are taken
from the compilation presented in Ref.~\cite{MACRO95}).  The SOUDAN
data were normalized to DIR for standard rock using the Crouch World
Survey, as described in Ref.~\cite{Kasahara95}.

Despite some spread of experimental points and a bump at $3\div4$
km w.e., one can see a reasonable agreement between the calculated
$\pi,K$-muon DIR and the data up to about 7 km w.e., but the last
point ($\sim 8.4$ km w.e.) is almost 2.5 times below the predicted
curve, as in the case of the NUSEX data.

\protect\subsection{Fr\'{e}jus}

In Figure~\ref{fig:Frejus}, we compare our calculations with the data
of the Fr\'{e}jus detector~\cite{Frejus89-90} (the underground
laboratory was located in a tunnel of the same name under the
Alps). The Alpine rock thickness has been converted into hg/cm${}^2$
of standard rock. We do not include in our assemblage the new and
very detailed data from the Fr\'{e}jus detector recently reanalyzed
in Ref.~\cite{Frejus96} (see also Ref.~\cite{Frejus94}).
The point is that the original data sample has been subdivided into
throughgoing, multiple and stopping muons. These subsamples are very
dependent of the features peculiar to the experiment and thus cannot
be directly compared with our calculations.

According to Ref.~\cite{Frejus96}, the neutrino-induced muon
background becomes dominant at $h \gtrsim 13$ km w.e. and the
measured mean background flux is ${\cal I}_\mu^\nu =
(3.67 \pm 0.66)\times 10^{-13}$\,cm${}^{-2}$s${}^{-1}$sr${}^{-1}$.
One can see that the intensity calculated without the PM contribution
and corrected for this background fits the Fr\'{e}jus data almost
everywhere, although in the vicinity of $h = 10$ km w.e. some hint of
an excess over the data (similar to the more evident excess in ERPM
and NUSEX) does take place. The PM contributions calculated with RQPM
and VFGS do not fall into Fr\'{e}jus data. However, in view of the
experimental uncertainties there is no telling that the RQPM
prediction is in serious conflict with the Fr\'{e}jus result.
Clearly the same is all the more true for the QGSM.

\protect\subsection{Gran Sasso Lab}

The recent data from the two largest underground detectors
MACRO~\cite{MACRO95} and LVD~\cite{LVD98} (located in the Gran
Sasso Laboratory) are presented in Figs.~\ref{fig:MACRO}
and~\ref{fig:LVD}, respectively. The data of MACRO are converted to
standard rock. The error bars include statistical uncertainty,
systematic uncertainty for the topographical map and the additional
estimated systematic scale uncertainty of $\pm 8\,$\%. Taken alone,
the statistical errors in the MACRO experiment are very small. The
main contribution to the absolute scale uncertainty comes from the
assumption of a homogeneous mountain instead of a layered structure.
The data of LVD are presented both for the Gran Sasso rock
and standard rock. Errors include both statistical and systematic
uncertainties. Notice that at $h \lesssim 5$ km w.e., the
statistical errors are less than the size of the circles in the
figure. 

The depths currently accessible for observation with the detector
MACRO are insufficient to study prompt muons. Thus, in
Figure~\ref{fig:MACRO} we present the calculated $\pi,K$-muon DIR
alone. In contrast, the LVD data (Figure~\ref{fig:LVD}) overlap the
total depth range where the PM contribution might be essential.
We use for our calculated curves the following value of the
neutrino-induced muon background:
${\cal I}_\mu^\nu = (2.98 \pm 1.15) \times 10^{-13}$\,%
cm${}^{-2}$s${}^{-1}$sr${}^{-1}$~\cite{LVD95}.

\clearpage

\begin{figure}[!h]
\center{\mbox{\epsfig{file=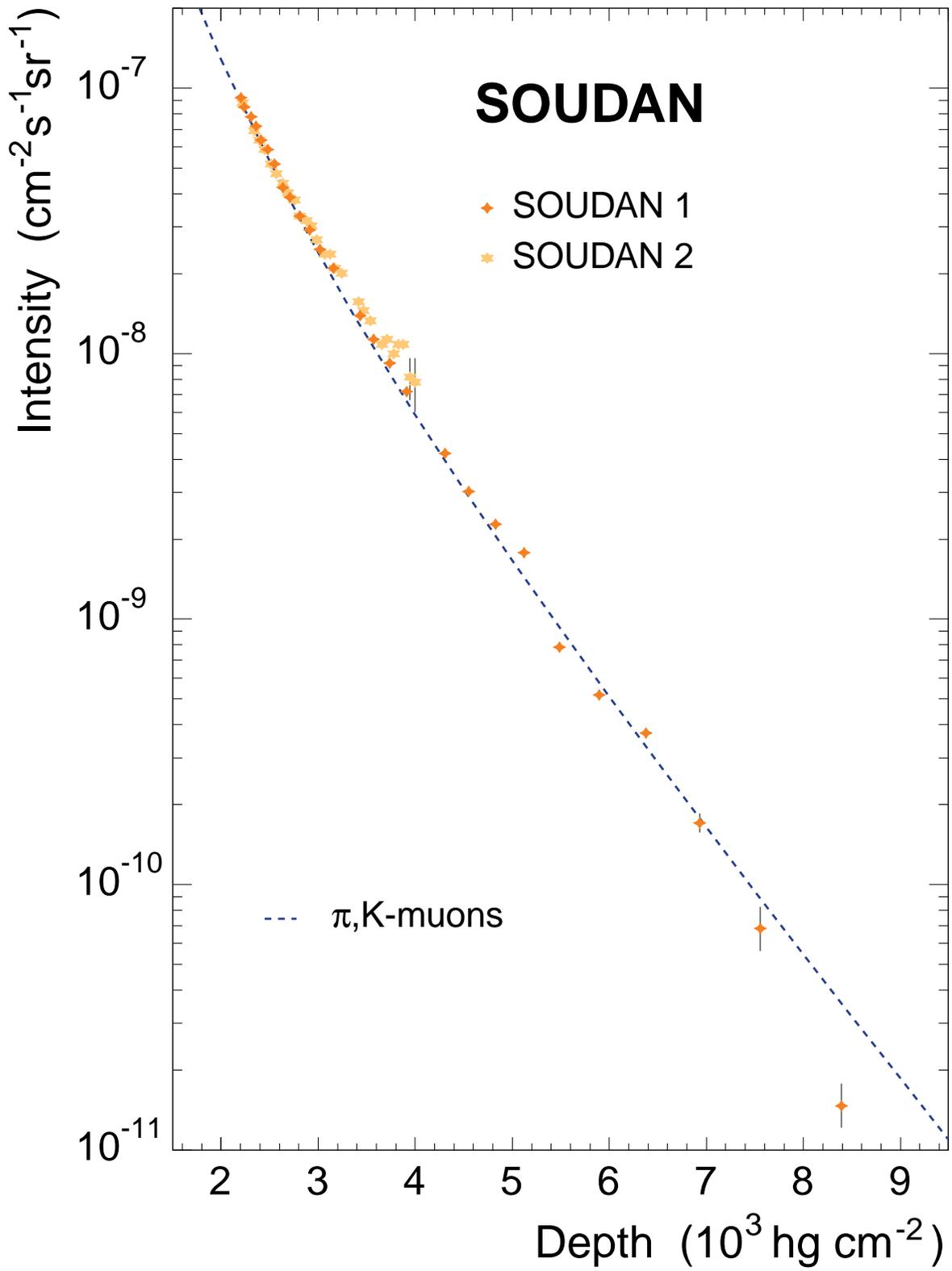,height=20.8cm}}}
\vspace{1.0cm}
\protect\caption[Muon DIR from the SOUDAN\,1 and SOUDAN\,2
                 underground experiments]
                {Muon intensity vs standard rock thickness measured
                 in the SOUDAN\,1 and SOUDAN\,2 experiments~%
                 \protect\cite{SOUDAN90,Kasahara95}. The curve is
                 for the predicted $\pi,K$-muon DIR.
\label{fig:SOUDAN}}
\end{figure}

\clearpage

\begin{figure}[!h]
\center{\mbox{\epsfig{file=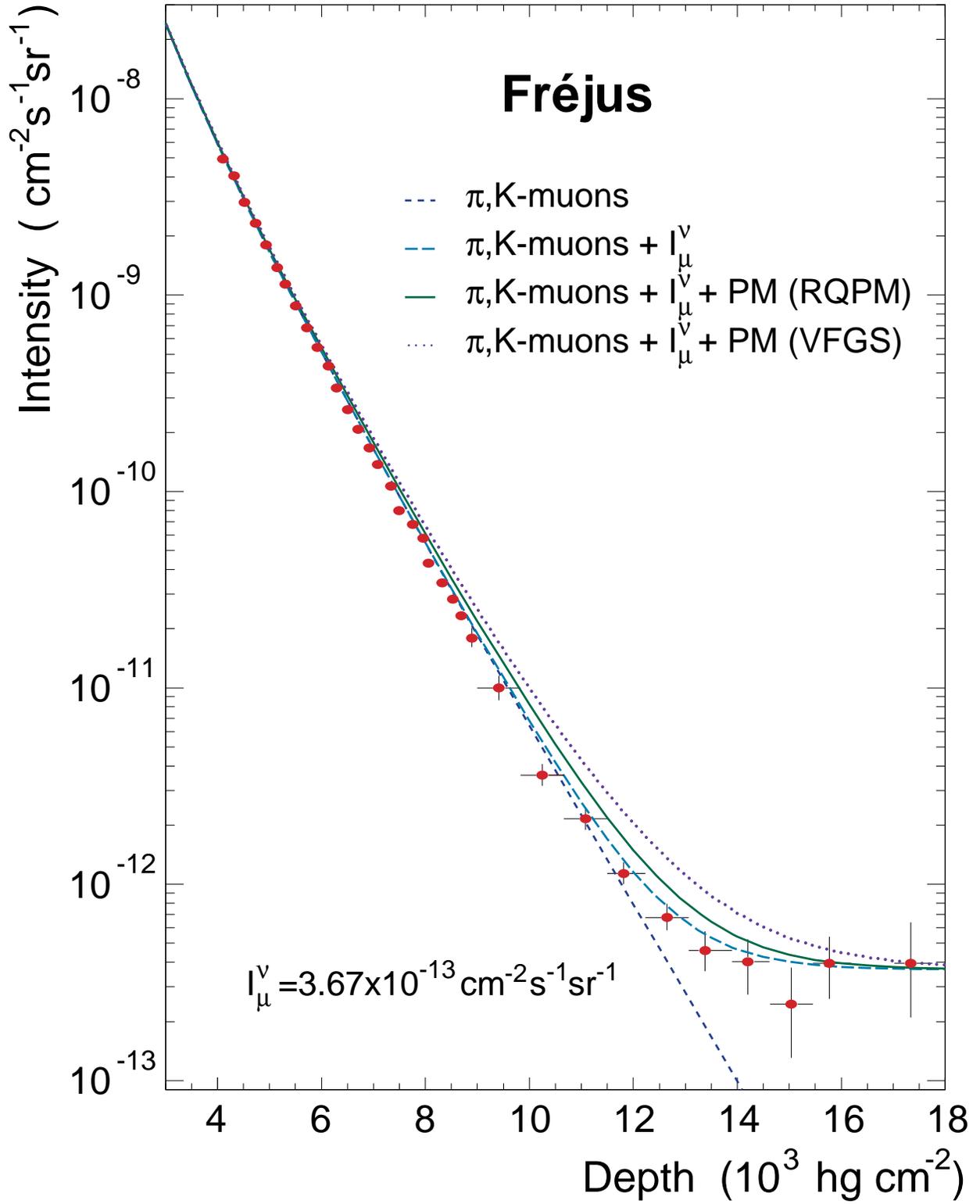,height=20.4cm}}}
\vspace{1.0cm}
\protect\caption[Muon DIR from the Fr\'{e}jus underground
                 experiment]
                {Muon intensity vs standard rock thickness measured
                 in the Fr\'{e}jus experiment~\protect\cite%
                 {Frejus89-90}. The dashed curve represents our
                 $\pi,K$-muon DIR, the solid curve represents the
                 same plus the neutrino-induced muon background,
                 ${\cal I}_\mu^\nu$, according to
                 Ref.~\protect\cite{Frejus96}. The other curves are
                 for the muon DIR with the PM contributions
                 calculated according to RQPM and VFGS plus
                 ${\cal I}_\mu^\nu$.
\label{fig:Frejus}}
\end{figure}

\clearpage

\begin{figure}[!h]
\center{\mbox{\epsfig{file=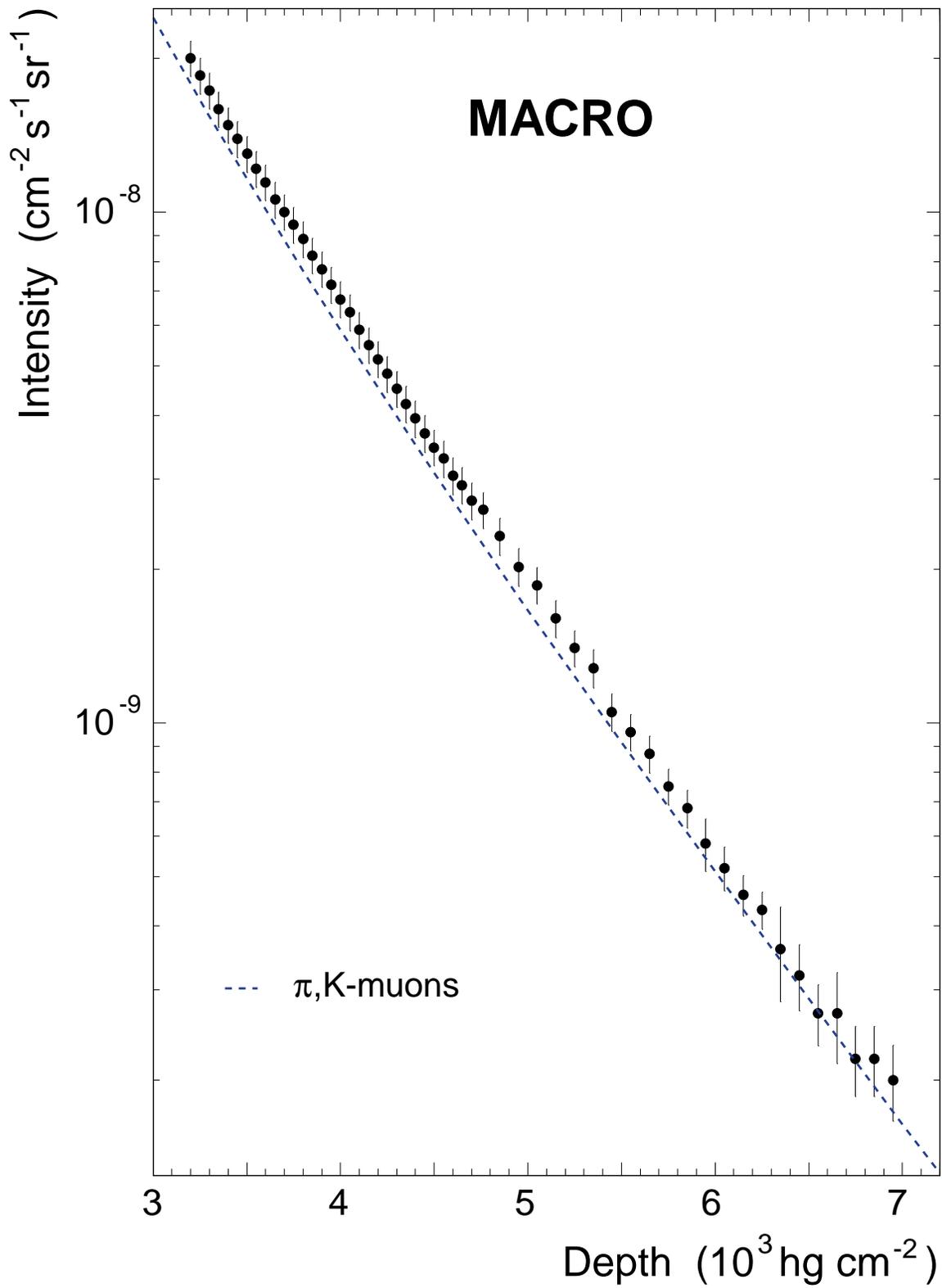,height=20.8cm}}}
\vspace{1.0cm}
\protect\caption[Muon DIR from the MACRO underground experiment]
                {Muon intensity vs standard rock thickness measured
                 in the experiment MACRO~\protect\cite{MACRO95}.
                 The curve is for the calculated $\pi,K$-muon DIR.
\label{fig:MACRO}}
\end{figure}

\clearpage
\begin{figure}[!h]
\center{\mbox{\epsfig{file=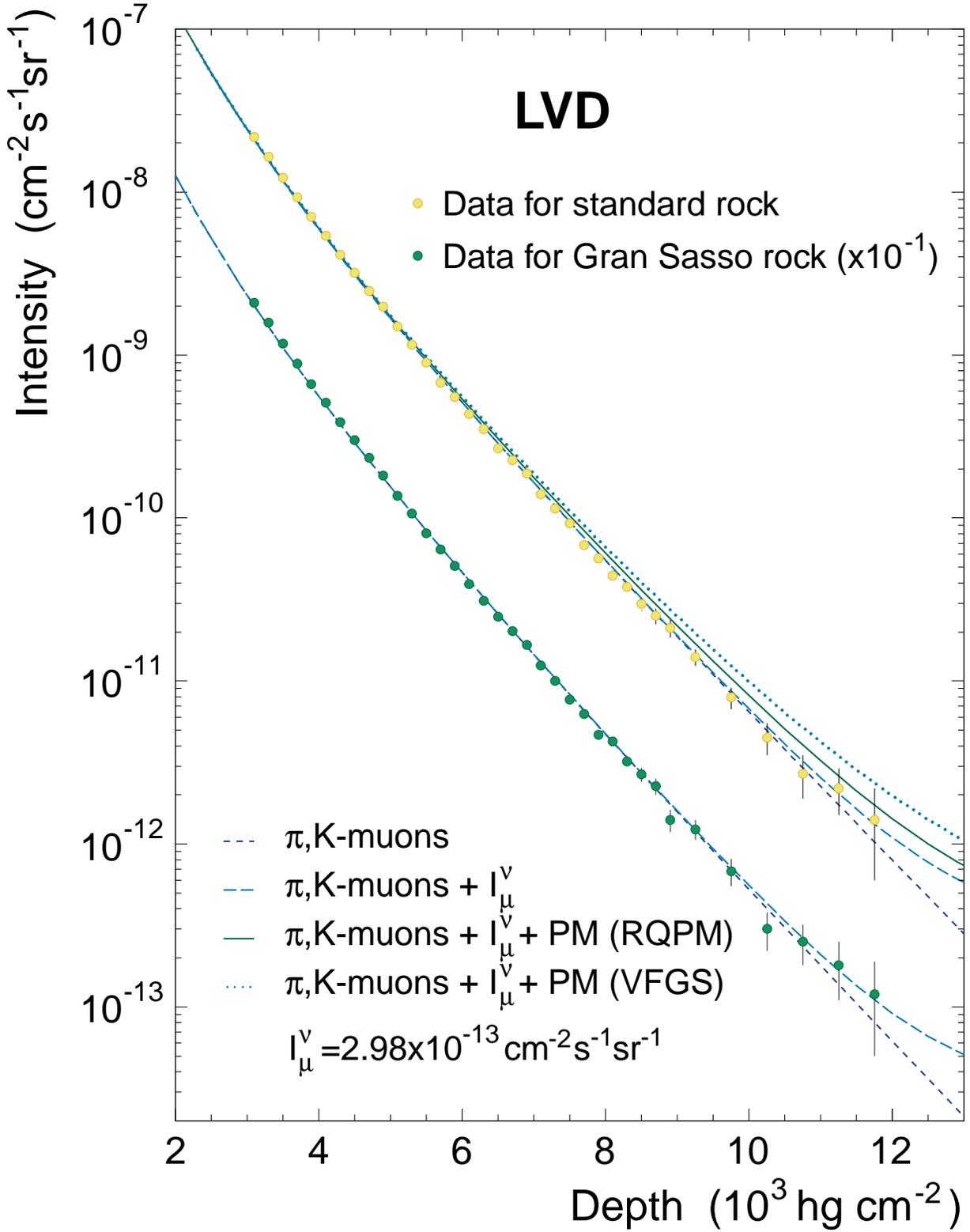,height=20.8cm}}}
\vspace{1.0cm}
\protect\caption[Muon DIR from the LVD underground experiment]
                {Muon intensity vs standard rock and Gran Sasso
                rock thickness~\protect\cite{LVD98}.
                The notation of the curves is the same as in
                 Figure~\protect\ref{fig:Frejus} but with the
                 neutrino-induced muon contribution
                 ${\cal I}_\mu^\nu$ deduced in
                 Ref.~\protect\cite{LVD95}.
\label{fig:LVD}}
\end{figure}

\clearpage

As Figure~\ref{fig:MACRO} suggests, in the range from 3200 to about
6000 hg/cm${}^2$ the MACRO data are systematically in excess of our
predicted muon intensity by about (8--10)\,\%, what is beyond the
total systematic error estimated in Ref.~\cite{MACRO95}. This fact
seems to be in dramatic contradiction with the sea-level muon
spectrum reconstructed from the MACRO underground data under
discussion (see Figure~\ref{fig:Dif}). Indeed, our sea-level
spectrum  of the $\pi,K$ muons is in good agreement with the MACRO
fit~(\ref{MACROfit}) of the sea-level spectrum from 600--700 GeV up
to 6--7 TeV and it {\em stands out above} the MACRO fit at higher
energies. However, as noted in Section~\ref{sec:SLM}, the overall
systematic error of the fit~(\ref{MACROfit}) is large enough
to explain this contradiction, at least formally.

At all depths, the LVD data are in excellent agreement with our
calculations for the conventional muon DIR. Therefore, the data
favor the models of charm production which predict a very low PM
contribution (QGSM, pQCD, DPM). However, the RQPM cannot be excluded
by the LVD result as yet. These conclusions concur with the
conclusions of Ref.~\cite{LVD97}. The consistency of the data with
our calculations for standard and Gran Sasso rocks provides an
important confirmation for the correctness of the new conversion
procedure to standard rock used in the LVD analysis~\cite{LVD98}.

\vspace{10mm}

\protect\subsection{Underwater data}

Some problems of the underground muon experiments can be overcome by
measurements underwater (and ``underice'') owing to unlimited (in
principle) detection volume, uniformity and well known composition of
the matter overburden.

We present the total (to our knowledge) assemblage of underwater data
in Fig.~\ref{fig:Water}. The measurements with compact closed
installations were performed in Suruga-bay, West Pacific (Higashi
{\em et al.}~\cite{Higashi66}), in Lake Geneva (Rogers and
Tristam~\cite{Rogers84}), in the Atlantic Ocean, Black, Mediterranean,
and Caribbean Seas during several expeditions of research ships
(Davitaev {\em et al.}~\cite{Davitaev70-74}, Fyodorov
{\em et al.}~\cite{Fyodorov85}). The measurements with open detectors
(strings with phototubes), the prototypes of future large-scale
neutrino telescopes, were performed in the Pacific Ocean off the West
coast of the island of Hawaii in 1987 (the DUMAND Short Prototype
String, Babson {\em et al.}~\cite{DUMAND90}), in the Mediterranean
Sea a short way off Pylos, during three expeditions in 1989, 1991
and 1992 (the NESTOR prototypes, Anassontzis
{\em et al.}~\cite{NESTOR93}), in Lake Baikal during two expeditions
in 1992 and 1993 (the stationary prototypes of the underwater
neutrino telescope NT-200, Belolaptikov
{\em et al.}~\cite{Baikal93,Baikal95}), and at the South Pole with
AMANDA (AMANDA-B4 experiment~\cite{AMANDA99}).

Our calculation for the $\pi,K$-muon DIR was done for sea water
with $\langle Z \rangle = 7.468$, $\langle A \rangle = 14.87$,
$\langle Z/A \rangle = 0.5525$, $\langle Z^2/A \rangle = 3.770$ and
$\langle \rho \rangle = 1.027$~g/cm${}^3$. At $h \lesssim 7$ km, the
difference with the DIR for pure $H_2O$ is less than 1\,\% and can
be neglected as compared to the theoretical and experimental
uncertainties. There are two predictions in Fig.~\ref{fig:Water}:
upper curve corresponds to muon threshold energy of 1 GeV and lower
one corresponds to 20 GeV.

At shallow depths (to $175$ m) there are two measurements with very
good statistics (Higashi {\em et al.}~\cite{Higashi66,Rogers84},
Rogers and Tristam~\cite{Rogers84}), but the results of Higashi
{\em et al.} are (except for the inclined data points at 105 m)
lower by 15 to 30\,\% than the result of Rogers and Tristam.
According to Ref.~\cite{Rogers84}, one reason for the discrepancy is
believed to be as follows. Higashi {\em et al.} normalized their data
to an intensity derived from earlier underground measurements and
measurements of the sea-level muon spectrum. The intensity chosen for
the normalization is not quoted in Ref.~\cite{Higashi66}, but was
almost certainly too low. Our prediction is in excellent agreement
with the absolute intensity obtained by Rogers and Tristam.
This provides good support of our nuclear cascade model at low
energies. However, the absolute measurements of Davitaev
{\em et al.}~\cite{Davitaev70-74} are systematically lower than our
prediction at $h\lesssim1$ km.

As for greater depths, $(1\div4)$ km, it can be concluded that
our prediction is in tolerable agreement with the data from the
DUMAND and NESTOR prototypes as well as with the data of Fyodorov
{\em et al.}; the discrepancy with a few specific data points is
within $(1-1.5)\sigma$ and is compatible with the overall data
scattering. The data of the Baikal
Collaboration~\cite {Baikal93,Baikal95} and the AMANDA
Collaboration~\cite {AMANDA99} are in very good
agreement with our curve.

As is evident from the foregoing, the present-day state of the
large-scale underwater projects does not permit to compete with the
underground detectors as yet. In particular, the (slant) depths
explored by the present-day underwater experiments are too small
to get useful information on the PM flux. It is hoped that the
situation will change in the immediate future.

\clearpage

\begin{figure}[!h]
\center{\mbox{\epsfig{file=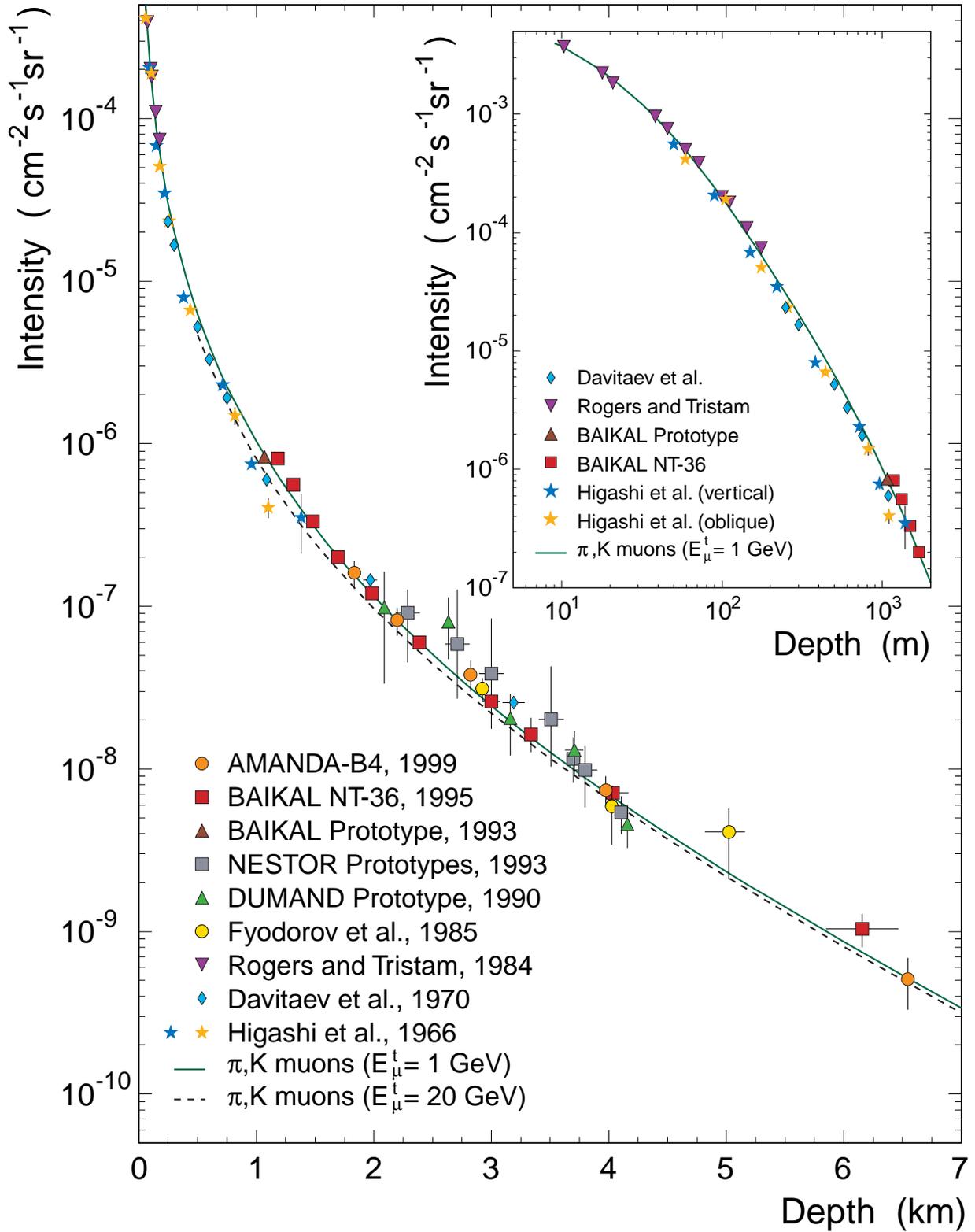,height=20.8cm}}}
\vspace{1.0cm}
\protect\caption[Muon DIR from the world underwater experiments]
                {Muon intensity vs depth in water. The data of
                 underwater experiments are from
                 Refs.~\protect\cite{Higashi66,Davitaev70-74,%
                 Rogers84,Fyodorov85,DUMAND90,NESTOR93,Baikal93,%
                 Baikal95,AMANDA99}. The curves are for the calculated
                 $\protect{\pi,K}$-muon DIR.
\label{fig:Water}}
\end{figure}

\clearpage

\protect\section{Conclusions}
\label{sec:Concl}

In this work we have attempted to study the vertical flux of
high-energy cosmic-ray muons ``from top to bottom'', that is from
the primary spectrum of cosmic rays to underground/water muon
intensity. Based upon the comparison of our calculations with
the present-day ground-level, underground and underwater
measurements, we have reached the following conclusions.

Below 1 TeV/c, the spread of the data on the vertical sea-level muon
spectra (differential and integral) measured in different experiments
runs up to about 50\,\% and it is as much as (25--30)\,\% even among
the data of absolute measurements. Our calculations in this momentum
range most closely fit the absolute 1984 data from the Nottingham
magnetic spectrograph~\cite{Rastin84}. Below 5--6 TeV/c, they agree
with the indirect data deduced in most of the underground experiments
(Artyomovsk~\cite{Collapse85}, Baksan~\cite{BNO90,BNO92},
MACRO~\cite{MACRO95}, Fr\'{e}jus~\cite{Frejus94}, MSU~\cite{MSU94})
{\em without reference to a charm production model}.

All available indirect sea level data (Baksan, KGF, Frejus, MSU) show
flattering of the sea-level muon spectrum at energies above
$\sim20$~TeV. The basic conclusion of
Refs.~\cite{BNO90,KGF90a,Frejus94,MSU94} is that this flattering is
due to the charm production in the atmosphere. This conclusion can
be deduced from the experimental and theoretical results
summarized in Figures~\ref{fig:Dif}b and \ref{fig:Int}b. First
one should note that almost all experimental works use a power law
muon spectrum with the spectral index similar (more or less) to our
one at low energies. Further, they assume this index energy
independent for conventional muons. According to these assumptions, a
prompt muon contribution is the main cause of the flattering.


However, the analysis shows definitely that the sea level data in
the aggregate cannot be quantitatively described by a single charm
production model. What this means is that the results of different
experimental groups are in rather poor agreement with one another.
Essentially all data on the differential spectrum shown in
Figure~\ref{fig:Dif}b lead to the conclusion about very high charm
production rate, as is predicted by the VFGS model or even higher.
At the same time, the data on the muon integral spectrum from
different groups require different rates of charm production: the
KGF data~\cite{KGF90a} do not contradict to the QGSM or RQPM, while
the Baksan data clearly favor the VFGS model.

At the depths from about 30 m w.e. up to 6--7 km w.e., essentially
all underground data on the muon DIR correlate with each other and
with the predicted intensity for conventional ($\pi,K$) muons,
to within 10\,\%. Hence it follows that our nuclear cascade model is
valid with the same precision from about 8 GeV up to 4--5 TeV of muon
energy at sea level, i.e. up to about 100 TeV/nucleon in the primary
spectrum. This precision is distinctly better than one might
expect with the current uncertainties in the input parameters
(including the primary spectrum model and the muon--matter
interaction cross sections). It is important that the underground
data at $h < 6-7$ km w.e. do more than correlate well, but also
have a very good statistical accuracy. Therefore, they may be of
utility, among other things, for a normalization of the atmospheric
neutrino flux in the range of neutrino energies most essential
for $\nu$-induced muons.

The present-day world underwater results, though moderately detailed,
provide a very important check upon the accuracy of the underground
experiments, since they are free of the uncertainties in the density
and composition of the matter overburden. The data obtained with the
prototypes of future large underwater neutrino telescopes,
especially with the Baikal Neutrino Telescope and with AMANDA,
are in good agreement with the underground data and with the
present calculations.

The situations with the underground data at $h \gtrsim 7$~km w.e. is
unsatisfactory in the same sense that it is with the ground-level
data at high energies. The data from KGF~\cite{KGF90a} and also
from Baksan~\cite{BNO87}
(measurements at $\vartheta=50^\circ-70^\circ$) demonstrate clear
{\em excess} over the predicted $\pi,K$-muon curve providing fair
indication of PM production.  However, our calculations show that the
KGF and Baksan sea-level data (Figures~\ref{fig:Dif},~\ref{fig:Int})
are inconsistent with their own (source) depth--intensity relations
(see Figures~\ref{fig:KGF},~\ref{fig:Baksan}). Namely, the Baksan DIR
is well explained by the RQPM rather than the VFGS model and the KGF
depth-intensity curve is rather well described by the VFGS model. On
the other hand, the corresponding sea--level spectra
(Figure~\ref{fig:Int}b) show appreciably higher charm production
rate in the case of Baksan and lower one in the case of KGF.
Probably, these inconsistencies arise due to some uncertainties in
the conversion procedures.

The data points from ERPM~\cite{ERPM,Crouch87} and especially from
NUSEX~\cite{NUSEX90} are {\em under} the $\pi,K$-muon curve. The
data of all the other underground experiments are between the
extremes represented by the Baksan/KGF and NUSEX. In particular,
the most recent LVD data~\cite{LVD98} are in good agreement
with our $\pi,K$-muon DIR and therefore they favor a low charm
production rate (as predicted by the QGSM or lower).

In closing, no undisputed conclusion about PM production can be
extracted from the world underground data.  Considering that the
statistical significances of the underground results at great depths
are comparable and quite tolerable, this situation is attributable to
the fact that certain of the experiments discussed have unrecorded
systematic uncertainties.

\acknowledgments

We acknowledge useful discussions with A.~A.~Lagutin,
P.~Le~Coultre, V.~A.~Kudryavtsev, \fbox{B.~Monteleoni},  P.~Papini,
W.~Rhode, L.~Rossi, O.~G.~Ryazhskaya, R.~Wischnewski, and
G.~T.~Zatsepin. We are grateful to Yu.~M.~Andreyev, N.~Ito,
E.~A.~Osipova, and G.~Sartorelli  for making available to us the
data of the Baksan, KGF, MSU, and LVD experiments.
We are thankful to Ch. Spiering and St. Hundertmark
for giving us details on muon zenith-angle distribution
and depth-intensity relation of the  AMANDA-B4 experiment.
We especially thank E.~S.~Zaslavskaya who had taken an active
part in the initial stage of this work. E.\,B., V.\,N., and S.\,S.
are indebted to INFN, Sezione di Firenze, where the part of this
work was made, for its hospitality and support.
The research of V.\,N., T.\,S., and S.\,S. was supported in part
by Ministry of Education of Russian Federation under Grants
No.~2-728 within the framework of scientific program
``Universities of Russia -- Basic Researches''.

\protect\appendix
\protect\section{Spectra of muons from inclusive decay of
                 $\protect{D}$ and $\protect{\Lambda_c}$}
\label{app:Decay}

\protect\subsection{$\protect{D\rightarrow\mu\nu_\mu X}$ decay}
\[
F^\mu_D(x) = \frac{1}{Z_D}\left[\frac{1}{6}(1-r_D^2)
\left(1-5r_D^2-2r_D^4\right)+r_D^4x-\frac{1}{2}(1-2r_D^2)x^2
+\frac{1}{3}x^3+r_D^4\ln\left(\frac{1-x}{r_D^2}\right)\right],
\]
\[
Z_D = \frac{1}{12}\left(1-r_D^4\right)
                         \left(1-8r_D^2+r_D^4\right)-r_D^4\ln r_D^2,
\]

Here $r = s_X^{\rm eff}/m_D^2$ and $s_X^{\rm eff}$ is the effective
invariant mass square in the decay. The best fit to the data on the
differential and total decay rates is achieved using
\[
\sqrt{s_X^{\rm eff}} = \left\{\begin{aligned}
0.63\;\text{GeV} \quad & \text{for $D^+$ and $D^-$}, \\
0.67\;\text{GeV} \quad & \text{for $D^0$ and $\overline{D}{}^0$}.
\end{aligned}\right.
\]
Therefore $r_{D^\pm} \approx 0.337$ and
$r_{D^0,\overline{D}{}^0} \approx 0.359$.

\protect%
\subsection{$\protect{\Lambda_c\rightarrow\mu\nu_\mu X}$ decay}
\[
F^\mu_{\Lambda_c}(x) = \frac{1}{Z_{\Lambda_c}}
                \sum_{1\leq i\leq j\leq 3}f_if_j\ae_{ij}(x), \qquad
Z_{\Lambda_c} = \sum_{1\leq i\leq j\leq 3}f_if_j\overline{\ae}_{ij},
\]
\begin{eqnarray*}
\ae_{11}(x) & = & \frac{11}{6}-3r_\Lambda-\frac{21}{2}r_\Lambda^2
                 -12r_\Lambda^3+\frac{9}{2}r_\Lambda^4+9r_\Lambda^5
                 +\frac{7}{6}r_\Lambda^6+6r_\Lambda^3(2+r_\Lambda)x
                 -\frac{3}{2}\left(3-2r_\Lambda
                 -5r_\Lambda^2\right)x^2                          \\
            &   &+\frac{8}{3}x^3+\frac{r_\Lambda^4\left(3+6r_\Lambda
                 +r_\Lambda^2\right)}{1-x}-\frac{r_\Lambda^6}
                  {(1-x)^2}+r_\Lambda^3\left(12+9r_\Lambda
                 +6r_\Lambda^2-r_\Lambda^3\right)
                  \ln\left(\frac{1-x}{r_\Lambda^2}\right),        \\
\ae_{12}(x) & = & \frac{5}{3}-6r_\Lambda-9r_\Lambda^2-24r_\Lambda^3
                 +3r_\Lambda^4+18r_\Lambda^5-\frac{5}{3}r_\Lambda^6
                 +24r_\Lambda^3x
                 -3\left(1-2r_\Lambda-r_\Lambda^2\right)x^2       \\
            &   &+\frac{4}{3}x^3+\frac{2r_\Lambda^4\left(3
                 +6r_\Lambda+r_\Lambda^2\right)}{1-x}
                 -\frac{2r_\Lambda^6}{(1-x)^2}+2r_\Lambda^3\left(12
                 +3r_\Lambda+6r_\Lambda^2-r_\Lambda^3\right)
                  \ln\left(\frac{1-x}{r_\Lambda^2}\right),        \\
\ae_{13}(x) & = &-\frac{1}{3}+3r_\Lambda^2-3r_\Lambda^4
                 -\frac{17}{3}r_\Lambda^6-12r_\Lambda^4x
                 +3(1-3r_\Lambda^2)x^2-\frac{8}{3}x^3                                   \\
            &   &+\frac{2r_\Lambda^4\left(3+r_\Lambda^2\right)}{1-x}
                 -\frac{2r_\Lambda^6}{(1-x)^2}
                 -2r_\Lambda^4\left(3+r_\Lambda^2\right)
                 \ln\left(\frac{1-x}{r_\Lambda^2}\right),         \\
\ae_{22}(x) & = & \frac{1}{(1+r_\Lambda)^2}\left[\frac{2}{3}
                 -\frac{4}{3}r_\Lambda-\frac{23}{3}r_\Lambda^2
                 -24r_\Lambda^3-21r_\Lambda^4
                 +\frac{40}{3}r_\Lambda^6+\frac{22}{3}r_\Lambda^7
                 -\frac{4}{3}r_\Lambda^8\right.                   \\
            &   & \left.+2r_\Lambda^3\left(6+9r_\Lambda+6r_\Lambda^2
                 -r_\Lambda^3\right)x-\frac{3}{2}\left(1-4r_\Lambda^2
                 -4r_\Lambda^3+r_\Lambda^4\right)x^2+\frac{4}{3}
                  \left(1+r_\Lambda-r_\Lambda^2\right)x^3\right.  \\
            &   &\left. -\frac{1}{2}x^4+\frac{r_\Lambda^4\left(3
                 +12r_\Lambda+14r_\Lambda^2+8r_\Lambda^3
                 +r_\Lambda^4\right)}{1-x}-\frac{r_\Lambda^6(1
                 +r_\Lambda)^2}{(1-x)^2}\right.                   \\
            &   &\left.+r_\Lambda^3\left(12+21r_\Lambda
                 +24r_\Lambda^2+10r_\Lambda^3+4r_\Lambda^4
                 -r_\Lambda^5\right)\ln\left(\frac{1-x}
                 {r_\Lambda^2}\right)\right],                     \\
\ae_{23}(x) & = & \ae_{13}(x),                                    \\
\ae_{33}(x) & = & \frac{11}{6}+3r_\Lambda-\frac{21}{2}r_\Lambda^2
                 +12r_\Lambda^3+\frac{9}{2}r_\Lambda^4-9r_\Lambda^5
                 +\frac{7}{6}r_\Lambda^6-6r_\Lambda^3(2-r_\Lambda)x
                 -\frac{3}{2}\left(3+2r_\Lambda
                 -5r_\Lambda^2\right)x^2                          \\
            &   &+\frac{8}{3}x^3+\frac{r_\Lambda^4\left(3-6r_\Lambda
                 +r_\Lambda^2\right)}{1-x}-\frac{r_\Lambda^6}
                  {(1-x)^2}-r_\Lambda^3\left(12-9r_\Lambda
                 +6r_\Lambda^2+r_\Lambda^3\right)
                 \ln\left(\frac{1-x}{r_\Lambda^2}\right),
\end{eqnarray*}
\begin{eqnarray*}
\overline{\ae}_{11} & = & \left(1-r_\Lambda^2\right)
                 \left(1-2r_\Lambda-7r_\Lambda^2-20r_\Lambda^3
                       -7r_\Lambda^4-2r_\Lambda^5+r_\Lambda^6\right)
-24r_\Lambda^3\left(1+r_\Lambda^2+r_\Lambda^3\right)\ln r_\Lambda,                                \\
\overline{\ae}_{12} & = & \left(1-r_\Lambda^2\right)
                 \left(1-4r_\Lambda-7r_\Lambda^2-40r_\Lambda^3
                       -7r_\Lambda^4-4r_\Lambda^5+r_\Lambda^6\right)
-24r_\Lambda^3\left(2+r_\Lambda+2r_\Lambda^2\right)\ln r_\Lambda, \\
\overline{\ae}_{13} & = & \overline{\ae}_{23} = 0,                \\
\overline{\ae}_{22} & = & \frac{1}{(1+r_\Lambda)^2}\left[\frac{2}{5}
                 -r_\Lambda-6r_\Lambda^2-28r_\Lambda^3-32r_\Lambda^6
                 +28r_\Lambda^7+6r_\Lambda^8+r_\Lambda^9\right.   \\
                    &   & \left. \;\;\;\;\;\;\;\;\;\;\;\;\;\;\;\;\;
          -\frac{2}{5}r_\Lambda^{10}-24r_\Lambda^3\left(1+2r_\Lambda
                 +3r_\Lambda^2+2r_\Lambda^3+r_\Lambda^4\right)
                        \ln r_\Lambda\right],                     \\
\overline{\ae}_{33} & = & \left(1-r_\Lambda^2\right)
                 \left(1+2r_\Lambda-7r_\Lambda^2+20r_\Lambda^3
                       -7r_\Lambda^4+2r_\Lambda^5+r_\Lambda^6\right)
+24r_\Lambda^3\left(1-r_\Lambda-r_\Lambda^2\right)\ln r_\Lambda.
\end{eqnarray*}
Here $r_\Lambda^2 = s_X^{\rm eff}/m_{\Lambda_c}$ and $s_X^{\rm eff}$
is the effective invariant mass square. The best fit to the
data on the differential and total decay rates is achieved using
$\sqrt{s_X^{\rm eff}} = 1.27\;\text{GeV}$. Therefore
$r_\Lambda \approx 0.551$. For the decay form factors averaged
over $q^2$ we have $f_1 \approx 0.991$, $f_2 \approx 2.170$, and
$f_3 \approx 0.805$.

\clearpage
\protect\section{Muon--matter interactions at high energies}
\label{app:MuInt}

Here we present with some comments a listing of the cross sections
for the muon--matter interactions and the formula for ionization
loss. In what follows, $Z$ and $A$ are the atomic number and atomic
weight of the target nucleus; $E$ and $E'$ are the respective
energies of initial and final muons; $v$ is the fraction of energy
lost ($E' = E(1-v)$); $\alpha$ and $r_e$ are the fine structure
constant and the classical electron radius; $m_e$ and $m_\mu$ are
the electron and muon masses; $c = 1$.

\subsection{Direct $e^+e^-$ pair production}
\label{app:ssec_PairProd}

The direct pair production cross section goes roughly as $1/v^2$ to
$1/v^3$ over most of the range, $v > 0.002$ (see, for example,
Ref.~\cite{vanGinneken86}).  Because of this, the energy loss through
pair production is usually considered as continuous. Nevertheless, as
it follows from our calculations, the fluctuation effect related to
this process is not negligible and it grows in magnitude with depth.
In the present work, only a part of the direct pair production cross
section, corresponding to relatively large energy losses ($v > v_0 =
2 \times 10^{-4}$), is included into the collision integral of the
muon transport equation while the energy losses caused by the range
$v < v_0$ were treated as continuous. The exact (in the
ultrarelativistic limit) numerical results for $d\sigma_p/dv$ have
been obtained by Kel'ner and Kotov and can be found in
Ref.~\cite{Bugaev70}. In this paper, we use the following simple
approximation of the exact results:
\[
v\frac{d\sigma_p}{dv} = \frac{16}{\pi}Z(Z+1)({\alpha}r_e)^2\,F(E,v),
\]
\[
F(E,v) = \frac{1.7\times10^{-4}(v+1.05\times10^{-4})}{v(v+0.006)^2}
         \left\{1-\frac{\exp\left[-0.025\ln^2(E/m_\mu)\right]}
                                  {1+0.323\ln(10^3v)}\right\}f(E,v),
\]
where
\[
f(E,v) = \frac{1+2v}{1+k/(vE)}, \qquad k = 0.02\;\mbox{GeV},
\]
for $10^{-3}\leq v \leq 0.2$ and $f(E,v) = 1,$ for
$v_0< v <10^{-3}$ or $v > 0.2$. A weak (logarithmic) $Z$-dependence
of the exact function $F(E,v)$ is neglected in this approximation.
The accuracy of the approximation is better than 10\,\% within the
most essential interval of $v$ ($v_0 < v < 0.1$). It provides a
reasonable (a few-percent) precision for the calculated muon DIR.

\subsection{Bremsstrahlung}
\label{app:ssec_Brems}

We use the formula derived by Andreev {\em et al.}~\cite{Andreev94}
with regard to the nuclear target structure and the exact
contribution to the cross section given by atomic electrons:
\[
v\frac{d\sigma_b}{dv} = \alpha\left(2r_eZ\frac{m_e}{m_\mu}\right)^2
 \left[\left(2-2v+v^2\right)\Psi_1\left(q_{\rm min},Z\right)
-\frac{2}{3}\left(1-v\right)\Psi_2\left(q_{\rm min},Z\right)\right],
\]
\[
\Psi_{1,2}\left(q_{\min},Z\right) =
                               \Psi_{1,2}^0\left(q_{\min},Z\right)
                              -\Delta_{1,2}\left(q_{\min},Z\right),
\]
\begin{eqnarray*}
\Psi_1^0\left(q_{\min},Z\right) & = & \frac{1}{2}\left(1+\ln
\frac{m_\mu^2a_1^2}{1+x_1^2}\right)-x_1\arctan\frac{1}{x_1}+
\frac{1}{Z}\left[\frac{1}{2}\left(1+\ln\frac{m_\mu^2 a_2^2}
                 {1+x_2^2}\right)-x_2\arctan\frac{1}{x_2}\right], \\
\Psi_2^0\left(q_{\min},Z\right) & = & \frac{1}{2}\left(\frac{2}{3}+
\ln\frac{m_\mu^2 a_1^2}{1+x_1^2}\right)+2x_1^2
\left(1-x_1\arctan\frac{1}{x_1}+\frac{3}{4}
                            \ln\frac{x_1^2}{1+x_1^2}\right)       \\
                     & + & \frac{1}{Z}\left[\frac{1}{2}
\left(\frac{2}{3}+\ln\frac{m_\mu^2 a_2^2}{1+x_2^2}\right)+
2x_2^2\left(1-x_2\arctan\frac{1}{x_2}+\frac{3}{4}\ln\frac{x_2^2}
                                    {1+x_2^2}\right)\right],
\end{eqnarray*}
\begin{eqnarray*}
\Delta_1\left(q_{\min},Z\right) & = & \ln\frac{m_\mu}{q_c}+
                       \frac{\zeta}{2}\ln\frac{\zeta+1}{\zeta-1}, \\
\Delta_2\left(q_{\min},Z\right) & = & \ln\frac{m_\mu}{q_c}
            +\frac{\zeta}{4}\left(3-\zeta^2\right)\ln\frac{\zeta+1}
                             {\zeta-1}+\frac{2 m_\mu^2}{q_c^2},
\end{eqnarray*}
\[
q_{\min}\simeq\frac{m_\mu^2 v}{2E(1-v)}, \quad x_i = a_i q_{\min},
\]
\[
a_1   = \frac{111.7}{Z^{1/3}m_e}, \quad
a_2   = \frac{724.2}{Z^{2/3}m_e}, \quad
\zeta = \sqrt{1+\frac{4m_\mu^2}{q_c^2}}, \quad
q_c   = \frac{1.9 m_\mu}{Z^{1/3}},
\]

From the foregoing equations it can be shown that, in the limit of
complete screening, i.e. for
\[
\gamma_Z(v,E)\equiv\frac{200 q_{\min}}{m_eZ^{1/3}}\simeq
\left(\frac{11}{Z}\right)^{1/3}
\left(\frac{1\,{\rm TeV}}{E}\right)\frac{v}{1-v}\ll 1
\]
(where $\gamma_Z$ is the degree of screening), the bremsstrahlung
cross section is a function of the variable $v$ only (scaling).
However for values of $v$ which are not too small (namely, at
$1-v \ll 1$) complete screening occurs only at very high energies,
$E \sim 10$ TeV. At lower energies, the cross section grows
logarithmically with $E$. It should be noted that the same estimate
($E \sim 10$ TeV) is also true as a limit of full screening for the
pair production cross section.

Now, let us discuss briefly the corrections to the Born
approximation. Recently, it was shown~\cite{Andreev97} that there are
two such corrections: in the region of small ($q \sim m_\mu^2/E$) and
large ($q \sim m_\mu$) momentum transfers which correspond to large
and small impact parameters, respectively. The correction from the
first region (large $q$) is just the same as in the case of electron
bremsstrahlung (it is the well-known correction of Davies, Bethe, and
Maximon~\cite{Davies54}). The essentially new result of
Ref.~\cite{Andreev97} is that the second correction has the opposite
sign and nearly compensates the first one. As a result, the Born
approximation formulas have rather good accuracy for the muon
bremsstrahlung even for very heavy targets. In the case of interest
for the present study, the corrections under discussion prove to be
completely negligible.

\subsection{Photonuclear Interaction}
\label{app:ssec_PhotoNucl}

For the photonuclear interaction of muon we use the generalized
vector dominance model (GVDM)~\cite{Bezrukov80}. Within the
vector meson dominance hypothesis, the differential cross section for
muon photonuclear interaction, $d\sigma_n/dv$, is expressed in terms
of the total cross section for virtual photon absorption by nucleons
and nuclei. The GVDM adequately describes the features of these
cross sections in the diffraction region (low 4-momentum transfers,
$Q^2$, and large photon energies, $\nu$): a growth with energy of the
cross section for nucleon photoabsorption and shadowing effects in
nuclear photoabsorption. According to Ref.~\cite{Bezrukov80},
\begin{eqnarray*}
\frac{d\sigma_n}{d v} &=& \frac{\alpha}{8\pi}A\sigma_{\gamma p}(\nu)
                 v\left\{H(v)\ln\left(1+\frac{m_2^2}{t}\right)
                 -\frac{2m_\mu^2}{t}\left[1-\frac{0.25m_2^2}{t}
                 \ln\left(1+\frac{t}{m_2^2}\right)\right]\right.\\
                      &+& \left.G(z)\left[H(v)\left(\ln
                 \left(1+\frac{m_1^2}{t}\right)
                 -\frac{m_1^2}{m_1^2+t}\right)-\frac{2m_\mu^2}{t}
              \left(1-\frac{0.25m_1^2}{m_1^2+t}\right)\right]\right\}.
\end{eqnarray*}
Here $\nu = v E$ is the virtual photon energy and
\[
H(v)=1-\frac{2}{v}+\frac{2}{v^2}, \qquad
G(z)=\frac{9}{z}\left[\frac{1}{2}+\frac{(1+z)e^{-z}-1}{z^2}\right],
\]
\[
z     = 0.00282A^{1/3}\sigma_{\gamma p}(\nu),\quad
t     = \frac{m_\mu^2 v^2}{1-v}, \quad
m_1^2 = 0.54\;{\rm GeV}^2, \quad
m_2^2 = 1.80\;{\rm GeV}^2.
\]

The differential cross section is proportional to the total cross
section for absorption of a real photon of energy $\nu=s/2m_N=vE$
by a nucleon, $\sigma_{\gamma N}$. In the present calculations, we
adopt the Regge-type parametrization for $\sigma_{\gamma N}$
from Ref.~\cite{Donnachie92},
\begin{equation}\label{Regge}
\sigma_{\gamma N} =
            \left[67.7s^{0.0808}+129s^{-0.4525}\right]\;\mu{\rm b}
\end{equation}
($s$ in GeV$^2$). This model gives the best fit to the accelerator
data. At $\sqrt{s} = 200$ GeV it differs by about $9$\,\%
from the parametrization of Ref.~\cite{Bezrukov80} used in our
previous calculations,
\begin{equation}\label{GVDM}
\sigma_{\gamma N} =
             \left[114.3+1.647\ln^2(0.0213\nu)\right]\;\mu{\rm b}
\end{equation}
($\nu$ in GeV). The disparity in the DIR resulting from the
difference in the models (\ref{Regge}) and (\ref{GVDM}) for
$\sigma_{\gamma N}$, is completely negligible up to $\sim10$~km w.e.
(independent of rock composition) and it is small at greater depths.
Namely, the muon intensities calculated with the use of the
parametrization (\ref{Regge}) exceed those calculated with the
parametrization (\ref{GVDM}) by 1.2, 2, 3, and 5\% at respectively
12, 14, 16, and 18~km w.e. of standard rock what is of no importance
for the interpretation of the current underground data.

The growth of $\sigma_{\gamma N}$ with the photon energy causes
$d\sigma_n/dv$ to depend on the muon energy, $E=\nu/v$. The
shadowing effect of nucleons inside a target nucleus gradually
compensating the energy dependence of $\sigma_{\gamma N}$, but a
logarithmic growth of $d\sigma_n/dv$ quantitatively remains up to
$E\sim10$~TeV and possibly in the asymptotics.

One should note here that, within the VDM approach, the growth of
$\sigma_{\gamma N}$ with the photon energy is resulted by the growth
of the hadron-nucleon cross section. However, non-VDM corrections to
$\sigma_{\gamma N}$ may, in principle, be not negligible. A part of
these corrections is caused by the pQCD  (``minijet'') contribution
to the $\gamma N$ total cross section being determined by the proton
perturbative structure function (rather than an intermediate vector
meson exchange). This correction is small because the pQCD cross
section is dominated by the ``VDM photon''~\cite{Forshaw91-92}. The
second non-VDM correction is dominated by direct photon-proton
reaction; it corresponds to the so-called ``unresolved photon''. The
magnitude of this correction strongly depends on the poorly known
behavior of the gluon structure function in the proton, $g_p(x)$, at
small $x$ and, of course, on the usual QCD parameter $p_T^{\min}$.
For example, if $g_p(x) \propto x^{-1.5}$, the correction from the
$\gamma g \rightarrow q\overline{q}$ subprocess behaves with the
photon energy as $\sqrt{s}$~\cite{Collins91} and depends on
$p_T^{\min}$ as $\left(p_T^{\min}\right)^{-3}$.

Available cosmic-ray data obtained with underground
detectors~\cite{Zatsepin89} (for $\nu\lesssim10$~TeV) and with EAS
arrays~\cite{Dumora92} (up to $10^3-10^4$ TeV) is in agreement with
formulas (\ref{Regge}) or (\ref{GVDM}). Nonetheless, the photonuclear
interaction rests one of the sources of uncertainties at very
high muon energies and, at the same time, a noteworthy subject for
investigation with the future large-scale underwater telescopes.
Luckily, this uncertainty is of little importance for the muon DIR.


\subsection{Ionization energy loss}
\label{app:ssec_Ion}

The ionization loss of a muon of energy $E$ is given by the
Bethe-Bloch stopping-power formula corrected to the density
effect~\cite{Sternheimer71} (see also Ref.~\cite{Lohmann85}),
\[
-\left(\frac{dE}{dx}\right)_{\rm ion}=\frac{C_0}{\beta^2}\frac{Z}{A}
\left[\ln\left(\frac{2 m_e p^2 W_{\max}}{m_\mu^2 I_Z^2}\right)
+\frac{W_{\max}^2}{4 E^2}-2\beta^2-\delta-U\right].
\]
Here $C_0 = 0.1535$ MeV\,g$^{-1}$\,cm$^2$, $p$ is the muon momentum,
$\beta = p/E$ is the muon velocity,
\[
W_{\max} = \frac{2m_e p^2}{m_\mu^2+m_e^2+2m_e E}
\]
is the maximum energy transfer from the muon to an atomic electron.
The function $\delta$ is the density-effect correction. Its
numerical values are given by the Sternheimer's fit
formula~\cite{Sternheimer71},
\[
\delta=\theta(X-X_0)\left[4.6052X+a\theta(X_1-X)(X_1-X)^m+C\right],
\]
where $\theta$ is the step function ($\theta(x) = 0$ at $x \leq 0$
and $\theta(x) = 1$ at $x > 0$), $X = \log(p/m_\mu)$. The values
$X_0$, $X_1$, $a$ and $m$ depend on the substance (for the specific
values we used the data of Ref.~\cite{Lohmann85} with some
modifications in the cases of Baksan and Kolar rocks);
$C = -\left[2\ln\left(I_Z/h\nu_p\right)+1\right]$, where $I_Z$ is
the mean excitation energy, $h\nu_p = 28.816\sqrt{\rho Z/A}$ is the
plasma energy (in eV) and $\rho$ is the density of the medium (in
g/cm$^3$).  The shell-correction term, $U = 2 C_K/Z+2C_L/Z+\ldots$,
is generally negligible for the energies at which the density-effect
correction $\delta$ is significant.


\begin{references}
\bibitem{footnoteKl3effect}
         It is pertinent to note that the $K_{\mu3}$ decays
         contribute significantly to the atmospheric neutrino flux;
         and what's more, the inclusion of $K_{\ell3}$ form
         factors noticeably affects the neutrino flavor ratio at
         high energies (see Ref.~\cite{Kl3}).
\bibitem{footnoteChargeRatio}
         Unfortunately, this fact is not exploitable for an
         experimental study of the prompt muon contribution. The
         analogous effect takes place for prompt neutrinos. The
         fitting formulas for the prompt neutrino flavor ratio are
         \[
         \frac{{\cal D}_{\nu}^{\rm pr}}{{\cal D}_{\overline{\nu}}
         ^{\rm pr}} = \left\{\begin{aligned}
         0.892-0.006\log^2\left(E/E_R\right)
                                         & \quad \text{for RQPM}, \\
         1.420+0.0025\left(E/E_R\right)^{1.08}
                                         & \quad \text{for QGSM},
         \end{aligned}\right.
         \]
         where $E_R=10$~TeV. These formulas are valid in the energy
         range $3\div10^3$~TeV at all zenith angles with an accuracy
         better than 2\,\%.
\bibitem{footnoteHorFlux}
         We have in mind the data for the vertical muon flux. Notice
         that our calculations for the near-horizontal flux are in
         good agreement with the direct data from the two large
         magnetic spectrometers MUTRON~\protect\cite{MUTRON84} and
         DEIS~\protect\cite{DEIS85} up to about 20~TeV/c
         irrespectively from the PM flux model (the near-horizontal
         muon flux is considerably less sensitive to the PM
         contribution than is the vertical spectrum).
\end{references}
\end{document}